\documentclass[12pt,a4paper]{article}
\usepackage[latin1]{inputenc}
\usepackage{dfttrob}  % removing this line and \dfttnum below results
\usepackage{times}
\usepackage[numbers,sort&compress]{natbib}
\usepackage{latexuseful2e}   % some useful definitions (not all used)
\usepackage{amsmath}
\usepackage{amssymb}
\usepackage{graphicx}
\usepackage{color}
\usepackage[ps2pdf,breaklinks=true,linktocpage=true,bookmarksnumbered=true,
pdftitle={Clan structure analysis and QCD parton showers in
multiparticle dynamics},
pdfsubject={An intriguing dialog between theory and experiment},
pdfauthor={A. Giovannini, R. Ugoccioni},
pdfkeywords={multiplicity distributions, correlations, collective 
variables}]{hyperref}

\dfttnum{DFTT 12/2004\\May 24, 2004}

\newcommand{\bFB}{b_{\text{FB}}}
\newcommand{\bFBxx}[1]{b_{\text{FB},\text{#1}}}

\begin{document}

\title{\bfseries Clan structure analysis
  and QCD parton showers\\
  in multiparticle dynamics.\\[0.1cm]
  An intriguing dialog between theory and experiment}
\author{\large A. Giovannini and R. Ugoccioni\\
 \itshape Dipartimento di Fisica Teorica, Universit\`a di Torino\\
	\itshape and INFN, Sezione di Torino,
	\itshape via Pietro Giuria 1, 10125 Torino, Italy}
\maketitle

\begin{abstract}
This paper contains a review of the main results of a search
of regularities in collective variables properties in multiparticle
dynamics, regularities which can be considered as manifestations of the 
original simplicity suggested by QCD. 
The method is based on a continuous dialog between experiment
and theory.
The paper follows the development of this research line, from its beginnings
in the seventies to the current state of the art, discussing how it
produced both sound interpretations of the most relevant 
experimental facts and intriguing perspectives for new physics signals in 
the TeV energy  domain.
\end{abstract}

\newpage

\section{INTRODUCTION}

The structure of the vacuum and confinement are still unsolved problems of 
Quantum Chromodynamics (QCD)
after many years from its introduction as the theory of strong interactions.
Sound experimental informations in order to approach the two problems can 
come from hadronic spectrum and multiparticle production data.
Attention in the present work is focused  on multiparticle production 
and concerns mainly collective variables properties of final charged particles
in full phase-space and restricted rapidity intervals, i.e., of collective
variables properties  in those awkward sectors  where perturbative QCD is hardly
applicable. 
Guiding line  is  the conviction that
complex structures which we observe at final hadron level might very well be, at
the origin of  their evolution,  elementary and have simple properties. These
characteristics  in the detected observables are revealed by  the 
occurrence of regularities which are expected to contain signals of the
original simplicity and to be expressed in terms of the minimum number of
physical parameters. 

This research line, along the years, has been inspiring and quite
successful for a phenomenological description, based on essentials of
QCD, of the main experimental facts in multiparticle dynamics.
This paper contains a summary of the results of this endeavour, which
might be quite stimulating in the approach to the TeV energy domain in
\pp\ and heavy ion collisions, and to the determination of their
possible substructures.

Multiparticle production has quite a long story and its understanding is 
indeed crucial  for strong interaction. 
The first observation of such events  goes  
back to  cosmic ray physics in the thirties of the 
past century:
the extraordinary and impressive fact had been the non-linearity of the 
phenomenon.

This unusual  experimental observation attracted  the attention of
many  theorists   
in the forties and early  fifties: in particular, the work of E. Fermi on the 
thermodynamical model  \cite{Fermi+Fermi:1+Fermi:2}
and of  L. Landau \cite{Landau} 
on the hydrodynamical model should be mentioned. 
Of course  the contributions   by 
J.F. Carlson and J.R. Oppenheimer \cite{Carlson}, 
H.J. Bhabha and W. Heitler \cite{Bhabha}, 
W.H. Furry \cite{Furry}, 
H.W. Lewis, J.R. Oppenheimer, S.A. Wouthuysen \cite{Oppenheimer},
together with the  pioneering work by  
N. Arley \cite{Arley:1+Arley:2} should not be forgotten.
Particular  aspects of the new experimental fact were described,  but the 
situation was considered not satisfactory from a theoretical point of view.
It was only W. Heisenberg who understood  that multiparticle 
production should be 
described in terms of a non-linear field theory of a new nuclear force
(which we call today  indeed strong interaction) 
\cite{Heisenberg:1+Heisenberg:2+Heisenberg:3+Heisenberg:4}. 

With the incoming  of the  multi-peripheral model
\cite{Fubini}, an important step 
was done in the understanding of the c.m.\  energy dependence of the average 
charged particle multiplicity in  high energy collisions in terms of a 
logarithmic function, a trend competitive with the square root rule proposed 
earlier on purely phenomenological grounds  \cite{Wataghin};  
$n$ charged particle multiplicity  
distribution (MD), $P_n$, was predicted to be 
Poissonian, when plotted vs.\ $n$,
suggesting an  independent particle production process.
Few years later (in 1966), 
P.K. MacKeown and A.W. Wolfendale noticed, in cosmic ray experiments,
remarkable violations in the dynamical  mechanism for
independent particle production,  by  observing quite large  
fluctuations of the pionization component in 
hadron showers originated   by primary hadrons at different primary 
energies \cite{Cosmic}.
They proposed to fit high energy  cosmic ray data on charged particle 
multiplicity distributions (MD's) in terms of a Negative Binomial (Pascal)
multiplicity distribution [from now on abbreviated as NB(Pascal)MD].
This  phenomenological 
distribution  is in fact characterised by an extra parameter in addition   
to the average charged multiplicity $\nbar$, i.e., the parameter  $k$
which is  linked to the dispersion $D$:  $k = \nbar^2 / (D^2 - \nbar)$.
$D^2 > \nbar$  implies  indeed  deviations from the Poissonian behaviour of the 
$n$-particle MD predicted by the multi-peripheral 
model to which the NB (Pascal) MD reduces for $k\to\infty$. 
The Authors gave also a sound phenomenological
expression for the  energy dependence of $k^{-1}$, 
showing that it is a finite number 
which increases  with the increasing of the energy of the primary hadron
toward an asymptotic constant value ($k^{-1} \approx 0. 4$). 

The discovery in the accelerator region,  in  the seventies, of 
the violations of multi-peripheral model predictions on $n$ charged particle
multiplicity distribution  in high energy  hadron-hadron collisions 
\cite{Derrick:1972} confirmed the 
cosmic ray physics findings.  The  parallel success of the 
NB (Pascal) MD   in describing in full phase-space in the accelerator region  
$n$ charged particle multiplicity distributions at different 
$p_{\text{lab}}$ 
in various collisions (53 experiments were successfully fitted
\cite{AGCim:2+AGCim:5,AGCim:3+AGCim:4}
led to guess that the distribution 
was a good candidate for representing  multi-peripheral model prediction
violations. 
Although  a germane attempt to justify the occurrence of the
distribution in terms of the so called generalised multi-peripheral 
bootstrap model \cite{AGCim:1+AGCim:6}
was quickly forgotten,  its phenomenological interest
remained in the field: the distribution was  rediscovered in
full  phase-space for non-single diffractive events  and extended to
(pseudo)-rapidity  
intervals  by UA5 Collaboration \cite{UA5:4}
at CERN  $p\bar p$ Collider c.m.\ energies, and
then successfully used  by NA 22 Collaboration \cite{NA22} at \roots{22} in 
\pp\ and $\pi^{\pm}p$ collisions and by HRS experiment \cite{HRS:1}
in \ee\ annihilation 
at \roots{29} in order to describe  $P_n$  vs.\ $n$ behaviour both in 
full phase-space and in symmetric (pseudo)rapidity intervals. 

The last two
experiments were of great importance: the first one established a low 
energy point in hadron-hadron collisions, the second one  extended to a new 
class of collisions the interest for the NB (Pascal) MD. In comparing NA22 with 
UA5 data it was found that  $\nbar$  increases and  $k$ parameter decreases in 
full phase-space as the c.m.\ energy becomes larger, whereas at fixed c.m.\ 
energy $\nbar$ and $k$ become larger  with the increase of (pseudo)rapidity 
interval. 
ISR, TASSO  and EMC collaborations data on $P_n$ vs.\ 
$n$ followed within a short time 
and confirmed the success of   their  description by means of
NB (Pascal) MD \cite{ISR:1,TASSO,EMC}.
The fact that so many experiments in a so wide energy range and
in symmetric (pseudo)-rapidity interval could  be fitted  by the
same $n$ charged particle  MD was  considered not accidental.
The impression was that one was facing an approximate  universal regularity.
 
A suggestive  interpretation of the regularity was then  proposed: it  
implied that the dynamical mechanism controlling 
multiparticle production in high energy collisions 
is a two-step process. To the independent (Poissonian) production of groups 
of ancestor  particles (called ``clan ancestors'') 
follows their decay according 
to a (logarithmic) hadron shower process (the particle MD within each clan).
Clans are by definition independently produced and, by assumption, exhaust
all existing particle correlations within each clan (the ``clan'' concept was  
introduced in high energy physics at the XVII International 
Symposium on Multiparticle Dynamics \cite{AGLVH:0}).  
This  interpretation 
gave  a sound  qualitative description of different classes of collisions in 
terms of the two new parameters, the average number of clans, $\Nbar$, and the 
average number of particles per clan, $\nc$, two  non-trivial functions 
of standard NB (Pascal) MD parameters $\nbar$ and $k$  (the
study of $\Nbar$ and $\nc$ constitutes what is known as clan structure
analysis, described in more detail in Section \ref{sec:I.3}). 

It turned out that within this analysis the average number of clans was larger 
in \ee\  annihilation than in \pp\ collisions, but the average number of 
particles per clan was smaller in \ee\  annihilation than in \pp. The 
situation was  intermediate between the last  two in deep inelastic scattering 
(the average number of clans was smaller than in \ee\  annihilation
and similar to   the average  number  in \pp\ collisions,  but
the average number of particle per clan was more  numerous than
in \ee\  annihilation.)  Larger  clans in \ppbar\ collisions with respect
to those in \ee\  annihilation   were  interpreted as an indication of a 
stronger colour exchange mechanism in the initial state  of the collision in 
hadron-hadron  collisions with respect to \ee\  annihilation.  The advent 
of QCD as the theory of of strong interactions ---the non-linear field theory 
foreseen by W. Heisenberg--- raised a new  question, i.e.,   how to reconcile
observed final $n$ charged particle MD's in various collisions, all
described by  NB (Pascal) MD's, with QCD expectations for  final parton MD's.

The problem was considered quite challenging and intriguing in view of the 
fact that, in solving QCD 
Konishi-Ukawa-Veneziano (KUV) parton differential evolution equations in 
the leading-log approximation
with a fixed cut-off regularization prescription, jets initiated by a 
quark and a gluon were found to be QCD Markov branching processes
controlled by quark bremsstrahlung and gluon self-interaction QCD vertices,
and final parton multiplicity distributions were in both cases found to be
again NB (Pascal) MD's \cite{AGQCD,KUV}. 
In order to solve the puzzle, in consideration of the lack of explicit QCD
calculations at final 
parton level, the suggestion was to rely on Monte Carlo calculations
based on Dokshitzer-Gribov-Lipatov-Altarelli-Parisi
(DGLAP) integral QCD  evolution equations (the integral version of the KUV 
differential QCD evolution  equations)  and on a convenient guesswork as 
hadronization prescription. It was found \cite{KittelNB,AGLVH:2}, by using   
the Jetset 7.2 Monte Carlo,  that NB (Pascal) MD  occurred both at
final hadron and  
parton level multiplicity distributions  for $q\bar q$ and $gg$ systems  at 
various c.m.\ energies and in symmetric rapidity intervals. Results at parton 
and hadron  levels turned out not to be independent and  their relation  
summarised in the so called `generalised' 
local  hadron(h) - parton(p) duality (GLPHD), i.e.,
$k_h = k_p$  and $\nbar_h =\rho \nbar_p$.
The name `generalised' came from the need to distinguish the present `strong'
 result  from the standard `weak' local hadron parton duality (LPHD) which 
requested  in  its first version  that only the second of the two equations be 
satisfied. Others unexpected regularities emerged when  Monte Carlo results 
were analysed  in terms of clan structure analysis
\cite{KittelNB,AGLVH:2}. 

Apparently the  
situation was quite well settled, but   as it very often happens in the 
continuous dialog between theory and experiment (the main characteristic of the 
field), new  experimental facts were just around the corner ready to question 
the universality of the  regularity, i.e., the statement
that final $n$ charged particle multiplicity  distributions in all classes of 
collisions both in full phase-space
(FPS) and in symmetric (pseudo)-rapidity intervals are
NB (Pascal) MD's. Experiments at top CERN  \ppbar\ Collider 
energy (UA5 Collaboration, \cite{UA5:3}) 
and in \ee\  annihilation at LEP energies 
(Delphi Collaboration \cite{DEL:1,DEL:2})  
showed  a shoulder structure in $P_n$ vs.\ $n$  plots 
for  the total sample of events both in FPS and in larger 
(pseudo)-rapidity intervals. In conclusion,  
the regularity was  violated as the 
c.m.\ energy of the collision increased. 
Since the regularity, in view also of the quite large 
experimental errors, has been always considered to be true in an approximate 
sense, a better analysis of the existing data, together with the new data on 
the shoulder structure of  $P_n$ vs.\ $n$ plot, 
led to guess that these new facts
were signals of substructures arising as the c.m.\ energy increased. 

Then an amusing situation happened: the regularities in terms of
NB (Pascal) MD could be restored at a more fundamental level of
investigation, i.e., at the level of the different classes of events
contributing to the total sample (for instance the classes of soft and
semi-hard events in \ppbar\ collisions and the classes of the two- and
the three-jet sample of events in \ee\ annihilation) 
\cite{DEL:4,Elba,hqlett:2}.

The result was that the weighted superposition mechanism of different
classes of events (each one described by a NB (Pascal) MD with characteristic
parameters $\nbar$ and $k$) in different collisions allowed to
describe quite successfully, in addition to the shoulder effect in
$P_n$ vs. $n$, also the observed $H_n$ vs.\ $n$ oscillations (where $H_n$
is the ratio of $n$-factorial cumulants, $K_n$, to $n$-factorial
moments $F_n$ and the truncation effect had been properly taken into
account \cite{hqlett:2}) and the general behaviour of the energy dependence 
of the strength of forward-backward (FB) multiplicity correlations (MC's)
\cite{RU:FB}.

Finally, extrapolations of collective variables for the different classes of
events based on our knowledge of the GeV energy domain became possible
in the TeV region accessible at LHC. Expected data  on  \pp\ collisions 
at LHC will test these predictions.

Clan concept emerges as fundamental in this new  context and its relevance
in the theoretical interpretation of the  above  mentioned  results 
leads to the conclusion that it would not  be  too bold to  ask to
experiment and theory respectively  the following two questions: 

Are clans real physical observable quantities? 

What is the  counterpart of clans  at parton level in QCD?

In addition,  in our approach,  signals of new physics at LHC in 
restricted rapidity intervals are foreseen   in the total
sample of events for  $P_n$ vs. $n$ (elbow structure) as a consequence
of the reduction of the average number of clan to few units in 
an eventual third class of events (described by a NB (Pascal) MD with $k < 1$).
The new class should  be added to the soft and semi-hard ones.  
All together, the above mentioned facts form what has been called
``the enigma of multiparticle dynamics'', which we would like to
disentangle in this paper.

The last word is once more  to experimental observation, of course, and is
challenging  for QCD! 
The problem in multiparticle phenomenology can be summarised in fact
in the following simple terms.

A rigorous application of QCD is limited to hard interactions only.
They  occur at very small distances and large momentum transfers among
quarks and gluons (the elementary constituents of the hadrons), and in very 
short times. Perturbative QCD can be applied with no limitations in
these regions. A situation which should be contrasted with the fact
that the majority of strong interactions are soft; they occur in
relatively large time interval  and distances among constituents, and
large transverse momenta. The search should be focused on how to build a bridge
between hard and soft interactions, and between soft interactions and
the hadronization mechanism, i.e., on  how to explore regions not accessible
to perturbative QCD, like those with high parton densities. Along this line 
of thought it is compulsory to isolate in these extreme regions the properties 
of the sub-structures or components  or classes of events contributing to the 
total sample of events, being aware of the fact that the only informations we 
can rely on in a collision ---as already pointed out---
are coming from the hadronic spectrum and 
multiparticle production. 
The present work (limited to multiparticle dynamics)
has been motivated by the need  to review  experimental facts  and  
theoretical ideas in the field, ideas   which, after more than thirty years 
from their introduction    in the accelerator region,  are still with us
today and arouse a  special interest 
\cite{rint:th,rint:nucl,rint:ee,rint:other}.
%\cite{recentinterest}. 

The  experimental facts we are referring to are indeed  very  important in
establishing collective variables behaviour in full phase-space and 
restricted rapidity intervals and represent the natural starting point
for the investigations of the new  horizon, opened at CERN and RHIC, in the TeV 
domain for hadron-hadron and nucleus-nucleus collisions. 

On the theory side,
the ideas we propose to examine are still quite  stimulating and matter of 
debate in  view of both their success in describing, with  good approximation,
important aspects of multiparticle phenomenology of high energy
collisions and of their connection with QCD parton showers. 
The search started a long time ago and took advantage of the
encouragement and great interest of many scientists;
of particular relevance has been the contribution of L\'eon Van Hove
who, we are sure, would be very glad to see how far some ideas
elaborated together went.

In  Section \ref{sec:I},  in order to enlarge the scope of our work from a 
phenomenological point of view, we decided 
%---after some  historic notes on 
%NB (Pascal) MD   origin and  name, and its  first applications 
%to science---   
to discuss the main statistical properties of the  collective variables   
of the class of MD's  to which NB (Pascal) MD belongs, i.e., the
class of Compound Poisson Multiplicity Distributions (CPMD).
 Within this framework particular attention is paid to clan concept and
its generalisations in view of its relevance in our approach to multiparticle
phenomenology. Clan structure  properties  are always  exemplified 
in the  case of the  NB (Pascal) MD.

In Section \ref{sec:II}, QCD roots of NB (Pascal) MD are examined by studying
jets at parton level as QCD Markov branching  processes in the framework
of QCD KUV equations. Then an attempt is presented
to build a model of parton cascading based on essentials of QCD
(gluon self-interaction) in  a correct kinematical framework. This
study led us to the generalised simplified parton shower model (GSPS)   with
two parameters. 

In Section \ref{sec:III}, global properties of collective variables   
in multiparticle 
dynamics in various experiments are discussed and their descriptions and
interpretations in terms of clan structure analysis  are presented. 
Attention is focused especially on $n$ charged particle MD's general
behaviour in FPS and in (pseudo)-rapidity intervals in \pp\ collisions
and \ee\ annihilation, on the oscillation of the ratio of $n$-order
factorial cumulants to factorial moments when plotted vs.\ $n$ and on
forward-backward multiplicity correlations.
This study is extended in \pp\ collisions to different scenarios in
the TeV energy domain and has been obtained by extrapolating our knowledge
of the collisions in the GeV region.
It is shown that the weighted superposition mechanism of different
classes of events, each one described by a NB (Pascal) MD provides a
satisfactory description of all above mentioned, more subtle aspects
of multiparticle phenomenology.
Clan structure analysis turns out to be quite important in this
respect. 
Its success raises a natural question on the physical properties of
clans.
New perspectives opened by the reduction of the average number of
clans in the semi-hard component of the scenarios with 
Koba-Nielsen-Olesen (KNO) scaling violation are examined.
Signals of new physics in \pp\ collisions in the TeV region
(ancillary to those expected for heavy ion collisions) are
discussed at the end.

\newpage

\section{COLLECTIVE VARIABLES AND CLAN STRUCTURE ANALYSIS}\label{sec:I}

In studying and testing perturbative QCD, 
two complementary approaches can be used:
one can select infrared-safe quantities in order to test pQCD in the
hard region, where calculations are well defined and non-perturbative
corrections are suppressed.
Alternatively, one can examine infrared-sensitive observables with the
aim of testing the validity of the theory in the long-range region,
where confinement becomes dominant.
It could be said that the latter approach determines the boundary
conditions that the former one must satisfy.
The present work takes the road of the infrared-sensitive observables,
which are discussed in detail in this Section 
\cite{Koba,Sergio:thesis,DeWolf:rep}.

\subsection{Observables in multiparticle production. 
The collective variables.}\label{sec:I.2}

In a $n$-particle production process in the collision of particles $a$ and $b$
\begin{equation}
	a + b \to c_1 + \dots + c_n ,
\end{equation}
by assuming for simplicity that all produced particles $c_i$ are of the same 
species, the \emph{$n$-particle exclusive distribution}   $P_n'(y_1,\dots,
y_n)$  in a sub-domain of phase-space (the 
$y_i$ with $i= 1\dots n$ are the particle (pseudo)-rapidities)
is described in terms of exclusive cross sections, i.e.,
\begin{equation}
  P_n'(y_1, \dots, y_n) =  (\sigma_{\text{inel}})^{-1} \frac{d^n {\sigma_n} 
    }{dy_1 \dots dy_n}
\end{equation}
 and is fully 
symmetric in its variables.
$P_n' ( y_1, \dots, y_n)$ is related to the probability of 
detecting $n$ particles at
rapidity variables $y_1,\dots, y_n$ with no other particle present in
the subdomain.

The corresponding \emph{integrated} observable in the rapidity
interval $\Delta y$, (i.e., one integrates over $y_i \in \Delta y$, 
$i=1,\dots,n$)
describes the  \emph{probability $P_n (\Delta y)$ of 
detecting $n$ particles} in the mentioned interval and is the $n$-particle  
exclusive cross section, $\sigma_n$, normalised to the total inelastic cross
section, $\sigma_{\text{inel}}$:
\begin{equation}
	 P_n ( \Delta y  ) = \frac{\sigma_n}{\sigma_{\text{inel}}}  =
   (n !)^{-1}  \int_{\Delta y} \dots \int_{\Delta y}
  P_n' (y_1,...,y_n) d y_1\cdots d y_n  .
\end{equation}

The \emph{$n$-particle inclusive distribution}, $Q_n(y_1, \dots , y_n)$, in the
reaction 
\begin{equation}
	 a + b \to c_1 +\dots + c_n + \text{anything}
\end{equation}
describes the finding of $n$ particles at  $y_1,\dots , y_n$, without paying
attention to other particles in the same sub-domain, in terms of the
$n$-particle inclusive cross-section density, i.e.,
\begin{equation}
	 Q_n (y_1, \dots, y_n) = ( {\sigma_{\text{inel}}}) ^ {-1}  
  \frac{d^n \sigma }{ dy_1 \cdots dy_n}
\end{equation}
(sometimes referred to in the literature as   $\rho ( y_1, \dots, y_n)$.)

The corresponding integrated observables  in $\Delta y$,
\begin{equation}
	 F_n ( \Delta y ) = \int_{ \Delta y}\dots \int_{\Delta y}
  Q_n(y_1, \dots, y_n) d y_1 \cdots d y_n  ,
\end{equation}
are  the  \emph{un-normalised $n$-factorial moments} 
of the multiplicity distribution
in $\Delta y$  (the so-called binomial moments). For instance
the average multiplicity in $\Delta y$, $\nbar (\Delta y)$, is
\begin{equation}
	F_1(\Delta y) =
	 \nbar (\Delta y)   = \int_{\Delta y}   Q_1 (y_1) d y_1 
  = \int_{\Delta y} \frac{ d {\sigma} }{d  y_1 }   d y_1 
\end{equation}
In the integration over the domain $\Delta y$  each particle will contribute
once to the $n$-particle event  and the event will be counted $n$ times.
Coming to the second order factorial moment, $F_2 ( \Delta y ) =
\avg{( n - 1) n} $, each ordered couple of
final particles will be counted once in one event and the event will be
counted $2\binom{n}{2}$ times.
 
  The $n$-particle inclusive distributions $Q_n(y_1, \dots, y_n )$ contain 
inessential   contributions due to combinations of inclusive distributions
of lower order. In addition, for large $n$, it is hard to measure all 
$Q_n(y_1, \dots, y_n)$ and a more 
essential information is demanded. Accordingly,  a new set of observables is 
introduced, the set of  
\emph{$n$-particle correlation functions} $C_n(y_1, ..., y_n)$
which reminds of cluster-expansion in statistical mechanics:
\begin{equation}\label{eq:I.8}
	 \begin{split}
		 Q_1 (y _ 1) &=  C_1 (y_1)\\
		 Q_2 (y_1, y_2) &= C_2 ( y_1, y_2) + C_1 ( y_1) C_1 ( y_2)\\
		 Q_3 (y_1, y_2, y_3) &=  C_3 ( y_1, y_2, y_3 ) + 
        C_1 ( y_1) C_2 ( y_2, y_3) +\\
        &\quad~ C_1 ( y_2) C_2 ( y_3, y_1) + C_1 (y_3) C_2 (y_1, y_2) +\\
        &\quad~ C_1(y_1) C_1(y_2) C_1 (y_3)\\
		 \ldots
	 \end{split}
\end{equation}

The reverse is of course always possible and one can express
$C_n(y_1,\dots, y_n)$ variables  in terms of $Q_n ( y_1,\dots, y_n)$ 
variables:
the $n$-particle correlation functions $C_n (y_1, \dots, y_n)$ 
give precious informations  on the production process:

\begin{equation}
		C_n ( y_1,\dots,  y_n)  = 0 \quad\Rightarrow\quad
		Q_n(y_1, \dots, y_n) = {\prod_{i=1}^n} Q_1( y_i) 
	\end{equation}
  Produced particles are  Poissonianly distributed.

\begin{equation}
		C_n (y_1,\dots, y_n)  >  0
	\end{equation}
  Produced particles follow a distribution wider than a Poisson
  distribution.

\begin{equation}
		C_n (y_1,\dots, y_n ) <  0
	\end{equation}
	Produced particles follow a distribution narrower than a Poisson
	distribution.

The corresponding integrated observable are called  
\emph{factorial cumulants} of the
multiplicity distribution and indicated in the literature with  $K_n$.
They  give informations on the shape of the multiplicity distribution when
plotted vs.\ $n$, i.e., on its dispersion  $\avg{ ( n- \nbar )^2 }$, 
skewness 
$\avg{( n - \nbar )^3}$, kurtosis  $\avg{(n - \nbar)^4}$, \ldots.

Both $F_n$ and $K_n$ are sensitive to events with many particles and control
the tail of the multiplicity distribution. In $K_n$ are subtracted the
$Q_{n-1}, ..., Q_1$ correlations present in $Q_n$, not those related to the 
fluctuations in the multiplicity, as lucidly discussed in \cite{correl}. 

Usually normalised differential collective  variables  
\begin{equation}
	\tilde Q_n = Q_n ( y_1,\dots, y_n)/ Q_1(y_1) \dots Q_1(y_n) ,
\end{equation}
\begin{equation}
	\tilde C_n  = C_n(y_1,\dots, y_n)/ C_1 (y_1)\dots C_1 (y_n)
\end{equation}
and the corresponding  normalised integral collective variables:
\begin{equation}
\tilde 	F_n  = F_n (\Delta y) / F_1 (\Delta y)\dots F_1 (\Delta y)  ,
\end{equation}
\begin{equation}
	\tilde K_n = K_n (\Delta y) / K _1 (\Delta y)\dots K_1 (\Delta y)
\end{equation}
are used.

In the literature  also \emph{$n$-particle combinants} $W_n (\Delta y)$ are
defined. Their expression is a function of $n$-particle
multiplicity distribution $P_n (\Delta y)$ according
to the following recurrence relation:
\begin{equation}
	W_n (\Delta y) =   \frac{P_n (\Delta y)}{P_0 (\Delta y) } 
  - \frac1n {\sum_{i=1}^{n-1}} i W_i(\Delta y)
  \frac{ P_{n-i} (\Delta y) }{ P_0(\Delta y) }  ,
\end{equation}
i.e.,  $W_n$ are ``finite combination" ratios of $P_n$ to $P_0$. The name
combinants  comes from this fact.

% what is this useful for??
Combinants  in their normalised form, $\tilde W_n$  are
\begin{equation}
	\tilde W_n (\Delta y)  =  \frac{W_n (\Delta y)  }{[W_1 (\Delta
	y)]^n }  .
\end{equation}

Notice that combinants are linked to factorial cumulants $K_n$
 i.e.
 \begin{equation}
	 W_n (\Delta y) = \frac{1}{n!}  {\sum_{i=0}^\infty}  
	 \frac{ (-1)^i K_{i+n} (\Delta y) }{i!}  .
 \end{equation}

Of particular interest is the relation which allows to express $P_0 (\Delta y)$
in terms of the combinants:
\begin{equation}
	P_0 (\Delta y) = \exp \left[ - \sum_{n=1}^\infty W_n \right]  .
\end{equation}

From the definition and properties of combinants it turns out that they
are sensitive to the \emph{head} of the
multiplicity distribution whereas  factorial cumulants and moments are
sensitive to the \emph{tail} of the multiplicity distribution.

The relation among collective differential variables
$P_n' (y_1,\dots, y_n)$, $Q_n (y_1,\dots, y_n)$ and $C_n (y_1,\dots, y_n)$
can be elegantly reformulated in terms
of \emph{functionals}, i.e., of  $G [z(y)]$ with  $z(y) = 1 + u(y)$ 
and $u(y)$ an  arbitrary
function of $y$.
Within this framework,  $G[z(y)]$ is defined as follows
\begin{equation}
	G[z(y)] =  P_0 +  \sum_{n=1}^\infty   (n!)^{-1}\int_{\Delta y_1}\dots
  \int_{\Delta y_n} P_n' (y_1,\dots, y_n) \prod_{i=1}^n z(y_i) d y_i ,
\end{equation}
i.e., the (exclusive) generating functional. It should be noticed that
\begin{equation}
	P_0 = 1 -  \sum_{n=1}^\infty (n!)^{-1}  \int_{\Delta y_1}\dots
  \int_{\Delta y_n} P_n'(y_1, \dots, y_n) d y_1\dots d y_n
\end{equation}
is the probability of events with zero particle multiplicity.

The knowledge of the functional $G[z(y)]$ allows to define two other
functionals  $G[u(y)]$ and $g[u(z)]$. They are linked by the  relation
\begin{equation}
  G[u(y)] = \exp\left\{  g[u(y)] \right\} ,
\end{equation}
where 
\begin{equation}
	G[u(y)] =   1 +{ \sum_{n=1}^\infty} (n!)^{-1} \int_{\Delta y_1}\dots
  \int_{\Delta y_n} Q_n(y_1,\dots, y_n) \prod_{i=1}^n u(y_i) d y_i ,
\end{equation}
is the (inclusive) generating functional, and 
\begin{equation}
	g[u(y)] = { \sum_{n=1}^\infty}    (n!)^{-1} \int_{\Delta y_1}\dots
  \int_{\Delta y_n} C_n(y_1,\dots, y_n) \prod_{i=1}^n u(y_i) d y_i .
\end{equation}

It is remarkable that, in view of  the existing connection 
among collective differential variables $P_n'$, $Q_n$, $C_n$ 
and their integrated
partners $P_n$, $F_n$, $K_n$,  the generating functions (GF's)
of the integrated variables, obtained by the substitution $z (y) \to z$, 
are  $G(z)$, $G(z+1)$ and $\log G(z+1)$
respectively. Accordingly, being   $G(z) =  \sum_n P_n z^n$ it 
follows
\begin{align}
	P_n &= \frac{ 1 }{ n! } \left. \frac{ \partial^n G(z) }{ \partial
		z^n} \right|_{z=0}, \label{eq:12}\\
	F_n  &= \left. \frac{ \partial^n G(z) }{ \partial z^n } 
	  \right|_{z=1}, \label{eq:13}\\
	K_n  &= \left. \frac{ \partial^n \log G(z) }{ \partial z^n } 
	  \right|_{z=1}. \label{eq:14}
\end{align}
                                       
In addition one gets
\begin{equation}
	G(z) = \exp{\sum_{n=1}^\infty }\frac{z^n K_n}{n!}
  = \prod_{n=1}^\infty \exp \left[ \frac{z^n K_n}{n!} \right] .
\end{equation}

Coming to multiplicity combinants $W_n (\Delta y)$  
their relation  with the exclusive generating function  $G(z;\Delta y)$ 
should be recalled:
\begin{equation}
  G(z;\Delta y) = \exp\left[ \sum_{n=1}^\infty W_n (\Delta y ) 
			(z^n - 1) \right] .
\end{equation}
This relation shows that combinants are the coefficients of the power
series expansion of the logarithm of the generating function.

The above general definitions will now be exemplified  for the NB (Pascal)
MD; explicit expressions are given below, and also plotted 
in Fig.~\ref{fig:546fps} for typical values of the parameters
(corresponding to the soft and semi-hard components which
describe \pp\   collisions data at 546  GeV c.m.\ energy \cite{Fug}.)

Being the GF of the NB (Pascal) MD
\begin{equation}
  G_{\text{NB}}(z) =  \left( \frac{ k }{k-(z-1)\nbar} \right)^k  
\end{equation}
(with  $k^{-1} =(D ^2 - \nbar )/ \nbar^2$, $D$ is the dispersion) 
it is found from Eq.s (\ref{eq:12}), (\ref{eq:13}), (\ref{eq:14}) that
\begin{equation}
	P^{\text{(NB)}}_n = \frac{ k (k+1)\cdots (k + n - 1) }{ n!}
	   \frac{ \nbar^n k^k }{ ( \nbar + k )^{k+n}},
\end{equation}
\begin{equation}
	F^{\text{(NB)}}_n   = \nbar^n  {\prod_{i=1}^{n-1}} ( 1 +  i/k) ,
\end{equation}
and
\begin{equation}
	K^{\text{(NB)}}_n =   \nbar^n {\prod_{i=1}^{n-1}}  i/k .
\end{equation}

In the literature also the collective  variables expressed in terms of
the ratio of factorial cumulants and moments \cite{Dremin:1993}, i.e.,
\begin{equation}
	H_n = K_n / F_n ,
\end{equation}
deserve great attention. For  the NB (Pascal) MD one gets
\begin{equation}
	H^{\text{(NB)}}_n = {\prod_{i=1}^{n-1}}  \frac{ i }{ (i+k) } .
	\label{eq:I.35}
\end{equation}
Accordingly, $H^{\text{(NB)}}_n$ depends on $k$ parameter only,
an important fact which will be of particular interest in
Section \ref{sec:II.hadronization}.

 \begin{figure}
  \begin{center}
  \mbox{\includegraphics[width=\textwidth]{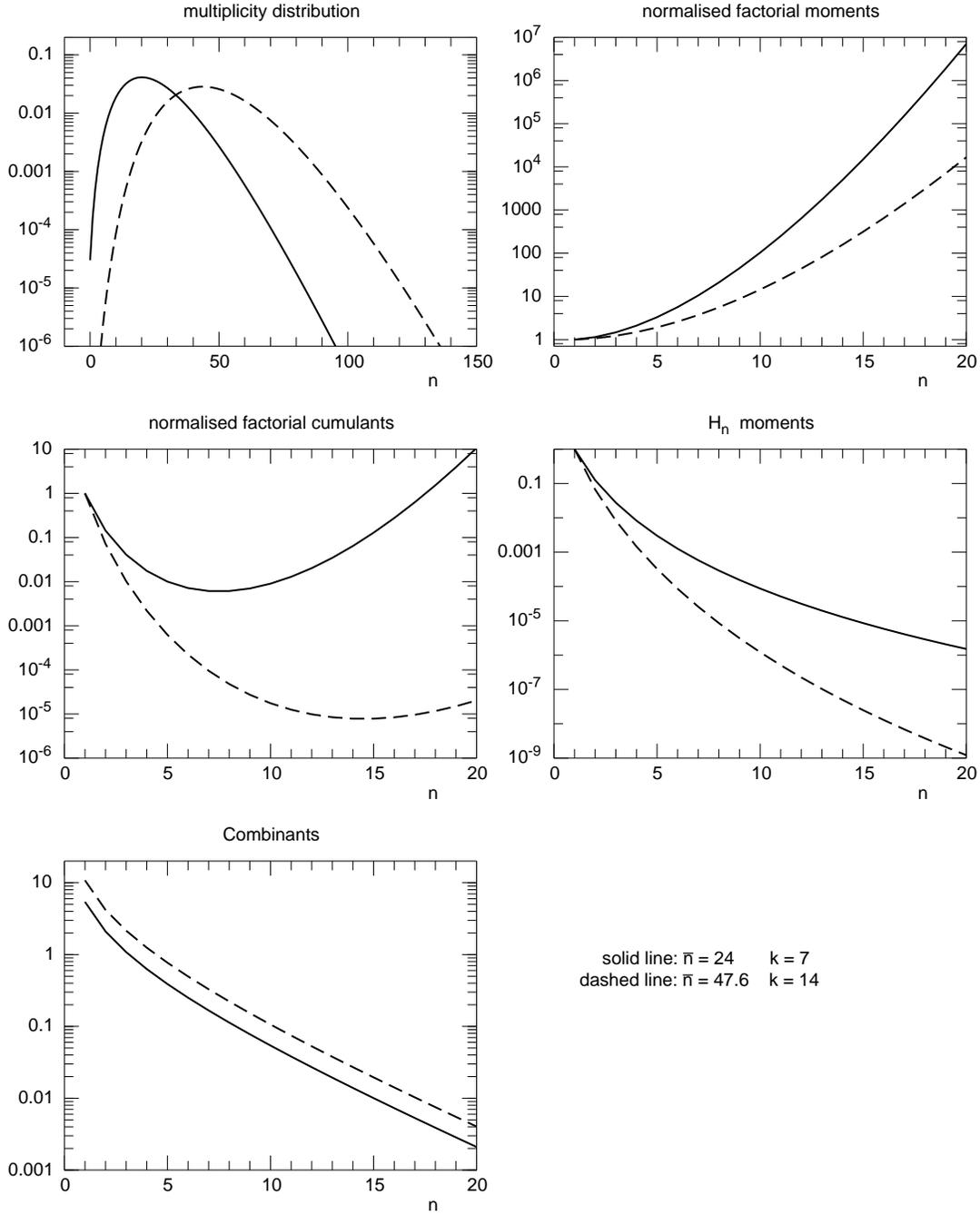}}
  \end{center}
  \caption[Global variables for the NBD]{General behaviour of 
  global variables $P_n$, $\tilde K_n$, $\tilde F_n$, 
	$H_n$, $W_n$  vs.\ $n$ for the 
  NB (Pascal) MD for two sets of parameters (the pairs of chosen values
  correspond respectively to the soft and semi-hard components which
  describe \pp\ collisions data at 546  GeV c.m.\ energy 
	as determined in \cite{Fug}.)}\label{fig:546fps}
  \end{figure}

\subsection{Clan concept and Compound Poisson Multiplicity 
Distributions.}\label{sec:I.3}

Clan concept, as already mentioned, has been introduced in order to interpret 
the wide occurrence of the NB (Pascal) MD  in multiparticle phenomenology
in different classes of collisions since 1987.
In terms of the NB (Pascal) MD generating function, 
$G_{\text{NB}} (\nbar, k; z) ={\sum_{n=0}^\infty} P_n^{(NB)} z^n$,
one  has
\begin{equation}
	G_{\text{NB}} (\nbar, k ; z ) = \exp \left\{
	   \Nbar  [G_{\text{log}}( \nbar, k ; z ) - 1]
		 \right\}          ,             \label{eq:1}
\end{equation}
where $\Nbar = k \ln (1 + \nbar / k)$ is the average number of clans,
$\nbar$  the average number of particles of the total  NB (Pascal) MD and $k$ 
the second characteristic parameter linked to the dispersion $D$ by 
$(D^2 - \nbar )/ \nbar^2 = 1/k$; the GF of the logarithmic MD,
$G_{\text{log}} ( \nbar , k ; z)$,     is given by 
\begin{equation}
	G_{\text{log}}  = 
         \frac{\ln (\nbar + k - \nbar z ) -  \ln ( \nbar + k)   }{
         \ln  k   - \ln (\nbar + k )} ,     \label{eq:2}
\end{equation}
and the average number of particles per clan, $\nc$,
turns out to be 
\begin{equation}
	\nc = \frac{ - \nbar/ k
       }{\ln k - \ln ( \nbar + k) } .   \label{eq:3}
\end{equation}
Notice that    
\begin{equation}
  \Nbar = \nbar / \nc  .   \label{eq:4}
\end{equation}

The interest of Eq.~(\ref{eq:1}) 
lies  in the fact that it  is indeed related to an 
important dynamical concept: the description of particle  production as a 
two-step process. 
To an initial phase in which  ancestor particles  are independently  produced  
(their multiplicity distribution is Poissonian as predicted for instance
by the multi-peripheral model)  it follows a second phase in which 
Poissonianly distributed particles (the ancestors) decay according to a given
new multiplicity distribution (for instance,
the logarithmic distribution given by Eq.~(\ref{eq:2}).)
An ancestor particle in this framework is very special: it is an intermediate 
particle source  and  together with its descendants is called in the literature
`clan' (`Sippe' in German language; see also the Oxford Dictionary entry).
Each clan contains at least one particle (its ancestor, the intermediate 
particle source) by assumption, and all correlations among generated particles 
are exhausted within the clan itself  by definition. The point is that 
$G(z)$ in Eq.~(\ref{eq:1}) is just by inspection 
the GF of  a compound Poisson 
multiplicity distribution (CPMD), a much  wider class of multiplicity 
distributions than the NB (Pascal) MD \cite{kendall,kotz}.
  Accordingly, the first step  of the production  process corresponding to 
independent emission of the particle sources is Poissonian 
for all the multiplicity distributions belonging to the class of CPMD:
what  fully characterises final  $n$-particle multiplicity distribution
in  this class are  the multiplicity distributions  of particles 
originated by the Poissonianly distributed  particle sources. Therefore  we 
propose to   generalise clan concept and its properties in terms of collective 
variables  to the full  class of compound Poisson multiplicity  distributions. 
The  generalisation is interesting since it gives a larger horizon to
multiparticle phenomenology in suggesting that particle multiplicity
distributions need not to be in general of logarithmic type as in the
case of the NB (Pascal) MD: in principle any true GF is 
allowed.\footnote{It should be 
pointed out that the generating function  $G(z)$  
(and the corresponding MD $P_n$)
belongs to the class of Infinitely Divisible Distributions (IDD)  GF's  if 
for every integer number $s$ there  exist $s$ independent random
variables  with the same GF $g_s (z)$ such that   $G(z)= [ g_s (z)]^s$ .
It is remarkable that  $s$  can be in general  a positive non integer value
and that all  GF's of discrete IDD's can be written as 
compound Poisson distribution GF's \cite{Feller}.}

In order to generalise clan concept to the class of CPMD, we suggest therefore
to write the GF of a generic  CPMD, $G_{\text{CPMD}} (z)$,  as follows 
\cite{Feller}
\begin{equation}
	G_{\text{CPMD}} (z)= \exp \left\{ \Nbar_g [g(z) -1] \right\},
	                                                          \label{eq:5}
\end{equation}
with  $G_{\text{CPMD}} (z)  = \sum_n   P^{\text{(CPMD)}}_n z^n$,
$g(z)$ is here the generic GF of the multiplicity 
distribution of particles produced by
the generalised clan ancestor and  ${\Nbar_g}$ is 
the average number of generalised clans. 
The   total  number $n$ of 
particles distributed among the $N_g$   generalised  clans  satisfies of course
the condition  $\sum_{i=1}^{N_g}  n^{(i)}_c = n$, with  
$n^{(i)}_c$  the number of particles within the $i$-th generalised clan.
The name ``generalised clans'' (g-clans)  will be given 
to clans when the GF $g(z)$ in the above equation is related to a generic  
multiplicity distribution $g_{n}$, i.e. $g(z)= \sum_n  g_{n} z^{n}$,
whereas the name ``clans'' refer properly  to the 
grouping of particles defined within
the NB (Pascal) MD, for which $g(z)$ is the logarithmic distribution. 

It is clear that since particles within each clan must contain at least one
particle (the clan  ancestor) particle distribution within each clan must 
satisfy the condition
\begin{equation}
	g(z)|_{z=0} = g_0 = 0      .                         \label{eq:6}
\end{equation}

The description in full phase-space can be extended to restricted regions of
phase-space,  i.e., for instance  to the set of  rapidity intervals $\Delta y$.
One has 
\begin{equation}
	G_{\text{CPMD}} (z, \Delta y) = \exp \left\{ \Nbar_{\text{g-clan}} (\Delta y) 
         [ g(z, \Delta y) - 1] \right\} ,         \label{eq:7}
\end{equation}
and, contrary to Eq.~(\ref{eq:6}), $g(z, \Delta y)|_{z=0} = g_0 (\Delta
y)$ is  different from 0; $g(z, \Delta y)|_{ z=0 }$ 
is the probability that a g-clan does not 
generates any particle in $\Delta y$.

In order to fulfill the request that at least one particle per  clan
will be produced in the interval $\Delta y$ a new renormalised GF 
$\tilde g(z,\Delta y)$  should be  defined 
\begin{equation}
	\tilde g(z,\Delta y)= \frac{g(z,\Delta y) - g_0 (\Delta y)}{
		1 - g_0 (\Delta y) }  .
\end{equation}

Accordingly,
\begin{equation}\label{eq:8}
	\begin{split}
	G_{\text{CPMD}} (z, \Delta y) &= \exp \left\{ \Nbar_{\text{g-clan}} (\Delta y)
	   [ \tilde g(z) - 1] \right\}\\ &=
    \exp \left\{ \Nbar_{\text{g-clan}} (\Delta y) 
    		\frac{g(z, \Delta y)-1}{1 - g_0(\Delta y) } \right\} .
	\end{split}
\end{equation}

It follows
\begin{equation}
  \Nbar_{\text{g-clan}}(\Delta y )=\Nbar_{\text{g-clan}}[ 1 - g_0(\Delta y) ]
\end{equation}
and
\begin{equation}
	\tilde g(z, \Delta y)|_{z=0}=0.
\end{equation}

It should be stressed that knowing the probability of generating zero
particles in the interval $\Delta y$, the average number of generalised
clans in the same interval can be determined from the knowledge of the
average number of generalised clans in FPS 
(this will be used in Section \ref{sec:II.kinem.gsps}.)

In conclusion, by considering only those clan which generates at least one
particle in the interval  $\Delta y$ the GF \
$G_{\text{CPMD}}(z, \Delta y)$ as given by
Eq.~(\ref{eq:8}) satisfies the standard clan definition also in the case of
generalised clans.

Now we will discuss two interesting theorems that characterise the class
of CPMD. 

i. A MD is a CPMD iff the probability of producing zero particles, $P_0$,
is larger than zero and all its combinants $W_n$ are positive definite.
The theorem  follows from the fact  that   combinants  $W_n$ are related to
the $n$-particle multiplicity distribution within a g-clan, $p_n$, 
by  the equation
\begin{equation}
	W_n = \Nbar_{\text{g-clan}} p_n ,
\end{equation}
and that the average number of g-clan $\Nbar_{\text{g-clan}}$ is related to
combinants $W_n$ by
\begin{equation}
  \Nbar_{\text{g-clan}} = - \log P_0 = \sum_{n=1}^\infty W_n  .
\end{equation}
\cite{Hegyi+Hegyi:1993}. 
The connection between $\Nbar_{\text{g-clan}}$ or  $P_0$ and the combinants is
particularly suggestive.

ii. All factorial cumulants of a CPMD are positive definite.
Being for a CPMD  $n$-order factorial cumulants, $\tilde K_n$, 
related to $n$-order
factorial particle  moments within the g-clan, $\tilde q_n$,  by 
\begin{equation}
	\tilde K_n = \Nbar_{\text{g-clan}} \tilde q_n,
\end{equation}
$ P_0 > 0 $ and $\Nbar_{\text{g-clan}}$ a finite number,  
positive definiteness of
$n$-order factorial moments, $\tilde q_n$, 
implies that also $\tilde K_n$ are positive  definite.
The theorem can be applied to $H_n$ variables, i.e., to the ratio of
$n$-order factorial cumulants to $n$-order factorial moments 
$\tilde K_n /\tilde F_n$.
The theorem is of particular interest in the study of $H_n$ vs.\ 
$n$ oscillations \cite{DreminHwa2}.

The mentioned relations  between the normalised $n$ order cumulants of
the total  
CPMD, $\tilde K_n$,  and  the normalised $n$ order factorial moments   
of the particle 
multiplicity  distribution within the g-clan, $\tilde q_n$,
can be easily extended 
from full phase-space to a generic rapidity interval $\Delta y$  as follows
\begin{equation}
  \tilde K_n(\Delta y) =  \tilde q_n(\Delta y) / { \Nbar_{\text{g-clan}}
  (\Delta y)}^{(n-1)}  .
\end{equation}
 
Along this line multiplicity combinants of a CPMD  $W_n (\Delta y)$ and
$n$ particle MD within the g-clan are related as in f.p.s .  
\begin{equation}
  W_n (\Delta y) = \Nbar_{\text{g-clan}} (\Delta y) p_n (\Delta y) ,
\end{equation}
with
\begin{equation}
  \Nbar_{\text{g-clan}} (\Delta y)  = \sum_{n=1}^\infty W_n (\Delta y) .
\end{equation}
It should be pointed out that,   if $P_0 = 0$ or one of the combinants $W_n$
or one of the factorial cumulants $\tilde K_n$    
is less than 0, then  the MD cannot be a CPMD.

\subsection{Hierarchical structure of factorial cumulants,
rapidity gap events and CPMD's}\label{sec:I.4}

From the discussion  on collective variables properties in CPMD's, 
it has been seen that a special
role is  played by $P_0 (\Delta y)$, i.e., by the probability of
detecting  zero particles in the rapidity interval $\Delta y$.

In fact, it is clear  from results of Sec.~\ref{sec:I.3}  
that, thanks to the knowledge
of $P_0 (\Delta y)$,  the $n$-particle multiplicity distribution
$P_n (\Delta y)$ can easily  be determined according to the following 
iterative equation:
\begin{equation}
	P_n (\Delta y) =( - \nbar{\Delta y} )^n \frac{ 1 }{n! }
	    \frac{ \partial^n }{\partial{\nbar}^n }P_0(\Delta y) ,
   \label{eq:I.53}
\end{equation}
and that there exists  an instructive connection of $P_0 (\Delta y) $
with factorial cumulants 
\begin{equation}
  P_0 (\Delta y) 
  =  \exp\left\{\sum_{n=1}^\infty [-\nbar (\Delta y )]^n \frac{1}{n!} 
	\tilde K_n(\Delta y) \right\}.
\end{equation}
where
\begin{equation}
  \tilde K_n(\Delta y) =  K_n(\Delta y) / [ K_1(\Delta y)]^n .
\end{equation}

Following the 
just mentioned relations, a new variable in terms of $P_0(\Delta y)$
can be defined which turns out to be  of great importance
in the study of $n$-particle correlation structure, i.e., the void function 
$V_0 (\Delta y$),
\begin{equation}
\begin{split}
	V_0 (\Delta y) &= { - 1/ \nbar (\Delta y)} \log P_0 (\Delta y)\\
	  &= \sum_{n=1}^\infty \frac{ [ - \nbar (\Delta y)]^{n-1} }{ n!}
	        \tilde K_n(\Delta y)
\end{split}
\end{equation}

In fact it can be shown that when plotted as the function of
$\nbar(\Delta y)\tilde K_2(\Delta y)$  the void function$ V_0 (\Delta y)$ scales
iff $n$-order normalised factorial cumulants, $\tilde K_n(\Delta y)$, 
can be expressed in 
in terms of second-order normalised factorial  cumulants, 
$\tilde K_2(\Delta y)$, i.e.,
\begin{equation}
	\tilde K_n(\Delta y) =  A_n [ \tilde K_2(\Delta y)]^{n-1}  \label{eq:9}
\end{equation}
 
The $A_n$ are energy and rapidity independent factors; they are determined
by the correlation structure of  $n$-particle MD. When
Eq.~(\ref{eq:9})  is satisfied 
one talks of ``hierarchical structure for cumulants.''

The  properties of the void function variable  can be easily
generalised to the class of  CPMD's.
It turns out indeed that
\begin{equation}
	V_{0,\text{CPMD}} (\Delta y)  = 1/ \nbar_{c,\text{g-clan}}(\Delta y) =
	    \frac{ \Nbar_{\text{g-clan}}(\Delta y) }{ 
				\nbar_{\text{CPMD}}(\Delta y) } ,
\end{equation}
being
\begin{equation}
  P_0^{\text{CPMD}} (\Delta y) = \exp [- \Nbar_{\text{g-clan}} (\Delta y)] .
     \label{eq:I.59p0}
\end{equation}

In view  of the  connection between normalised cumulants, $\tilde K_n$, and the
normalised $n$-particle  correlation functions, $\tilde C_n$, the hierarchical 
structure prescription in a symmetric rapidity interval $\Delta y$
on $\tilde K_n$ can be translated in terms of  $\tilde C_n$ variables
 \begin{equation}
	 \tilde C_n(y_1,\dots, y_n; \Delta y) = 
	 \sum_{\alpha_t} A_{n, \alpha_t} \sum_{\sigma} 
   \tilde C_2(y_{i_1}, y_{i_2}; \Delta y ) \dots \tilde C_2(y_{i_{n-1}},
   y_{i_n}; \Delta y )  \label{eq:10}
 \end{equation}
where the sum over $\sigma$ denotes all non symmetric relabelings of the
$n$ particles; $A_{n, \alpha_t}$ are 
coefficients independent on the c.m.\ energy
and rapidity interval $\Delta y$; $\alpha_t$ labels the different ways of
connecting the $n$ particles among themselves \cite{Peebles}.

\begin{figure}[p]
  \begin{center}
  \mbox{\includegraphics[width=0.9\textwidth]{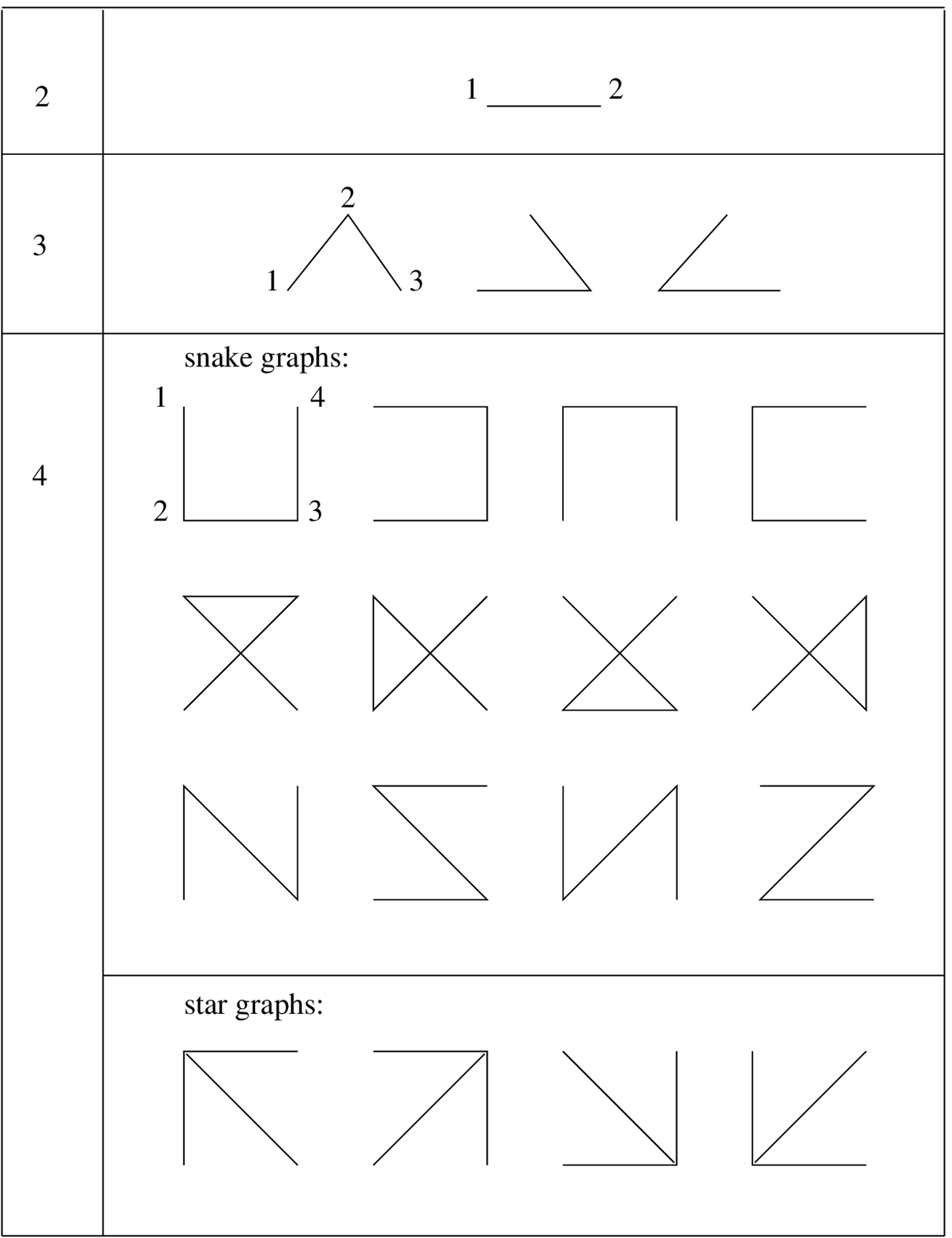}}
  \end{center}
  \caption[Snakes and stars correlations]{Graphical representation 
  of correlation functions of
  different order (left column) in the hierarchical scheme.
  At order 4, the LPA and the VHA begin to differ:
  the LPA includes only the 'snake' graphs, while the VHA
  includes also the 'star' graphs \cite{Void}.}\label{fig:snakes}
  \end{figure}

     A  graphical description of the possible connections turns out to be
quite useful. 
A two-particle normalised correlation function $C_2 (y_i, y_j ; \Delta y)$
is associated to an edge linking particles of rapidity $y_i$ and rapidity $y_j$;
a graph with $(n-1)$ edges is generated for each  
$\tilde C_n(y_1,\dots, y_n; \Delta y)$
configuration of  distinct particles as  described by  Eq.~(\ref{eq:10}).
The sum over $\alpha_t$  should be read as the sum over all topologically
distinct connected $n$-graphs (see Figure~\ref{fig:snakes}). 
Among the different models of correlations functions let us mention two of them:

i) the linked pair ansatz model (LPA) \cite{Carrut}.
In this model a particle cannot appear more than twice in a product and
$\sigma$ relabeling reduces to a standard permutation, in conclusion 
only $\alpha_t$ = ``snake" graphs are allowed (see Figure~\ref{fig:snakes})
\begin{equation}
	\tilde C_n(y_1, ..., y_n; \Delta y)|_{\text{LPA}}  =    
	A_{n,{\text{snake}}}   \sum_{\text{permutations}}
  \tilde C_2(y_{i_1}, y_{i_2}; \Delta y ) \dots \tilde
  C_2(y_{i_{n-1}}, y_{i_n}; \Delta y )  ;
\end{equation}
                                    
ii) the Van Hove ansatz model (VHA) \cite{LVH:3}.
In this model, in addition to ``snake" graphs, also graphs with  a particle
linked to three other particles are allowed, the so called ``star"
graphs ($\alpha_t$ = ``snake", ``star") and Eq.~(\ref{eq:10}) can be
re-expressed with a recursion relation:
\begin{equation}
	\tilde C_{n+1}(y_1,\dots, y_{n+1}; \Delta y)|_{\text{VHA}} = 
    n [ \tilde C_{n}(y_1,\dots, y_n; \Delta y) 
        \tilde C_2(y_i, y_{n+1}; \Delta y)    ]_S ,
\end{equation}
where $[...]_S$ means symmetrization over all particles.

In general, by integrating  Eq.~(\ref{eq:10}) ---in particular, both
the LPA and VHA cases---  over the central rapidity interval $\Delta
y$, assuming translation invariance of the $\tilde C$ \cite{DeWolf:rep},
one obtains Eq.~(\ref{eq:9}),  
which defines hierarchical structure for
cumulant moment; thus it is apparent that hierarchical structure for
cumulants is not sufficient to  discriminate among different ansatz
for correlation  functions.

In order to test hierarchical structure for normalised cumulant moments,
the void function $V_0 (\Delta y)$ can be used. Since 
\begin{equation}
	V_0 (\Delta y) = \sum_{n=1}^\infty  \frac{A_n}{ n!} [ - \nbar (\Delta y)
    \tilde K_2 (\Delta y )]^{n-1}   ,     \label{eq:11}
\end{equation}
energy and rapidity dependence are confined here in the product
$\nbar(\Delta y) \tilde K_2 (\Delta y)$, i.e., the void function scales for
hierarchical cumulant moments  with $\nbar(\Delta y) \tilde K_2 (\Delta y)$.
On the contrary, if scaling holds, 
i.e., $V_0 = V_0( \nbar(\Delta y) \tilde K_2 (\Delta y) )$,
Eq.~(\ref{eq:11}) follows with
\begin{equation}
	A_n = n ( - 1)^{n-1} \left. \frac{ d^{n-1} V_0 (\Delta y)  }{ 
		 d [ \nbar(\Delta y) \tilde K_2 (\Delta y) ]^{n-1}}
	   \right |_{\nbar \tilde K_2 = 0 }
\end{equation}
energy and  rapidity independent.

For a CPMD's,  $P_0(\Delta y)$  and $V_0(\Delta y)$ variables are completely 
equivalent to $\Nbar_g (\Delta y)$ and $\nbar_{c,\text{g-clan}}
(\Delta y)$, being 
\begin{equation}
  \Nbar_g (\Delta y) = - \log P_0 (\Delta y)
\end{equation}
and
\begin{equation}
			\nbar_{c,\text{g-clan}} (\Delta y) =[ V_0 (\Delta y)]^{ -1} .
\end{equation}

In case of the NB (Pascal) MD, 
\begin{equation}
	\tilde K_n (\Delta y) |_{\text{NB}} = (n-1) !  [ k(\Delta y) ]^{-n+1} ,
\end{equation}
being 
\begin{equation}
  \tilde K_2 (\Delta y) |_{\text{NB} } = 1 / k(\Delta y)
	  \qquad\text{and}\qquad
	A_n = (n-1) ! .
\end{equation}

The scaling function $V_0 (\Delta y)$  turns out to be
\begin{equation}
	V_0 (\Delta y) |_{\text{NB}}  = \frac{ k(\Delta y) }{ \nbar (\Delta y)}
	  \log \left( 1 +  \frac{ \nbar (\Delta y) }{k(\Delta y) }
		\right) .
\end{equation}

%RICORDARSI REDUCED MOMENTS AND NORMALIZED

\begin{figure}
  \begin{center}
  \mbox{\includegraphics[width=\textwidth]{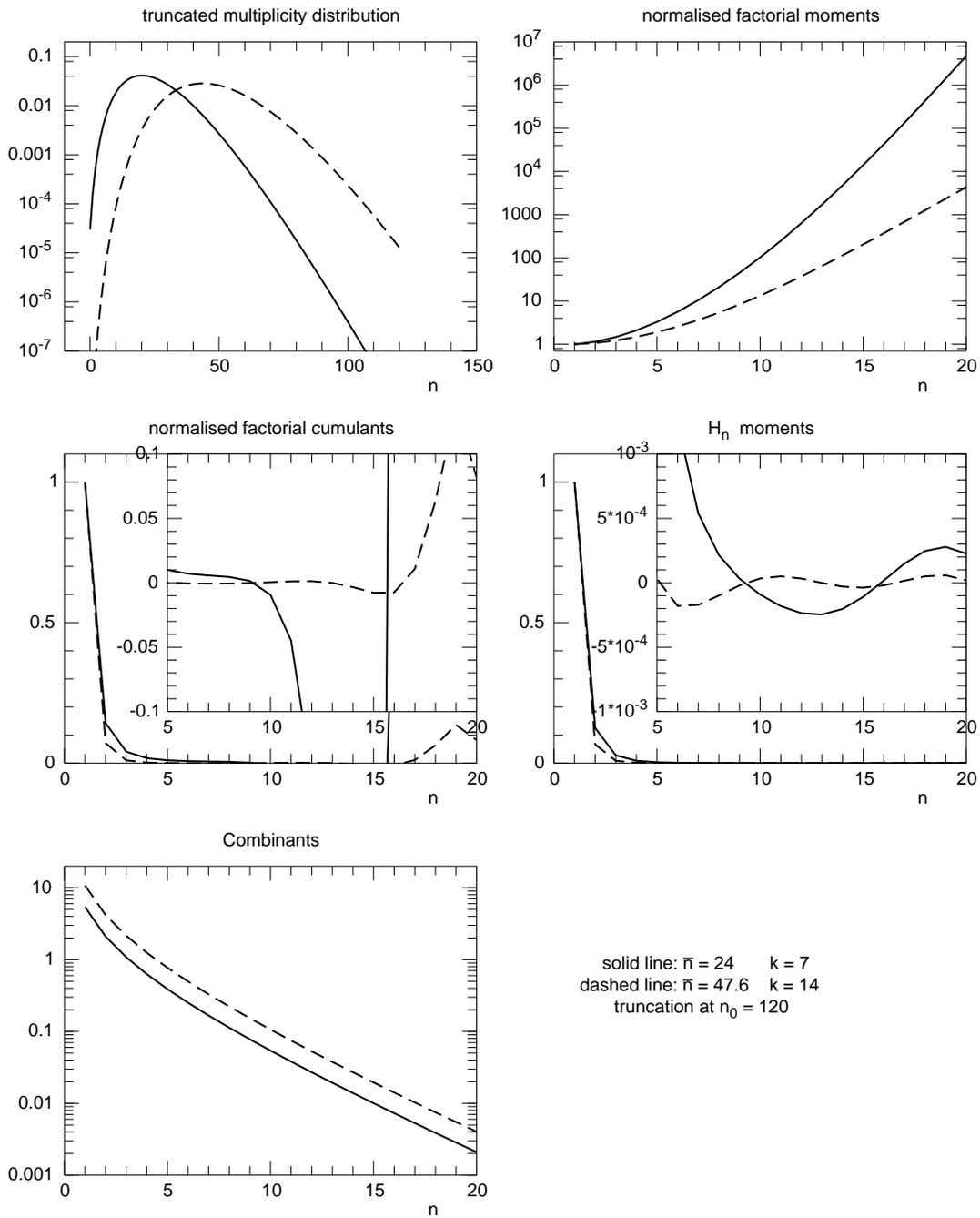}}
  \end{center}
  \caption[Global variables for the truncated NBD]{General behaviour
		of global observables  
		$P_n,\tilde K_n,\tilde F_n,H_n$ and $W_n$
		as in Fig.~\ref{fig:546fps} but for  truncated 
	  NB (Pascal) MD's.}\label{fig:trunc}
  \end{figure}

\begin{figure}
  \begin{center}
  \mbox{\includegraphics[width=\textwidth]{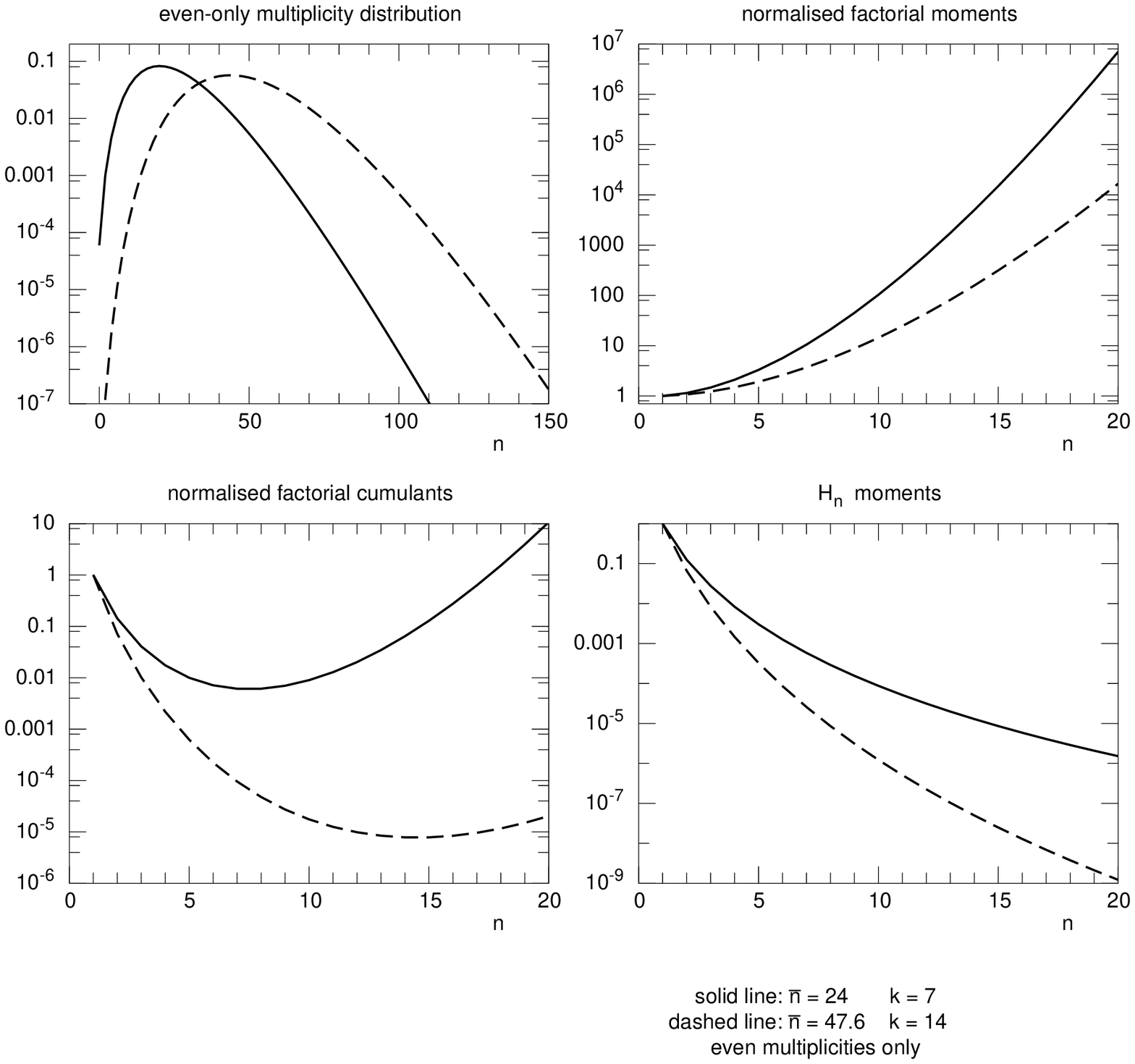}}
  \end{center}
  \caption[Global variables for the even-only NBD]{General behaviour
	of  global observables $P_n,\tilde
	K_n,\tilde F_n$ and $H_n$
	as in Fig.~\ref{fig:546fps} but for even-only
	NB (Pascal) MD's (combinants are not shown as they
	are not relevant in this context.)}\label{fig:even}
  \end{figure}

A second important application of $P_0 (\Delta y)$
resides with jet production in hadron-hadron
collisions: when jet production is associated with a colour-singlet
exchange process, one expects a signature of rapidity gap events
\cite{Bjorken2},
while the production associated with colour-octet exchange (i.e., the
one which is the focus of this review) will produce a much lower rate
of empty rapidity intervals \cite{pzero}.
The difference between the two processes could lie in the different
internal structure of clans \cite{Huang:1998gd}: 
in the colour-octet exchange, clans decay
with a logarithmic distribution, as discussed in the previous section;
in the colour-singlet exchange, clans decay with a geometric
distribution. 
In other words, in both cases Eq.~(\ref{eq:I.59p0}) remains valid, but
with different values of $\Nbar_{\text{g-clan}}$.
It was found that such a scheme can describe well 
the excess of low multiplicity events in the colour-singlet exchange
process \cite{Huang:1998gd}.

\subsection{CPMD's, truncation and even-odd effects.}\label{sec:I.5}  % 1.5

The maximum  number of observed  particles in the $n$ charged particle 
multiplicity distribution, $P_n$, never exceed a given number $n_0$, which of 
course increases with the c.m.\ energy of the collision.
Theoretically, at least from energy-momentum conservation,
this number is related to the available energy and the
mass of the lightest charged particle, the pion.
In experiments, there is in addition a practical limit (usually much
lower than the theoretical one) related to the luminosity and total
cross-section, i.e., dependent on the statistics.
In order to guarantee that the probabilities sum to one,
from the original (non-truncated) distribution $P_n$ a new
MD is defined, $\tilde P_n$,
\begin{equation}
	\tilde P_n = \begin{cases}
		      A P_n  &   \text{if $n \leq n_0$}\\
					0&         \text{if $n > n_0$}
				\end{cases} ,
\end{equation}
such that $\sum_{n=0}^{n_0} \tilde P_n = 1$.
$A$ is the normalisation factor. It follows from their definition
that factorial moments $F_n$ are equal to zero for $n > n_0$, 
whereas factorial 
cumulants $K_n$ are different from zero for any $n$. 
It should be pointed out that  
a truncated MD is not a CPMD and that at least one of its  combinants
$W_n$ is negative (for $n>n_0$.) 
Accordingly, factorial cumulants $F_n$ and the ratio of
factorial cumulants to factorial moments, $H_n$, of a truncated MD are
not positive definite.
Notice that combinants  of the truncated MD are not affected by the truncation
process for $n < n_0$ (combinants are sensitive indeed  to the head of the 
distribution and only to ratios $P_n/P_0$.)
The truncation process becomes relevant for  factorial cumulants, $K_n$,  and
for the ratio of factorial cumulants to factorial moments, $H_n$.
This fact reflects the peculiar property of $K_n$ and $H_n$ 
to be sensitive to the tail of the MD.
As an illustrative example, in Fig.~\ref{fig:trunc} we present the
same observables as in 
Fig.~\ref{fig:546fps}, for the case of a truncated NB (Pascal) MD.
Of particular interest are the zeros of the generating function of a
truncated MD  (the GF in this case is a polynomial of order
$n_0$).

Another conservation law which has to be obeyed is charge
conservation: in annihilation events, only even multiplicities
can be produced in FPS, while in narrow intervals this restriction is
negligible (the so-called even-odd effect).
One must also in this case re-normalise the MD:
\begin{equation}
	\tilde P_n = \begin{cases}
		      A' P_n^{\text{(NBD)}}&   \text{if $n$ is even}\\
					0&               \text{if $n$ is odd}
				\end{cases}
\end{equation}
so that $\sum_{n=0}^{\infty} \tilde P_{2n} = 1$.
For completeness, in Fig.~\ref{fig:even} we present the same
observables as in Fig.\ref{fig:546fps}
for a complete NB (Pascal) MD in which only even
multiplicities are non-zero.

\newpage
%%%%%%%%%%%%% %%%%%%%%%%% %%%%%%%%%%%%%%

\section{COLLECTIVE VARIABLES AND QCD PARTON SHOWERS}\label{sec:II}

\subsection{Parton showers in leading log 
approximation}\label{sec:II.leading-log}

Because our aim is to study the emission of soft gluons, fixed-order
calculations in the coupling constant $\alpha_s$ are insufficient and a
resummation of perturbative diagrams is needed.
The standard approach \cite{pQCD} is to select the set of diagrams which
dominate  the perturbative series at high energies; since in the
expansion one encounters physical amplitudes proportional to
$\alpha_s^n(Q^2) \log^m(Q^2)$ with $m \le n$, the leading terms are
those in which $m=n$, which give the so-called \emph{leading log
approximation} (LLA).

\begin{figure}
  \begin{center}
		\input{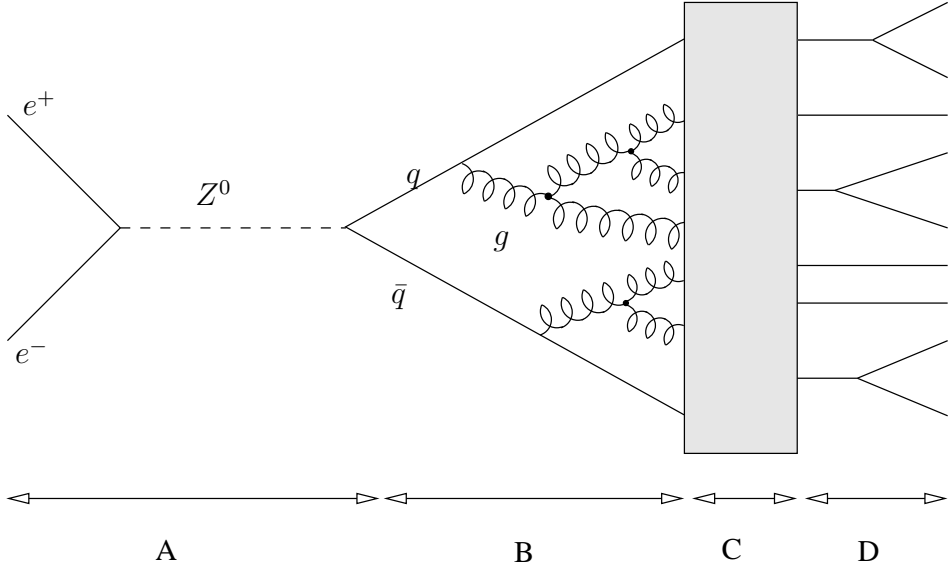}
  \end{center}
  \caption[Schematic view of electron-proton annihilation into
		hadrons]{Schematic view of electron-proton annihilation into
		hadrons.
    (A) electroweak step (scale: $10^{-17}$ cm); (B) hard processes,
		described by perturbative QCD (scale: $10^{-15}$ cm); (C) soft
		non-perturbative processes (hadronization, scale: $10^{-13}$ cm);
		(D) decay of resonances into observable hadrons (mostly 
		pions).}\label{fig:QCDPS_t}
  \end{figure}

In general, the factorisation theorem for collinear singularities
allows to build a parton model description of the production process.
For example, in the case of \ee\ annihilation into hadrons, the common
wisdom suggests the following scheme (see Fig.~\ref{fig:QCDPS_t}):
the electron and the positron annihilate into a virtual particle (a
photon or a $Z^0$) which then decays into a quark-antiquark pair; this
part is governed by the electroweak interaction theory. 
Perturbative QCD (pQCD) then describes the emission of further partons
(mostly gluons) until the virtualities involved become too small: in
this ``soft region'' where perturbation theory cannot be applied
the produced partons merge with very soft gluons
to form hadrons, often large mass resonances which then
decay according to standard model rules.

Let us now go into more details: the single-inclusive
cross section at hadron level $\sigma(\text{\ee}\to h X)$ can be
expressed in terms of a perturbative elementary cross-section
$\sigma_p(\text{\ee} \to q_i \bar q_i)$ for a hard process
at scale $Q^2$, and of a fragmentation
function $D_i^h(z,Q^2)$ which is interpreted as the probability of
finding a hadron $h$ in the fragmentation of a parton of flavour $i$,
carrying a fraction $z$ of the parton momentum:
\begin{equation}
	\frac{d \sigma(\text{\ee}\to h X)}{d z} =
	\sum_i \sigma_p(\text{\ee} \to q_i \bar q_i) D_i^h(z,Q^2) ,
\end{equation}
where the sum runs over all flavours.
Fragmentation functions, like structure functions in deep inelastic
scattering, are
universal and not calculable with perturbative methods. However, they
can be expressed in terms of a $Q^2$-dependent partonic fragmentation
function, $D_i^j(z,Q^2,Q_s^2)$, and a universal hadronic function 
at a fixed (soft) scale $Q_s$, $H_j^h(y,Q_s)$,
\begin{equation}
	D_i^h(z,Q^2) = \sum_j \int_0^1 \frac{d y}{y} D_i^j(z/y, Q^2,Q_s^2)
	   H_j^h(y,Q_s) .
\end{equation}
Furthermore, the evolution with the hard scale 
can be calculated and is given by the
Dokshitzer-Gribov-Lipatov-Altarelli-Parisi 
(DGLAP) equations:
\begin{equation}
	\frac{ d D_i^j(z,Q^2) } {d \log Q^2} =
	   \frac{\alpha_s(Q^2)}{2\pi} \sum_k \int_z^1 \frac{dx}{x}
		    D_k^j(z/x, Q^2, Q_s^2) P_{jk}(x) ,
\end{equation}
where $P_{jk}(z)$ is the DGLAP elementary kernel for the emission of
parton $k$ from parton $j$, with parton $k$ carrying a fraction $z$
of $j$'s momentum. 
These kernels can be computed with elementary perturbative 
methods \cite{pQCD}.

It is important to stress the simple physical meaning of DGLAP
evolution equations in LLA: in the axial gauge, this approximation
corresponds to select dressed ladder diagrams without interference
terms and with strong ordering in the virtuality and transverse
momentum of the offspring partons.
It is this feature which allows to interpret the production process in
a probabilistic language, in terms of a branching process. In this
language, the elementary DGLAP kernel $P_{ab}(z)$ 
gives the probability density
for a parton $a$ to emit a parton $b$ which carries a momentum
fraction $z$.

Taking into account also the virtuality, we are led to a
bi-dimensional elementary probability for the splitting of parton $a$
producing parton $b$ with momentum fraction $z$:
\begin{equation}
	 \frac{ \alpha_s(Q^2) }{ 2\pi } \frac{ dQ }{ Q } 
	   P_{ab}(z) dz .
\end{equation}
In order to normalise this probability, one has to re-sum all virtual
corrections, i.e., in the parton shower language, to take into account
the probability that no parton is emitted at virtualities larger than
$Q$; such probability is given by the Sudakov form factor:
\begin{equation}
	S_a(Q|W) = \exp\left\{
	          - \int_{Q^2}^{W^2}   \frac{ \alpha_s(K^2) }{ 2\pi }  
						   \left[
								     \sum_b
                       \int_{z_{\text{min}}}^{z_{\text{max}}}
										   P_{ab}(z) dz
						  \right] \frac{ dK^2 }{ K^2 } 
	         \right\} .
\end{equation}
Here the integration limits $z_{\text{min}}$ and 
$z_{\text{max}}$ depend on the
kinematics of the splitting: this means that in general they 
(and consequently the whole expression in brackets) depend
on the virtualities of both offspring partons and
not only on $K$: this makes a closed form
for $S_a(Q|W)$ impossible to find, unless some simpler approximation 
(like the fixed cut-off introduced below) is used.

Thanks to this probabilistic partonic picture, 
the result for single-inclusive distributions can be
generalised using the ``jet-calculus'' rules \cite{KUV} for $n$-parton
fragmentation functions.

It should be noticed that both collinear
and infrared singularities 
are present in the LLA expression of DGLAP kernels:
the former ones are avoided by imposing a soft cut-off on the evolution
($Q_s > 0$).
The simplest way of curing infrared divergences is to impose a fixed
(i.e., virtuality independent)
cut-off on $z$, say $z_{\text{min}} \equiv \epsilon' = 1 - z_{\text{max}}$; 
then one can
simply interpret the integral of the regularized kernels as
elementary splitting probabilities
\begin{align}
	A &\equiv \int_{\epsilon'}^{1-\epsilon'} P_{gg}(z) dz =
	     \frac{C_a}{\epsilon} = \frac{ N_c }{ \epsilon }\\
	\tilde A &\equiv \int_{\epsilon'}^{1-\epsilon'} P_{gq}(z) dz =
       \frac{ C_F }{ \epsilon} = \frac{ N_c^2-1 }{2\epsilon N_c }\\
	B &\equiv N_f \int_{\epsilon'}^{1-\epsilon'} P_{qg}(z) dz =
	     \frac{ N_f }{ 3 }
\end{align}
with $\epsilon = (-2\ln\epsilon')^{-1}$.

We will use the so-called ``jet thickness'' $\Y$ as evolution
variable:
\begin{equation}
	\Y = \frac{ 1 }{ 2\pi b} \log\left( 
	  \frac{ \alpha_s(Q^2) }{ \alpha_s(W^2) } \right) 
    =\frac{ 1 }{ 2\pi b} \log\left( 
		    \frac{ \log (W^2/\Lambda^2) }{ \log (Q^2/\Lambda^2) }
		\right) ,
\end{equation}
from virtuality $W$ down to $Q$,
where $b = (11N_c - 2N_f)/12\pi$ and we used the leading order expression
\begin{equation}
	\alpha_s(Q^2) = \frac{ 1 }{ b\,\log(Q^2/\Lambda^2)} .
\end{equation}
Then the probability $P_q(Q|W) dQ$
that a quark of virtuality $W$ splits at
a virtuality in the range $[Q,Q+dQ]$ (by emitting a gluon) is given by
\begin{equation}
	P_q(Q|W) dQ =  e^{-\tilde A \Y} \tilde A d\Y ,  \label{eq:II.1}
\end{equation}
and the probability $P_g(Q|W) dQ$ that a  gluon splits (by 
either emitting another gluon or a quark-antiquark
pair) by
\begin{equation}
	P_g(Q|W) dQ =  e^{-(A+B) \Y} (A+B) d\Y .   \label{eq:II.2}
\end{equation}
Neglecting conservation laws, the last two equations imply that the
splitting probability is constant for each $d\Y$ interval: this
allows to classify the process as Markovian, and therefore to write
the appropriate forward and backward
Kolmogorov equations for the probabilities to create
$n_q$ quarks and $n_g$ gluons from an initial quark,
$P_q(n_q,n_g;\Y)$, or an initial gluon, $P_g(n_q,n_g;\Y)$, at thickness
$\Y$ \cite{AGQCD}. The corresponding non-zero transition probabilities in
an infinitesimal interval $d\Y$ are as follows:
\begin{equation}
		\begin{split}
			(n_q,n_g) \to (n_q,n_g) \quad&=\quad 
			1 - An_g d\Y- \tilde A n_q d\Y - B n_g d\Y ;\\
			(n_q,n_g) \to (n_q, n_g+1) \quad&=\quad 
			An_g d\Y + \tilde A n_q d\Y ;\\
			(n_q,n_g) \to (n_q+2, n_g-1) \quad&=\quad B n_g d\Y .
		\end{split}
\end{equation}
It is simpler to use the generating functions (GF's):
\begin{equation}
	G_a(u,v;\Y) \equiv \sum_{n_q,n_g} u^{n_q} v^{n_g} P_a(n_q,n_g;\Y)
\end{equation}
where $a=q,g$.
One obtains the following differential equations \cite{AGQCD}:
\begin{align}
	\frac{ dG_g }{ d\Y } &= A (G_g^2 - G_g) + B (G_q^2 - G_g) ,\\
	\frac{ dG_q }{ d\Y } &= \tilde A G_q (G_g-1) .
\end{align}

At low energy, the production of quark-antiquark pairs is negligible,
and one can study solutions setting $B=0$: by looking only at the
evolution of the number of gluons, the above equations decouple and
the result is
\begin{align}
	G_g(u,v;\Y) &= v [ v + (1-v)e^{A\Y} ]^{-1}  \label{eq:QCD:Gg}\\
	G_q(u,v;\Y) &= u [ v + (1-v)e^{A\Y} ]^{-\tilde A/A}
\end{align}
The gluon multiplicity distribution
(MD) in a gluon-initiated shower is found to be a shifted
geometric distribution with average multiplicity $e^{A\Y}$ while the
gluon MD in a quark initiated shower is a NB (Pascal) MD with average
multiplicity $\nbar = \tilde A(e^{A\Y}-1)/A$ and parameter 
$k = \tilde A/A$, which for $\varepsilon$-regularization is just
$(N_c^2 - 1)/2N_c^2$ (i.e., 4/9 with 3 colours).
Clans in the quark jet case can be defined as in Section~\ref{sec:I.3}
and we obtain 
\begin{equation}
	\Nbar = \tilde A\Y ;\quad\quad \nc = \frac{ e^{A\Y}-1 }{ A\Y }.
\end{equation}
The two vertices: gluon production from a quark, controlled by
parameter $\tilde A$, and gluon emission from a gluon, controlled by
parameter $A$, have been separated and found to correspond
respectively to clan production and gluon showers inside clans,
in contrast to the standard parameterisation $\nbar$, $k$ where,
as shown, the two vertices are mixed.
This result leads to approximate clans to QCD bremsstrahlung 
gluon jets; gluons inside a clan follow on average a logarithmic 
distribution, which in turn can be written as a weighted average
of the geometric distribution given in Eq.~(\ref{eq:QCD:Gg})
(see Appendix \ref{sec:A.4}.) 

To summarise,
the essential conditions for the QCD interpretation of the occurrence
of NB (Pascal) MD here discussed are: i) the independent emission of
bremsstrahlung gluons, ii) the dominance of the vertex $g\to g+g$ over
$g \to q+\bar q$, and iii) weak effects of coherence and conservation laws.
The weakness of this approach is that it is very hard to introduce in
the QCD Markov branching process the dependence on rapidity and
therefore investigate MD's far from full phase-space.
We discuss in the next section some solutions to this problem.

%%%%%%%%%%%%%%%%%%%%%%%%%%%%%%%%%%%%%%%%%%%%%%%%%%%%%%%%%%%%%%%%%%%%%%%%%%%%

\subsection{The kinematics problem and possible
	answers.}\label{sec:II.kinematics} 
In the treatment of the previous section, 
correct kinematics has not been taken into
account: for example, at each splitting, phase-space is limited, as
one should make sure that in $a\to b+c$, the virtualities sum
up correctly: $Q_a \geq Q_b + Q_c$. Furthermore, energy and momentum
conservation are not ensured.
In order to address these problems, several methods have been
developed: we will briefly mention the most popular of them, then
discuss the simplified parton shower (SPS) and
the generalised simplified parton shower (GSPS) models, which, although
overlooked by the common wisdom, deserve in our opinion particular
attention as an attempt to work in a fully correct kinematical
framework by using essential features of QCD.

\subsubsection{DLA and MLLA and Monte Carlo}\label{sec:II.kinem.dla}
In order to include more kinematics in the shower evolution,
perturbations theory can be improved \cite{pQCD,Dremin:2000ep,Khoze:1997dn}. 
In the Double Log Approximation
(DLA), for example, one takes into account the phase-space for the
emission of a soft gluon; however, recoil is not 
considered for the parent parton, i.e., energy is not conserved in
each individual emission. This can be justified in a first
approximation by the hypothesis that emitted gluons are required to be
soft, which is certainly applicable at asymptotic energies. Notice
that DLA still allows a probabilistic interpretation of parton
production as a cascade.

A noteworthy result in DLA is the prediction of 
Koba-Nielsen-Olesen (KNO) scaling \cite{KNO}  for the MD of
gluons in a gluon jet; the high multiplicity tail is given by
\begin{equation}
	f(n/\nbar) \propto \exp ( -Cn/\nbar ) ,
\end{equation}
where $C\approx 2.55$, calculated in pQCD. This KNO scaling form does
not depend on the running of $\alpha_s$, being insensitive to all
details of the evolution but the branching structure of the process
($\alpha_s$ of course appears in the energy dependence of $\nbar$.)

In order to bring into play also recoil effects, thus trying to solve
the problem of the too prolific parton production of DLA, one arrives
at the Modified Leading Log Approximation (MLLA) equations.
Recoil effects modify the argument of evolution equations, taking the
energy lost in the creation of a new parton into account.
The equations become then differential equations with retarded terms
(difference-differential equations). The probabilistic interpretation
of the branching is still retained.

By considering only gluon production in a gluon-initiated jet,
the DLA prediction has been corrected with pre-asymptotic terms, and
the KNO scaling result is now violated at finite energies
\cite{Dokshitzer}. 
The new result for the tail is
\begin{equation}
	f(n/\nbar) \propto \exp [ - (C'n/\nbar)^\mu ] ,
\end{equation}
where $\mu = (1-\gamma)^{-1}$ and $C' = C \gamma^\gamma
(1-\gamma)^{1-\gamma} / \Gamma(1 + \gamma)$ and $\gamma =
d\log\nbar/d\log Q$ is the anomalous dimension.
MLLA predictions are much closer to experimental data than DLA ones,
but the overall description is still poor, especially near the
maximum. The tail is still close to an exponential (like that of a
NB (Pascal) MD). 

\begin{figure}
  \begin{center}
  %\mbox{\includegraphics[width=0.8\textwidth]{plot_qcdHq.ps}}
  \mbox{\includegraphics[width=0.5\textwidth]{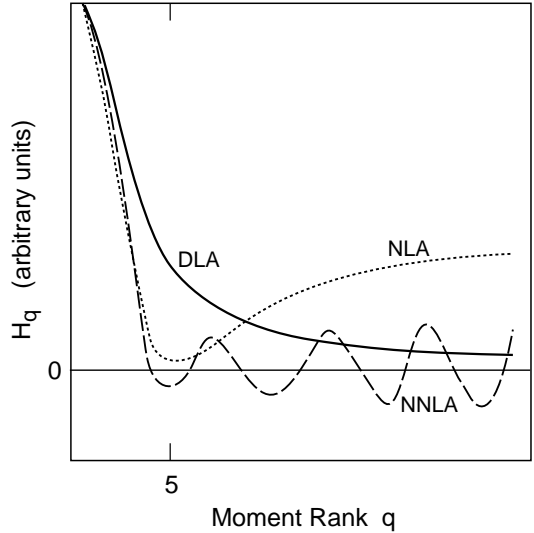}}
  \end{center}
  \caption[Hq ratio in NBD,DLA,MLLA]{Ratio of cumulant to factorial
  moments, $H_q$, as function 
  of the order $q$, computed in pQCD, using different 
  approximations \cite{SLD}:
	notice that `oscillations' only appear in
  NNLA (dashed line).}\label{fig:HqDremin}
  \end{figure}

In addition to study the c.m.\ energy dependence of the MD, much can
be learned from a study of its moments as a function of the order,
at fixed energy.
It was indeed found in \cite{Dremin:1993} 
that it is possible to obtain analytic
formulae within MLLA, even with the addition of higher order terms
in an expansion in $\gamma$, in the limit of frozen $\alpha_s$, 
for the ratio $H_q$ of factorial cumulants to
factorial moments.
Numerical solutions \cite{Dremin:1993} with running $\alpha_s$ have
confirmed the 
oscillating behaviour of $H_q$ as a function of the order $q$
found analytically for solutions to the QCD equations for the
generating functions. 
See Fig.~\ref{fig:HqDremin}.
Oscillations of this type have been found experimentally,
but before comparing to the pQCD predictions one has to take into
account the truncation of the tail in the data (which was shown also
to produce comparable oscillations \cite{hqlett}) and the effect of
hadronization (actually, there is no effect when using 
generalised local parton-hadron duality, see
Section~\ref{sec:II.hadronization}.)

In all analytical calculations however, while energy conservation is
taken into account, momentum conservation is handled only by the angular
ordering prescription; one then resorts to numerical methods,
usually in terms of a Monte Carlo simulation of a parton shower:
the evolution of partons can be traced
step-by-step at each LLA vertex,
imposing all conservation laws and required phase-space restrictions.
Furthermore, not only multiplicities, but every aspect of the reaction may be
simulated, although usually at the cost of adding extra parameters.
It is remarkable that excellent results
have been obtained in \ee\ annihilation into hadrons but not in
minimum-bias \pp\ collisions, where general features like MD's are
still poorly described.

The Monte Carlo approach has been however particularly useful to obtain
phenomenological ideas to be used in analytical calculations
(see, e.g., Section~\ref{sec:II.hadronization} on hadronization.)

\subsubsection{The SPS model}\label{sec:II.kinem.sps}
Another possibility to introduce kinematics constraints in QCD
shower evolution is to isolate the fundamental features of pQCD
on the basis of which a simplified analytical model can be developed
in a correct kinematical framework.
This is what has been attempted with the Simplified Parton Shower
(SPS) model \cite{SPS}, described here in some detail.

% solved problem: Q_s was used for the cut-off in virtuality aka soft scale!
We consider an initial parton of maximum allowed virtuality $W$ which
splits at virtuality $Q$ into two partons of virtuality $Q_0$ and
$Q_1$.  We require $Q \ge Q_0 + Q_1$ and $Q_0, Q_1 \ge 1$ GeV; this
implies that any parton with virtuality less than 2 GeV cannot split further.
We define the probability for a parton of virtuality $W$ to split at $Q$,
$p(Q \vert W)$, which is normalised by a Sudakov form factor, as in
Eq.~(\ref{eq:II.1}) or (\ref{eq:II.2}). Since we are interested in the structure
of the solution, we will discuss only one type of parton, introducing
a free parameter which we call $A$, and use
\begin{equation}
	p(Q \vert W)dQ = \frac{A}{Q} \frac{(\log Q)^{A-1}}{(\log W)^A} dQ 
	   =d\left( \frac{\log Q}{\log W} \right)^A  .
		 \label{eq:sps:(9)}
\end{equation}
The probability for an ancestor parton of maximum allowed virtuality $W$
to generate $n$ final partons, $P_n(W)$, and the probability for a
parton which splits at virtuality $Q$ to generate $n$ final partons,
$R_n(Q)$, with $P_n(Q) = R_n(Q) = \delta_{n1}$ for $Q < 2$ GeV, lead to the
following generating functions:
\begin{equation}
	f(z,W) = \sum_{n=1}^{\infty} P_n(W) z^{n-1} ,
	\label{eq:sps:(1)}
\end{equation}
\begin{equation}
	g(z,Q) = \sum_{n=2}^{\infty} R_n(Q) z^{n-2} ,
	\label{eq:sps:(2)}
\end{equation}
where the series has been shifted w.r.t.\ the standard definition in
consideration of the fact that there is always at least one parton in the
cascade; this definition will simplify some of the formulae to follow.
The two generating functions are linked by
\begin{equation}
	f(z,W) = \int_1^2 p(Q \vert W) dQ + \int_2^W p(Q \vert W) z g(z,Q) dQ .
	\label{eq:sps:(3)}
\end{equation}
The joint probability density ${\mathcal{P}}(Q_0Q_1|Q)$ for 
a parton of virtuality $Q$ to 
split into two partons of virtuality $Q_0$ and $Q_1$ is defined by  
\begin{equation}
	{\mathcal{P}}(Q_0Q_1|Q) = p(Q_0 \vert Q) p(Q_1 \vert Q) K(Q) 
           \theta(Q - Q_0 - Q_1) ,
	\label{eq:sps:(4)}
\end{equation}
where $K(Q)$ is a normalisation factor and the conditions on the
virtualities are shown explicitly.
The dynamical content of the model in {\it virtuality} 
is expressed by the following equation
\begin{equation}
    R_n(Q) = \sum_{n'=1}^{n-1} \int_1^{\infty} dQ_0 
		   \int_1^{\infty} dQ_1 {\mathcal{P}}(Q_0Q_1|Q) 
			 R_{n-n'}(Q_0)  R_{n'}(Q_1)  ,
			 \label{eq:sps:(5)}  
\end{equation}
which gives the probability for a parton which splits 
at virtuality $Q$ to  generate $n$ final partons in terms of 
the joint probability density, Eq.~(\ref{eq:sps:(4)}),
and of daughter parton's respective showers.

\begin{figure}
  \begin{center}
		\scalebox{0.7}{\input{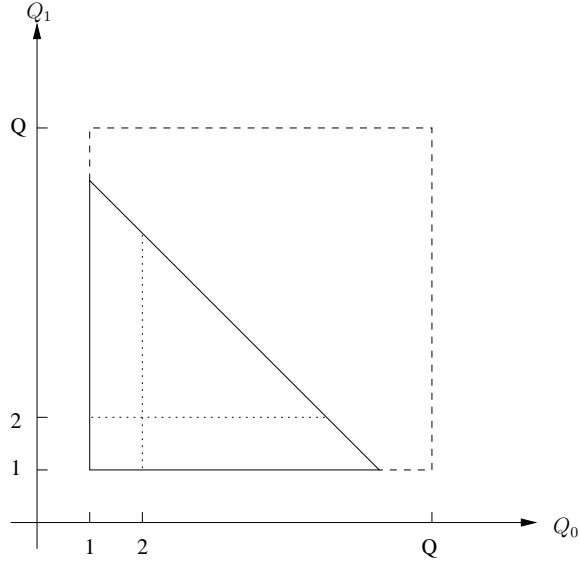}}
  \end{center}
  \caption[Virtuality allowed in SPS]{Kinematically allowed virtuality
  domain in the SPS model for a parton splitting at virtuality $Q$
  producing two offspring partons of virtualities $Q_0$ and $Q_1$,
  with $Q_0+Q_1<Q$. he dashed line indicates the extension of phase
  space due to the violation of the virtuality conservation law; the
  dotted lines indicate the domain in which only one of the two
  offsprings can split further \cite{Sergio:thesis}.}\label{fig:SPS:triangolo}
  \end{figure}

Eq.~(\ref{eq:sps:(5)}) can be reformulated for the corresponding 
generating function by dividing the domain of integration in 
three sub-domains; one obtains 
\begin{equation}
	\begin{split}
		g(z,Q) &= \int_1^2 dQ_0 \int_1^2 dQ_1 {\mathcal{P}}(Q_0Q_1|Q) \\
		&\quad + 2 \int_1^2 dQ_0 \int_2^{\infty} dQ_1 
    {\mathcal{P}}(Q_0Q_1|Q) z g(z,Q_1) \\
		&\quad + \int_2^{\infty} dQ_0 \int_2^{\infty} dQ_1 
    {\mathcal{P}}(Q_0Q_1|Q) zg(z,Q_0) zg(z,Q_1) .
		\label{eq:sps:(6)}
	\end{split}
\end{equation}
The above three sub-domains correspond to the possible different 
situations in which the two generated partons can be found,  i.e., 
neither of them splits, only one parton splits, or both partons split
(see Figure \ref{fig:SPS:triangolo}.)
This general scheme is valid for any choice of splitting function $p(Q \vert
W)$, but in the present case this function factorizes and
Eq.~(\ref{eq:sps:(3)}) simplifies into the differential equation
\begin{equation}
	\frac{\partial f(z,W)}{\partial W} = p(W \vert W) [z g(z,W) -
	f(z,W)] .
	\label{eq:sps:(8)}
\end{equation}

This equation can be easily solved in two extreme cases, obtained
respectively by restricting or by relaxing phase-space constraints.

In the first case we allow to generate only very soft partons:
$\theta(Q - Q_0 - Q_1) \to \theta(Q-Q_0)\theta(2-Q_1)$.
This results in bremsstrahlung-like emission and a Poisson
distribution of generated partons:
\begin{equation}
	f(z,W) = e^{\lambda(W)(z-1)} ,
\end{equation}
where
\begin{equation}
	\lambda(W) = \int_2^W p(Q|Q) dQ = A \log \left( \frac{ \log W }{ \log 2}
	\right) 
\end{equation}
gives the average multiplicity of generated partons (in addition to
the initial ancestor).

The second case is obtained by decoupling the virtualities as in
$\theta(Q - Q_0 - Q_1) \to \theta(Q-Q_0)\theta(Q-Q_1)$.
This is exactly the case examined for the LLA with fixed cut-off, 
Eq.~(\ref{eq:QCD:Gg}), and
indeed the result is a geometric distribution
\begin{equation}
	f(z,W) = \left[ 1 - (z-1)(e^{\lambda(W)}-1) \right]^{-1} ,
\end{equation}
with average multiplicity of generated partons $e^{\lambda(W)}-1$.

Further analytical solutions of the SPS model were not possible, but
numerical solutions using Monte Carlo methods show that the
NB (Pascal) MD describes the MD's rather well \cite{SPS}, which is not
surprising considering that the NB (Pascal) MD interpolates between
the Poisson ($1/k=0$) and the geometric ($k=1$) distributions.

% Attenzione alle distribuzioni traslate!
Moreover, by differentiating  the above equations for
generating functions in $z=1$, one can obtain equations for factorial
moments which are linear in the moments themselves, although they
contain all moments of lower order. This feature, common to other
evolution equations, is a consequence of the parton shower
structure. But because the SPS model implements a correct kinematical
framework, at small virtualities the equation simplify and a
recursive solution can be achieved. Such a solution has to be
numerical, but it was found \cite{Sergio:thesis} that the virtuality
evolution is consistent with pQCD results and experimental
observations for values of the parameter $A$ between 1.5 and 2.
The behaviour of high order factorial moments qualitatively agrees
with experimental data and with the most detailed pQCD calculations
and is still consistent with NB (Pascal) predictions.

For describing the {\it rapidity} structure of the model, we propose
to use the singular part of the QCD kernel controlling gluon branching:
\begin{equation}
  p(y_0|Q_0Q_1Qy) \propto 
	   P(z_0)dz_0 \propto \frac{dz_0}{z_0 ( 1 - z_0)}
		 \label{eq:sps:(10)} 
\end{equation}
Here $z_0$ is the energy fraction carried away by the produced parton 
in the infinite momentum frame.

The limits of variation of $z_0$ are fixed by the exact kinematical relations 
\begin{equation}
	B - \sqrt{B^2 - \left( \frac{Q_0}{Q} \right)^2} \le z_0 \le B + 
	   \sqrt{B^2 - \left( \frac{Q_0}{Q} \right)^2} ,
		 \label{eq:sps:(11)}
\end{equation}
where $B = \frac{1}{2} [1 + ({Q_0}/{Q})^2 - ({Q_1}/{Q})^2]$ is 
the scaled parton energy in the centre of mass system 
and $\sqrt{B^2 - (Q_0/Q)^2}$ its 
maximum scaled transverse momentum. In this way the scaled 
transverse momentum (w.r.t.\ the parent direction)
of the parton with energy fraction $z_0$
\begin{equation}
	\frac{|\pt_0|^2}{Q^2} = \sqrt{B^2 - \left( \frac{Q_0}{Q} \right)^2} - 
	   (z_0 -B)^2 
	\label{eq:sps:(12)}
\end{equation}
and its rapidity
\begin{equation}
	y_0 = y + \frac{1}{2} \log \frac{ z_0}{2B - z_0}
	\label{eq:sps:(13)}
\end{equation}
are uniquely determined.
Rapidity of the second parton of virtuality $Q_1$ 
is obtained by energy-momentum conservation:
\begin{equation}
  y_1 = y + {\tanh}^{-1} \ \left[ \frac{B}{B+1} {\tanh}\ |y_0-y|
  \right] .
	\label{eq:sps:(14)}
\end{equation}
Notice that only the first step has to be treated differently in 
rapidity because it corresponds to the degrading from the maximum 
allowed virtuality $W$ to the virtuality of the first splitting $Q$. In 
this case the rapidity of the ancestor is fixed by conservation laws and 
is given by 
\begin{equation}
	y = {\tanh}^{-1} \ \sqrt{1 - \left( \frac{Q}{W} \right)^2} .
	   \label{eq:sps:(15)}
\end{equation}

Again an analytical solution was not possible, but good fits to the
Monte Carlo implementation of the SPS model in
rapidity intervals were obtained with the 
NB (Pascal) MD.

\subsubsection{The GSPS model: generalised clans}\label{sec:II.kinem.gsps}

At this level of investigation, it should be clear that still open 
problems are the lack of the analytical 
solution of Eq.~(\ref{eq:sps:(6)}) and, in more general terms, 
the lack of a complete analytical 
study of the parton evolution process in rapidity.

\begin{figure}
  \begin{center}
  \mbox{\includegraphics[width=0.8\textwidth]{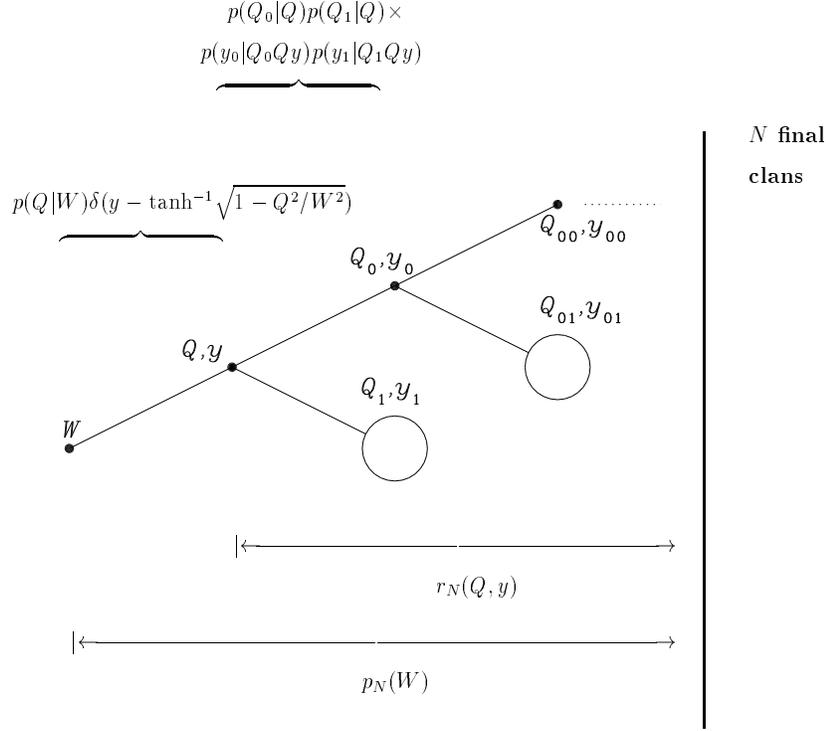}}
  \end{center}
  \caption[GSPS, clan structure of]{The structure of the Generalised
		Simplified Parton Shower  
model (GSPS) for clan production in virtuality $Q$ and rapidity $y$. 
The production process is decoupled at each splitting both in virtuality 
and in rapidity, see Eqs.~(\ref{eq:sps:(18)}) and (\ref{eq:sps:(20)}).
Each blob represents a clan. Notice that here  $Q_0 \le Q$,  
$Q_1 \le Q$, $|y_0 - y| \le \log {Q/Q_0}$,
$|y_1 - y| \le \log {Q/Q_1}$ \cite{GSPS}.}\label{fig:bubbles}
  \end{figure}

In order to solve part of these problems, we proposed to 
incorporate in the SPS model the idea of \emph{clans}; 
we called this version of the model 
``Generalised Simplified Parton Shower'' (GSPS) 
model \cite{GSPS,GSPS:2}. Accordingly, 
we decided to pay attention for each event  to the ancestor which, 
splitting $n$ times, 
gives rise to $n$ subprocesses (one at each splitting, 
see Fig.~\ref{fig:bubbles}) and 
we identify them with clans. Therefore in this model for 
a single event the concept of clan at parton level 
is not a statistical one, as it was in the SPS model:
in the present picture the clans are independent active parton sources 
and their number in each event coincides with the number 
of splittings of the ancestor, i.e., with the number of steps in the 
cascade.

Notice that each clan generation is independent of previous history 
(it has no memory); thus the process is Markovian. Furthermore each 
generation process depends on the evolution variable only and is 
independent of the other variables of the process like the number of 
clans already present and their virtualities.
In the original version of the SPS model, the splitting function of 
the first step, $p(Q_0|Q)$, was different from the splitting 
function of all the other steps, $\beta(Q_0|Q)$, obtained by integrating 
the joint probability function $\P$,
\begin{equation}
	\beta (Q_0 \vert Q) = \int_1^{Q-Q_0} dQ_1 \P .
	\label{eq:sps:(16)}
\end{equation}

In order to generalise the SPS model as it stands 
we assume that 
virtuality conservation law is locally violated (although conserved globally) 
according to 
\begin{equation}
	1 \le Q_0 \le Q, \qquad 1 \le Q_1 \le Q .
	\label{eq:sps:(17)}
\end{equation}
The upper limit of integration of Eq.~(\ref{eq:sps:(16)}) becomes 
$Q$, the normalisation factor $K(Q)$ reduces to 1 and 
the process becomes homogeneous in the evolution variable since 
\begin{equation}
	\beta(Q_0|Q) = p(Q_0|Q) \int_1^Q p(Q_1|Q) dQ_1 = p(Q_0|Q) .
	\label{eq:sps:(18)}
\end{equation}
The approximation described by this equation, therefore,  
can be interpreted as the effect of local 
fluctuations in virtuality occurring at each clan emission.

This violation of the virtuality 
conservation law spoils of course the validity of the energy-momentum 
conservation law, which, in the SPS model, uniquely determines 
the rapidity 
of a produced parton, given its virtuality and the virtuality and 
rapidity of its germane parton --see Eq.~(\ref{eq:sps:(14)}). 
In the GSPS model 
the two produced partons at each splitting are independent 
both in virtuality and in  rapidity; however, 
the rapidity of each parton is bounded by the extension of phase-space 
fixed by its virtuality and the virtuality of the parent parton:
\begin{equation}
	|y_i - y| \le \log \frac{Q}{Q_i} .
	\label{eq:sps:(19)}
\end{equation}
In conclusion, by weakening locally conservation laws, we 
decouple the production process of partons at each splitting. 
Consequently, the GSPS model allows to follow just a 
branch of the splitting, since 
each splitting can be seen here 
as the product of two independent parton emissions.
This consideration will be particularly useful in discussing the 
structure in rapidity of the model; in fact, it is implied that 
the DGLAP kernel given in Eq.~(\ref{eq:sps:(10)}) should be identified with
\begin{equation}
	p(y_0|Q_0Qy)  dy_0  p(y_1|Q_1Qy) dy_1 \propto 
	\frac{dz_0}{z_0}  \frac{dz_1}{z_1} .
		\label{eq:sps:(20)}
\end{equation}

Then 
one gets the multiplicity distribution  $p_N(W)$:
\begin{equation}
	p_N(W) % = \int_1^2 dQ_0 p_N(Q_0|W) 
	= e^{-\lambda(W)} \frac{[\lambda(W)]^{N-1} }{ (N-1)!} ,\qquad N > 0 .
		\label{eq:sps:(30)}
\end{equation}
Eq.~(\ref{eq:sps:(30)}) is a shifted Poisson distribution in the
number of clans $N$, with average number of clans given by:
\begin{equation}
	\bar N(W) %\equiv \sum_{N=0}^{\infty} N p_N(W)
	= \sum_{N=1}^{\infty} N e^{-\lambda(W)} \frac{[\lambda(W)]^{N-1}}{(N-1)!} 
	= \lambda(W) + 1   .
	\label{eq:sps:(31)}
\end{equation}
We stress that this result has been obtained {\it a priori} in the 
present  
generalised version of the model, differently from 
what has been done previously in \cite{AGLVH:1} where the independent 
production of clans was introduced 
{\it a posteriori} in order to explain 
the occurrence of NB (Pascal) regularity.

Since clans are by definition not correlated (i.e., only particles
belonging to the same clan are correlated, while particles belonging to
different clans are not)
the MD of clans in rapidity intervals is obtained by binomial
convolution
\begin{equation}
	p_N(\Delta y, W) = \sum_{N'=N}^\infty 
	  \binom{N'}{N} \pi^N (1-\pi)^{N'-N} p_{N'}(W)  ,
\end{equation}
where $\pi(\Delta y,W)$ is the probability that one clan is produced
within the interval $\Delta y$ by an ancestor parton of maximum
virtuality $W$.
The average number of clans in the interval $\Delta y$ is therefore
\begin{equation}
	\Nbar(\Delta y,W) = \pi(\Delta y,W) \Nbar(W).
\end{equation}
In terms of generating functions
\begin{equation}
  F_{\text{clan}}(z;\Delta y) = \left[ z \pi(\Delta y,W) +
		1 - \pi(\Delta y,W)\right] e^{\pi(\Delta y,W)\lambda(W)(z-1)} .
\end{equation}
Presenting the sum of two Poissonian distribution (the first a 
shifted one), this relations has a nice physical meaning: 
the two terms correspond to the 
probability of having the ancestor within or outside the given rapidity 
interval.
When $\Delta y$ is very small, $\pi(\Delta y,W)$ 
tends to zero  and the unshifted Poissonian dominates. Thus, it can be 
stated that in the smallest rapidity intervals the clan multiplicity is 
to a good approximation Poissonian and the full MD belongs 
to the class of Compound Poisson Distribution. 
$\pi(\Delta y,W)$ tends to 1 for large $\Delta y$  and 
the exact full phase-space shifted-Poisson distribution is approached.
Finally, when $\pi(\Delta y,W)\lambda(W)$ is sufficiently large, the 
shifted-Poissonian dominates at large $N$ (the tail of the distribution)
while the unshifted one dominates at small $N$ (the head of the 
distribution).
These facts might have some consequences in interpreting the anomalies 
found in NB behaviour for small $N$ and the deviations from NB behaviour 
in large rapidity intervals, which are controlled by the behaviour of the 
distribution at large $N$.

The calculation of $\Nbar(\Delta y,W)$ can then be carried out
analytically, although the procedure is very tedious. The key
observation is that, thanks to the decoupling, the elementary
splitting function for an ancestor parton of virtuality $Q$ to produce a
clan of virtuality $Q_1$ has the same functional form of the
probability that the next splitting of the ancestor itself happens at
virtuality $Q_0$. Therefore, the ancestor and the clan at the last
step of the shower evolution can be interchanged.
We will skip the details of the calculation, to be found in \cite{GSPS},
and proceed to illustrate the results.

\begin{figure}
  \begin{center}
  \mbox{\includegraphics[width=0.5\textwidth,angle=-90]{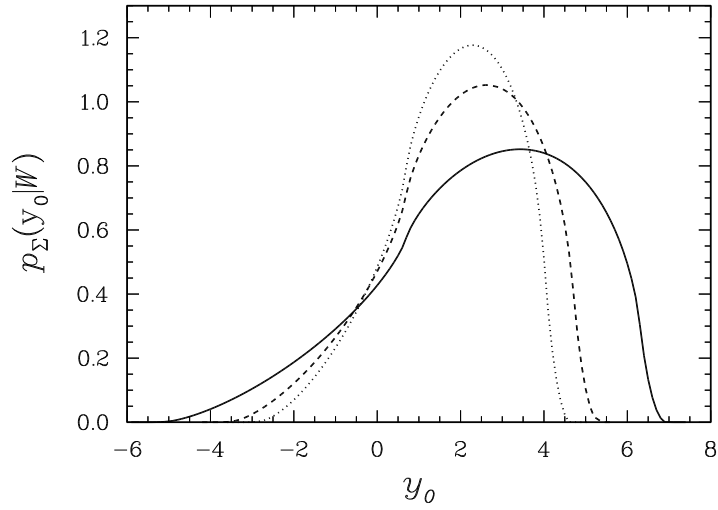}}
  \end{center}
  \caption[Clan density in rapidity]{Clan density in rapidity, i.e.,
  number of clans per unit 
  rapidity, in a single parton shower, 
  according to the GSPS model with parameter $A=2$. 
  The three curves refer to different initial virtualities 
  (dotted line: $W=50$ GeV; dashed line: $W=100$ GeV; solid line:
  $W=500$ GeV)
  and are normalised each to its own average
  number of clans in full phase-space \cite{GSPS}.}\label{fig:XX}
  \end{figure}

\begin{figure}
  \begin{center}
  \mbox{\includegraphics[width=0.5\textwidth,angle=-90]{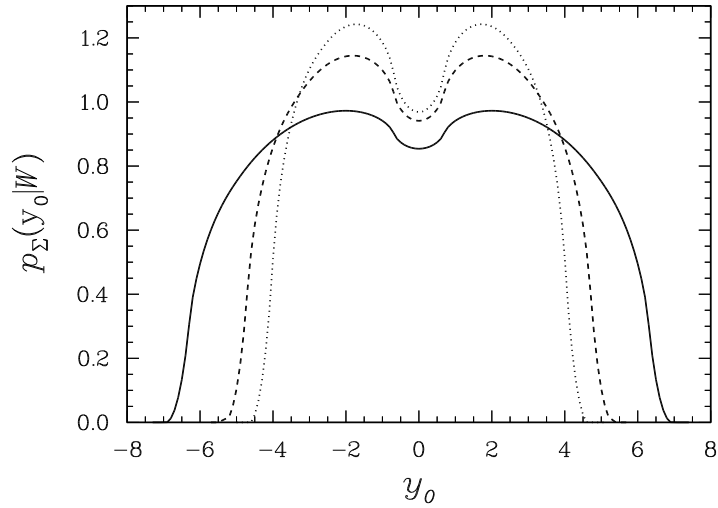}}
  \end{center}
  \caption[Clan density in rapidity, b2b]{Clan density in rapidity resulting
  by the addition of two 
  back-to-back showers, at different initial virtualities in the GSPS
  model \cite{GSPS}. All parameters as in Fig.~\ref{fig:XX}}\label{fig:XX1}
  \end{figure}

In Fig.~\ref{fig:XX}
the clan density  is shown  as a function of the width of the interval
$\Delta y \equiv [-y_c,y_c]$  for $A=2$ for maximum allowed virtualities 
$W = 50$ GeV, $W=$ 100 GeV and $W= 500$ GeV. 
The contribution of one-parton showers turns out to be 
negligible for this choice of $A$.
Notice that the height of the distribution is 
decreasing (simply because the full phase-space value increases only
as a double log)
and the width increasing with the energy (phase-space grows
logarithmically). 
Convolution of clan density for two back-to-back 
parton showers is shown in Fig.~\ref{fig:XX1}. 
Notice that  the 
central dip at $y \simeq 0$ is slowly removed by increasing the energy of the 
initial parton. It should be kept in mind that the structure of
Figures \ref{fig:XX} 
and \ref{fig:XX1} refers to clan production; the different behaviour for 
parton production inferred from data
is not in contradiction with this behaviour since we have 
still to include in our scheme parton production within a single clan.

\begin{figure}
  \begin{center}
  \mbox{\includegraphics[width=0.5\textwidth,angle=-90]{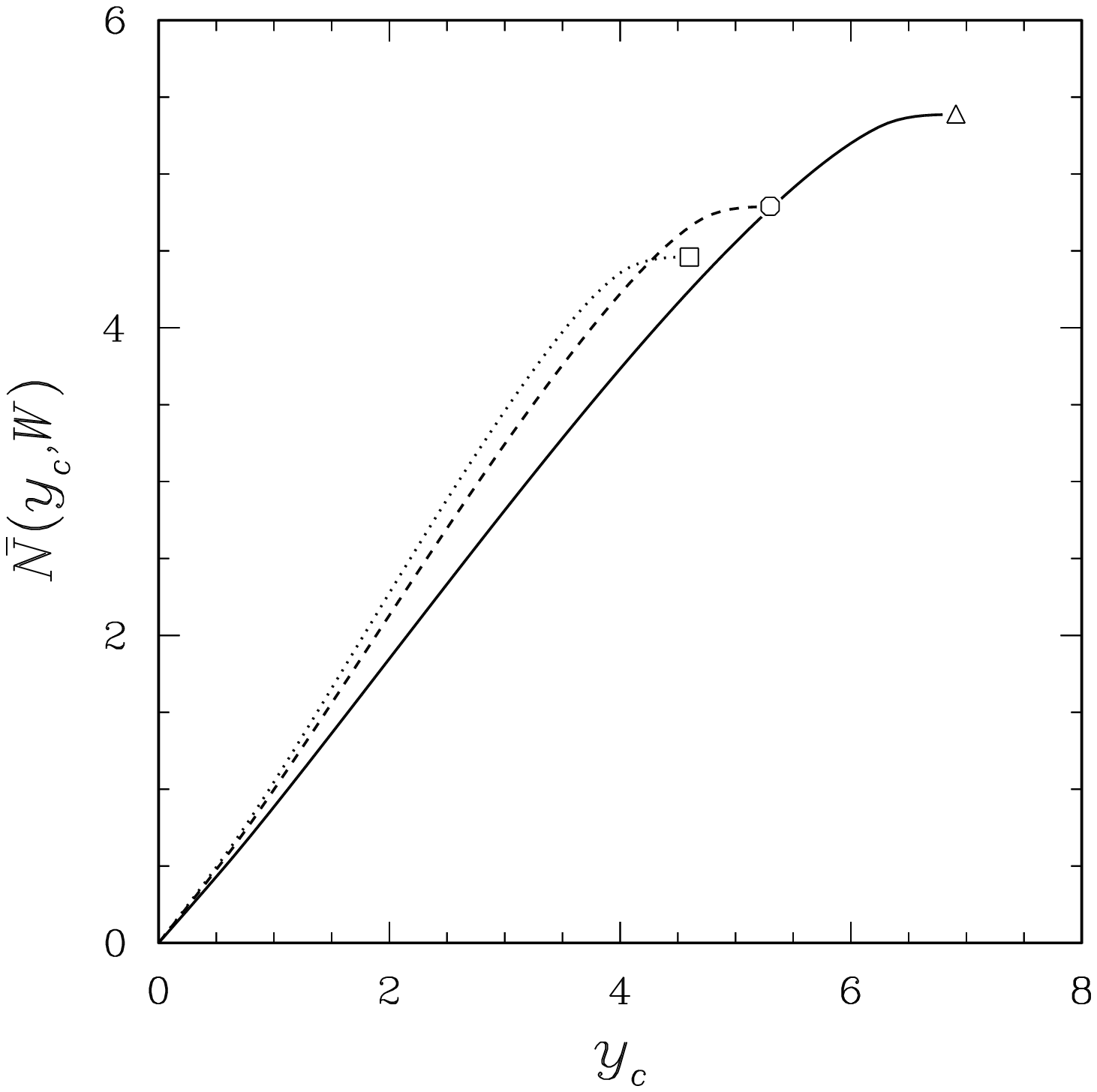}}
  \end{center}
  \caption[Average number of clans in the GSPS model]{Average number
		of clans in a central rapidity interval  
		$[-y_c,y_c]$ as a function of the interval's half-width $y_c$
		for a single shower in the GSPS model with $A=2$.
    The three curves refer to different initial virtualities 
    (dotted line: $W=50$ GeV; dashed line: $W=100$ GeV; solid line:
    $W=500$ GeV) \cite{GSPS}.}\label{fig:XX2}
  \end{figure}

In Figure \ref{fig:XX2} the average number of clans $\Nbar(\Delta y,W)$
is given as a 
function of rapidity width $y_c$ for the same $W$ values of 
Figure \ref{fig:XX}. 
Limitations on the rapidity intervals are determined by the 
available phase-space corresponding to the different initial parton 
virtualities. 

\begin{figure}
  \begin{center}
  \mbox{\includegraphics[width=0.5\textwidth,angle=-90]{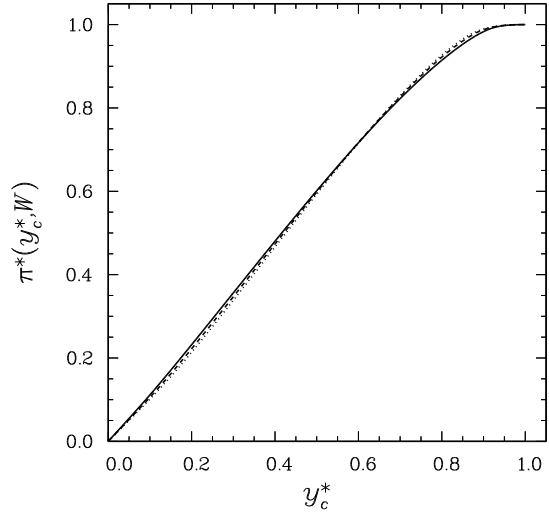}}
  \end{center}
  \caption[Average number of clans, rescaled]{The same curves as in
  Fig.~\ref{fig:XX2}, rescaled in the 
  abscissa to the respective full phase-space 
  (FPS) rapidity and in the ordinate to the
  respective FPS average number of clans, i.e.,
  Eq.~(\ref{eq:sps:Nbar.rescaled}) 
  vs.\ $y_c^* = y_c/y_{\text{fps}}$ \cite{GSPS}.}\label{fig:XX3}
  \end{figure}

Accordingly, 
the GSPS model predicts, for the average number of clans 
at parton level in a single shower (jet):

\begin{enumerate}
\item[{\it a)}]
a rise with the rapidity width $y_c$, for 
different initial parton virtualities $W$,
which is very close to linear 
for $1 < y_c < y_{\text{fps}}$; 
the rise is still linear but with a somewhat different slope for $y_c <1$.
A characteristic bending occurs finally for 
rapidity width $y_c \lesssim y_{\rm fps}$;

\item[{\it b)}]
approximate (within 5\%) energy independence in a fixed rapidity interval 
$y_c$ for $W$ below 100 GeV. 
For higher virtualities, deviations from energy independence become 
larger; they are within 20\% when comparing $\bar N(y_c,50\ {\text{GeV}})$ 
and $\bar N(y_c,500\ {\text{GeV}})$. It should be noticed that the average 
number of clans slowly decreases with virtuality;  
this behaviour has been already observed in Monte Carlo simulations for 
single gluon jets \cite{Single}.
\end{enumerate}

In addition to the above results which are consistent with our 
expectations on clan properties in parton showers, the model shows energy 
independent behaviour (see Figure \ref{fig:XX3}) 
by normalising the average number of 
clans produced in a fixed rapidity interval $|y| \leq y_c$ to the 
corresponding average number in full phase-space, and by expressing this 
ratio as a function of the rescaled rapidity variable $y_c^* \equiv y_c/
y_{\rm fps}$:
\begin{equation}
	\pi^*(y_c^*,W) \equiv \frac{\bar N(y_c^* \  y_{\rm fps},W) }{ \bar
	N(W)}   \label{eq:sps:Nbar.rescaled}
\end{equation}
This new regularity turns out to be stable for different choices of the 
parameter $A$. In Figure \ref{fig:XX3} a clean linear behaviour 
is shown for the 
above ratio corresponding to the parameter value $A$=2. 

Having completed the analytical treatment of the number of clans, we
now proceed to the final partons level.
In order to study the average number of clans in the rapidity interval
$\Dy$, $\bar N(\Dy,W)$, in the previous treatment
we limited our discussion to the first step of parton shower
evolution in the GSPS model. It is clear that if one wants to
calculate the average number of partons per clan in the same interval,
$\bar n_c(\Dy,W)$, one has to analyse the second step of parton shower
evolution, i.e., to study the production of partons
inside clans.
In order to do that, inspired by the criterion of
simplicity and previous findings, we decided
to maintain inside a clan
the structure of the model seen in the first step.
The only difference lies in the
introduction of a new parameter, $a$, controlling the length of the
cascade inside a clan, in the expression of the probability that a parton
of virtuality $Q$ emits a daughter parton in the virtuality range
[$Q_0, Q_0+dQ_0$], i.e.,
\begin{equation}
	p_a(Q_0|Q) dQ_0 = d\left(\frac{\log Q_0 }{ \log Q}\right)^a \; .
	\label{eq:psplit2}
\end{equation}
The GSPS model is thus a two-parameter model: $A$ and $a$,
controlling the length of the cascade in step 1 and 2 respectively.
This is equivalent to introducing the SPS structure for a single clan;
however, by maintaining the decoupling in each splitting (a la DLA)
one keeps the possibility to solve the model exactly.

The MD in a clan of virtuality $Q$ is therefore
\begin{equation}
	g_{\text{fps}} (z,Q ) =\begin{cases}
	\frac{z}{1 +(\bar n_c(Q )-1)(1-z)}  &Q  \ge 2\ {\text{GeV}}  \\
   z &  Q < 2 \ {\text{GeV}}
	 												 \end{cases}
\end{equation}
where
\begin{equation}
	\bar n_c(Q) = e^{\lambda_a(Q)}\; ,
	\qquad \lambda_a(Q) = \int_2^Q p_a(Q'|Q')dQ' \; .
\end{equation}
The solutions correspond to a shifted
geometric distribution ($Q \ge 2$ GeV)
and to a clan with only one
parton ($Q < 2$ GeV). The bound 2 GeV is a consequence of the fact
that in the GSPS model the virtuality cut-off is fixed at 1 GeV
(a parton with virtuality $Q  < 2 $ GeV cannot split any further
by assumption).
Notice that this finding agrees with the clan model discussed in
\cite{AGLVH:3}, where the logarithmic MD for partons inside average clans is
interpreted as the result of an
average over geometrically distributed single clans of different
multiplicity, i.e., initial virtuality (See also Appendix \ref{sec:A.4}.)

We now calculate the
generating function of partons MD in a rapidity interval $\Dy$
inside a clan of virtuality $Q$ and rapidity $y$: this is done
through a binomial convolution on the corresponding generating function
in full phase-space:
\begin{equation}
	g_{\Dy}(z,Q,y) =\begin{cases}
	\frac{1 + (z-1) \pi_a(\Dy,Q,y)}
	{1 +\pi_a(\Dy,Q,y)(\bar n_c(Q)-1)(1-z)}  &
	Q  \ge 2 \ {\text{GeV}} \\
  1 + (z-1) \pi_a(\Dy,Q,y) & Q  < 2 \ {\text{GeV}}
									\end{cases}
\end{equation}
where $\pi_a(\Dy,Q,y)$ is the probability that a clan of initial
virtuality $Q$ and rapidity $y$ produces a daughter parton inside the
interval $\Dy$. 
Notice that this approximation neglects rapidity correlations among
particles in the same clan. However, it allows exact analytical
solutions (although long and cumbersome to handle).
The average number of partons per clan is then
\begin{equation}
	\nc(\Dy,Q,y) = \pi_a(\Dy,Q,y) e^{\lambda_a(Q)} .
\end{equation}

\begin{figure}
  \begin{center}
  \mbox{\includegraphics[width=0.6\textwidth]{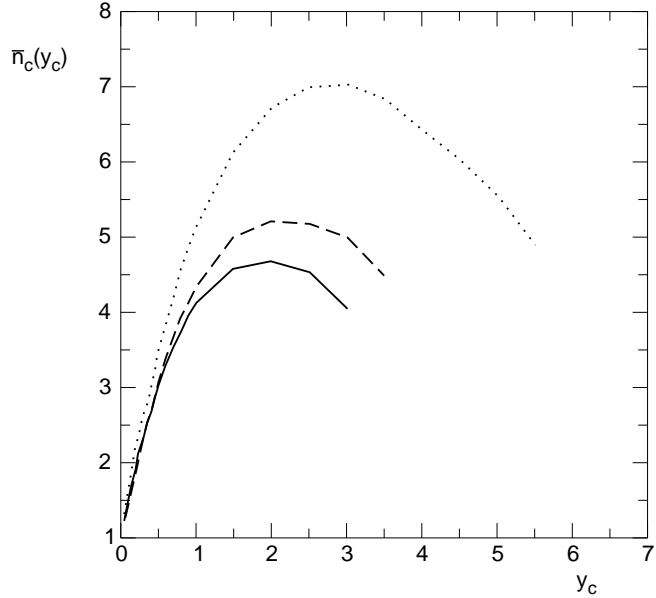}}
  \end{center}
  \caption[Average number of partons per clan in GSPS]{Average number
		of partons per clan in a central rapidity interval  
		$[-y_c,y_c]$ as a function of the interval's half-width $y_c$
		for a single shower in the GSPS model with $A=2$, $a=1$.
    The three curves refer to different initial virtualities 
    (dotted line: $W=50$ GeV; dashed line: $W=100$ GeV; solid line:
    $W=500$ GeV) \cite{GSPS:2}.}\label{fig:ZZ1}
  \end{figure}

\begin{figure}
  \begin{center}
  \mbox{\includegraphics[width=0.7\textwidth]{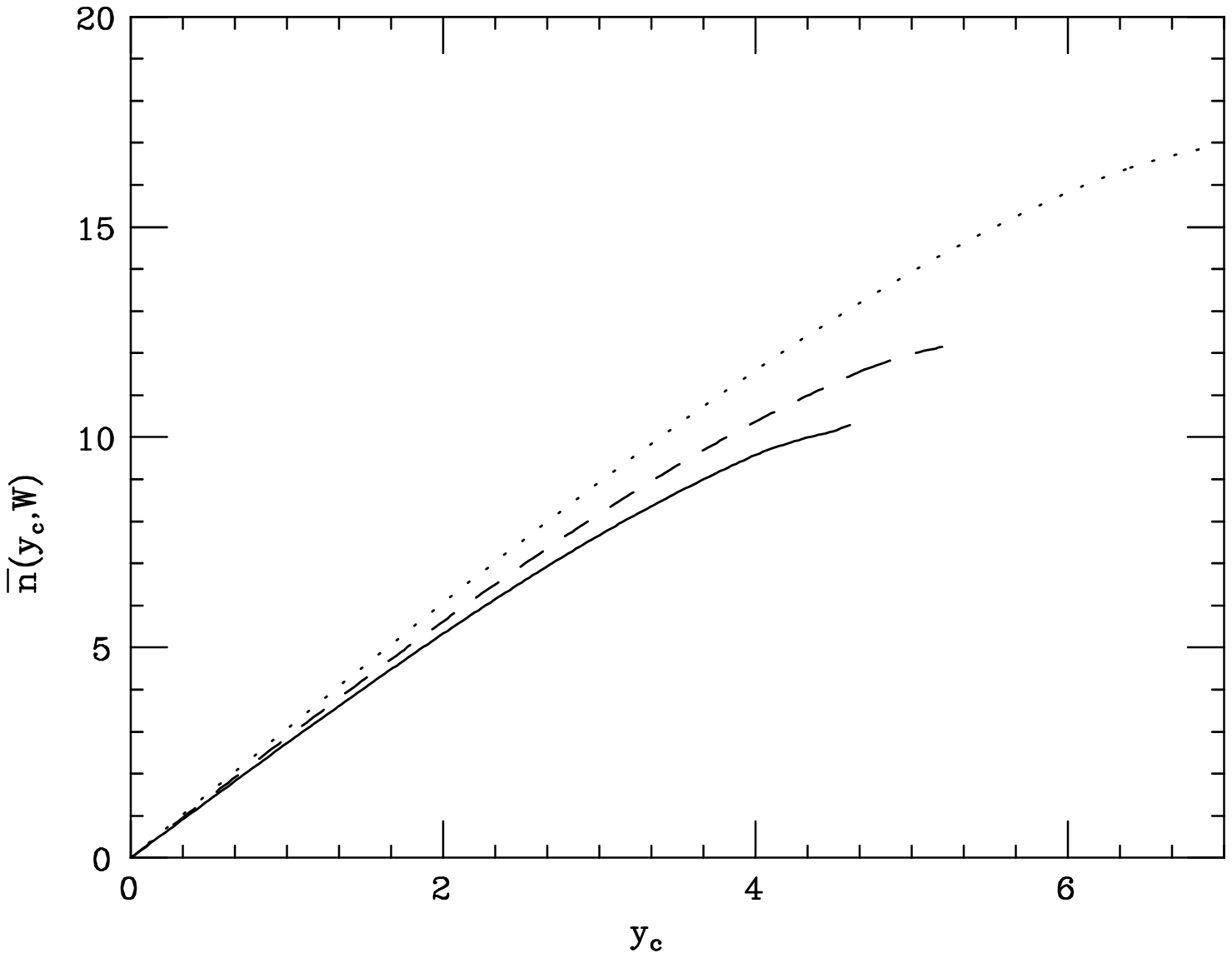}}
  \end{center}
  \caption[Average number of partons in GSPS]{Average number of
		partons in a central rapidity interval  
		$[-y_c,y_c]$ as a function of the interval's half-width $y_c$
		for a single shower in the GSPS model with $A=2$, $a=1$.
    The three curves refer to different initial virtualities 
    (dotted line: $W=50$ GeV; dashed line: $W=100$ GeV; solid line:
    $W=500$ GeV) \cite{GSPS:2}.}\label{fig:ZZ2}
  \end{figure}

\begin{figure}
  \begin{center}
  \mbox{\includegraphics[width=0.7\textwidth]{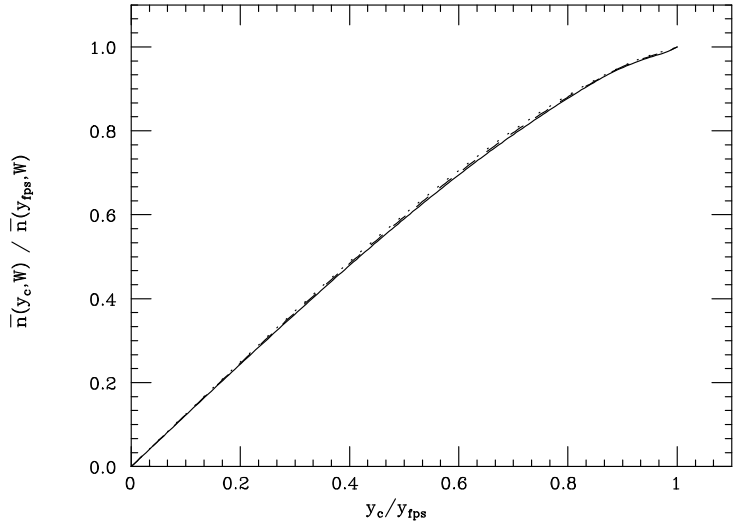}}
  \end{center}
  \caption[Average number of partons, rescaled]{Same curves as in
	Fig.~\ref{fig:ZZ2}, rescaled 
	in abscissa to the respective FPS rapidity, and in ordinate to the
	respective average FPS multiplicity \cite{GSPS:2}.}\label{fig:ZZ3}
  \end{figure}

When the above equation is averaged over the probability 
that a parton
of maximum allowed virtuality $W$ produces a clan of virtuality $Q$
and rapidity $y$, we obtain the average number of partons in an
average clan generated in a shower of virtuality $W$.
As probability over which we average, one can use the bi-dimensional clan
density in virtuality and rapidity normalised by the average number of
clans. 
Results of the calculations of the average number of particles per clan
as a function of the rapidity interval $\Dy$ and
of the maximum allowed virtuality $W$ with $A=2$ and $a=1$
are shown in Fig.~\ref{fig:ZZ1} for $W$= 50 GeV (solid line), $W$=100 GeV
(dashed line) and $W$= 500 GeV (dotted line).
The trend fully coincide with the behaviour of clans structure
parameters obtained by analysing quark and gluon jets MD's \cite{Single}.
The result of the analytical calculation   of the average number of
partons in the shower, $\nbar(\Dy,W)$
is shown in Figure \ref{fig:ZZ2} with the same parameters.
The average  parton multiplicity grows   almost linearly
with rapidity for relatively small $\Dy$ intervals and then it is slowly
bending for $\Dy$ intervals approaching f.p.s., where it reaches its
maximum.
It is interesting to remark that the normalised average number of partons in
the shower, $\nbar(\Delta y, W)/\nbar({\text{fps}}, W)$
scales in virtuality  as a function
of the rescaled rapidity interval, $\Dy/{\text{fps}}$,
see Figure \ref{fig:ZZ3}.
This scaling in $W$ is found to depend on the parameter $a$, as different
values of $a$ give different scaling curves,
differently from the scaling found for
for $\bar N(\Dy,W)/\bar N({\text{fps}},W)$ which is independent
of the mechanism inside clans.

In conclusion, the GSPS model is a parton shower model which was built
assuming QCD-inspired dependence of the splitting functions in
virtuality and in rapidity, with Sudakov form factors for their
normalisation; in addition to these ingredients 
(which were called ``essentials of QCD''), the characterising feature
was introduced of distinguishing explicitly the two steps of clan production
and subsequent decay, allowing at each step in the cascade 
local violations of
energy-momentum conservation laws but requiring its global validity.
This model was found to have an important predictive power in 
regions not accessible to pQCD.
Analytical calculations of the virtuality and rapidity dependence of
the average number of clans and of the average number of
particles per clan have been carried out.
Results are consistent with what is known on clan properties in single
quark and gluon jets, disentangled at hadron level 
by using jet finding algorithms and
analysed at parton level assuming that 
generalised local parton-hadron duality (discussed in the next
Section) is applicable.

%%%%%%%%%%%%%%%%%%%%%%%%%%%%%%%%%%%%%%%%%%%%%%%%%%%%%%%%%%%%%%%%%%%%%%%%%%%%

\subsection{Hadronization prescriptions}\label{sec:II.hadronization}

Having performed computations at parton level, the problem arises on
how to make the connection with the measured hadron level.
Such a process is of course non-perturbative in nature, and usually
approached through various models: the string model \cite{LundModel} 
and the cluster model \cite{Marchesini:1988cf} 
are widely used in Monte Carlo calculations; statistical
hadronisation models \cite{Becattini:1996if+Becattini:1997rv} 
are now starting to know considerable success.

It has been noticed however that many predictions of perturbation
theory can reproduce experimental results down to low virtuality
scales, and often give the correct energy evolution except for an
overall normalisation.
This behaviour can be expressed in terms of pre-confinement: the
perturbative evolution is continued to low virtualities while partons
rearrange themselves in their evolution to form colour singlet
clusters which hadronize subsequently at a soft scale of the order of
the perturbative cut-off to the shower. This picture lead to the Local
Parton-Hadron Duality (LPHD, `weak' duality) prescription: 
single-particle inclusive
distributions at hadron level are taken proportional to the
corresponding distribution at parton level:
\begin{equation}
	Q_1^{(h)}(y)  = \rho Q_1^{(p)}(y)  ,  \label{eq:LPHD}
\end{equation}
where in general $\rho\approx 2$.
It is a way to investigate to what extent pQCD can directly reproduce
experimental data up to a rescaling factor. In particular, integrating
Eq.~(\ref{eq:LPHD}) one obtains for the average multiplicities:
\begin{equation}
	\nbar^{(h)} = \rho \nbar^{(p)} .
\end{equation}

In \cite{AGLVH:2}, it was noticed in that Monte Carlo calculations of MD's, the
NB (Pascal) MD provided a good fit both at parton and hadron level,
with the same 
parameter $k$. This lead to the formulation of Generalised LPHD
(GLPHD; `strong' duality) which brings into play higher order
inclusive distributions:
\begin{equation}
	Q_n^{(h)}(y_1,\dots,y_n) = 
	\rho^n Q_n^{(p)}(y_1,\dots,y_n) .
\end{equation}
In general, one expects $\rho > 1$.
Integrating over rapidity, one obtains the proportionality of
(un-normalised) factorial moments:
\begin{equation}
	F_n^{(h)} = \rho^n F_n^{(p)} ;
\end{equation}
recalling Eq.~(\ref{eq:13}) which links factorial moments with generating
functions, one obtains the relation:
\begin{equation}
	G^{(h)}(z) = G^{(p)}(1-\rho + \rho z) .  \label{eq:GLPHD:gf}
\end{equation}
This relation leaves unchanged all the observables that do not depend
on the average 
multiplicity. It can easily be seen, e.g., that normalised factorial
moments are the same at parton and hadron level, being $F_1^{(h)} =
\rho F_1^{(p)}$. The same is true for 
normalised factorial cumulant moments,
since they can be expressed as a sum over factorial moments (recall
the cluster expansion that links inclusive distributions to
correlation functions, Eq.~(\ref{eq:I.8}). Of course, it follows that
also the ratio
$K_n/F_n = H_n$ is invariant under the GLPHD transformation:
\begin{align}
	K_n^{(h)} &= \rho^n K_n^{(p)} ;\\
	H_n^{(h)} &= K_n^{(h)}/F_n^{(h)} =  K_n^{(p)}/F_n^{(p)} = H_n^{(p)} .
\end{align}

When applying Eq.~(\ref{eq:GLPHD:gf}) to MD's which are infinitely
divisible at parton level, one still obtains (barring pathological
cases) distributions which are CPMD's:
\begin{equation}
	G^{(h)}(z) = \exp\{ \Nbar^{(p)}[ g^{(p)}(1-\rho + \rho z) - 1] \} .
\end{equation}
However, at $z=0$ one obtains $g^{(p)}(1-\rho) \ne 0$, which
violates the condition that there are no empty clans. One has to
redefine the MD within clans \cite{LVH:1,Finkel}:
\begin{equation}
	g^{(h)}(z) = \frac{ g^{(p)}(1-\rho + \rho z) - g^{(p)}(1-\rho) }{ 
		1 - g^{(p)}(1-\rho)} .
\end{equation}
This is equivalent to the following transformation on the clan
parameters:
\begin{align}
	\Nbar^{(h)} &= \Nbar^{(p)} \left[ 1 - g^{(p)}(1-\rho) \right] , \\
	\nc^{(h)} &= \nc^{(p)} \frac{ \rho }{1 - g^{(p)}(1-\rho) } .
\end{align}
Because (in reasonable situations) $g^{(p)}(1-\rho) \leq 0$, one has
$\Nbar^{(h)} \geq \Nbar^{(p)}$. It is interesting to remark that hadronic
clans do not coincide with the partonic ones: hadronisation creates
in general new clans (or breaks partonic ones) \cite{AGBecattini}. 
Indeed one also has $\nc^{(h)} \geq \nc^{(p)}$.
The exact sharing of the multiplicity increase between $\Nbar$ and
$\nc$ depends on the actual shape of $g^{(p)}(z)$.

As an example, in the case of the NB (Pascal) MD we obtain that GLPHD
is equivalent to the following simple requirement:
\begin{equation}
	\nbar^{(h)} = \rho \nbar^{(p)} \quad,\qquad k^{(h)} = k^{(p)} :
\end{equation}
notice the NB(Pascal) shape is not lost, only one parameter changes when
going from the partonic to the hadronic level. In this example, both
the average number of clans and the average number of partons
(particles) per clan increase during hadronization:
\begin{align}
	\Nbar^{(h)} &= k^{(p)} \ln\left( 1 + \rho\nbar^{(p)}/k^{(p)} \right),
	\\
	\nc^{(h)} &= \frac{ \rho \nbar^{(p)} }{ k^{(p)} 
		\ln\left( 1 + \rho\nbar^{(p)}/k^{(p)} \right)} .
\end{align}

It is worth pointing out that GLPHD has some drawbacks: relation
(\ref{eq:GLPHD:gf}) resembles a convolution with a binomial
distribution, except that $\rho > 1$. On one hand this fact prevents a
probabilistic interpretation of GLPHD: it is impossible to define 
event by event a probability distribution for obtaining $n$ particles 
from $m$ partons satisfying Eq.~(\ref{eq:GLPHD:gf}). On the other hand
it allows to study the distributions that are invariant in
form under transformation (\ref{eq:GLPHD:gf}): they are all MD's whose
GF depends on $z$ and $\nbar$ only via the product $\nbar(z-1)$, i.e.,
which satisfy the following differential equation:
\begin{equation}
	\nbar \frac{ \partial G(z) }{ \partial \nbar } = 
	  (z-1) \frac{ \partial G(z) }{ \partial z }.
\end{equation}
This equation can be formally solved and gives, at the level of
probabilities,  Eq.~(\ref{eq:I.53}) ---see \cite{Sergio:thesis}.
%% This equation can be formally solved and gives at the level of
%% probabilities the following relation:
%% \begin{equation}
%% 	P_n = \frac{ (-\nbar)^n }{ n!} \frac{ \partial^n P_0 }{ \partial
%% 	  \nbar^n } ,
%% \end{equation}
%% where only $\nbar$ is allowed to vary in $P_0$ while all other
%% parameters are taken fixed with respect to $\nbar$ variations.

It should be clear by now 
that GLPHD is a very strong prescription, certainly too
strong: there are effects present at hadronic level only (e.g.,
resonances' decays) which affect hadronic correlations and not partonic
ones; on the contrary, GLPHD fixes all correlations already at parton
level (except for their overall strength). Nonetheless, GLPHD is very
useful as a tool to investigate the predictive power of purely
perturbative calculations: recall for example the case of $H_q$
moments oscillations.

\newpage
%%%%%%%%%%%%% %%%%%%%%%%% %%%%%%%%%%%%%%

\section{COLLECTIVE VARIABLES REGULARITIES IN MULTIPARTICLE 
   PRODUCTION: DATA AND PERSPECTIVES}\label{sec:III}

The attempt to achieve a unified, QCD-inspired  description of multiparticle 
production in all classes of collisions, in full phase-space and in its limited 
intervals,   both at final hadron and parton level  and  from the soft
sectors up to the hard ones, including high parton density regions, is   
the challenge in the field.
The main motivation of the present section  is the conviction 
---already stated in the introduction--- that  complex 
structure which we observe might very well be, at the origin, simple, and that 
such initial simplicity manifests itself  in terms of   regularities of  final 
particle multiplicities.
With this aim,  it  is  instructive and stimulating ---in our
opinion---  both from a theoretical and an  experimental point of view, to  
follow the advent of the NB (Pascal) MD regularity and of  KNO scaling 
violation in multiparticle production (facts which haven't yet been fully 
appreciated in all their implications),  then to see  
the  sudden  failure of the regularity as a consequence of  the shoulder 
structure observed in $n$ charged particle MD's $P_n$ when plotted vs.\
$n$,    and finally
its reappearance at a deeper level of  investigation, i.e., in  the description
of the  substructures or classes of events   of the various  collisions.

What makes even more attractive this development is the 
finding   that, under certain reasonable  assumptions,  the 
NB (Pascal) MD occurs also (see Section \ref{sec:II}) at 
final parton level:  the generating function 
of the NB (Pascal) MD  is  the solution of the
differential QCD evolution Equations  which can be understood as a  
Markov branching process  initiated by a parton and controlled in its 
development by QCD elementary probabilities.
It should be noticed   that the same regularity appears also  
in final $n$-parton 
multiplicity distributions in $q\bar q$ and $gg$ systems in Jetset 7.2 Monte 
Carlo calculations based on DGLAP equations. 
In addition, by using  a convenient guesswork as hadronization prescription, 
the NB (Pascal) MD describes within the 
same Monte Carlo generator the hadron level, which turns out not to be 
independent from the parton level but linked to it by  strong GLHP
duality.
 All these results can be interpreted, as we shall see, in terms of
clan structure analysis and suggest that the dynamical mechanism
responsible of multiparticle production in all classes of collisions
is independent intermediate gluon sources (the clan ancestors)
emission followed by QCD parton shower formation.

\subsection{An unsuspected regularity in particle production 
in cosmic ray physics}\label{sec:III.cosmic}

\begin{figure}
  \begin{center}
  \mbox{\includegraphics[width=0.7\textwidth]{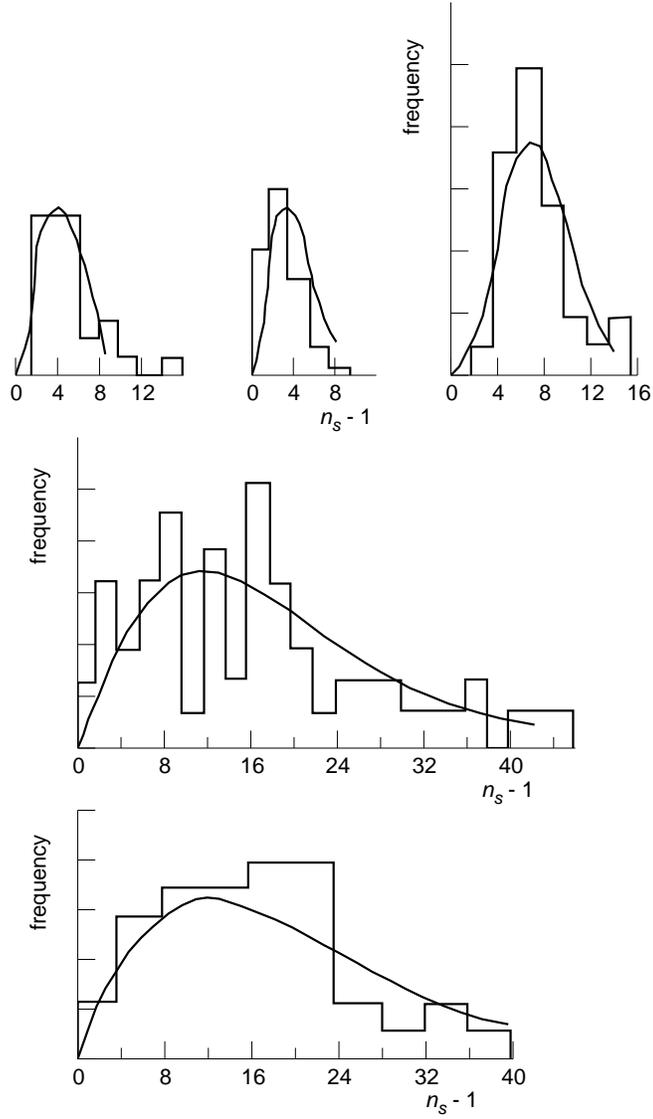}}
  \end{center}
  \caption[Cosmic rays MD data]{MD of produced particles from a
  variety of experiments in 
  cosmic ray vs the number of charged tracks $n_s$. The curves
  represent NB (Pascal) MD's (a) $E_0=30$ GeV, (b) $E_0=25$ GeV, (c)
  $E_0=300$ GeV, (d) $E_0=1.5\cdot10^4$ GeV, (e) $E_0=10^4$ GeV, where
  $E_0$ is the primary nucleon energy \cite{Cosmic}.}\label{fig:wolfendale}
  \end{figure}

After the discovery of multiparticle production (groups of ``mesotrons'') in 
cosmic rays  in the thirties  and the important theoretical work
in the forties and early fifties in order to explain the highly non linear 
new phenomenon,  the contribution of the  1966  paper by 
P.M. MacKeown and A.W. Wolfendale \cite{Cosmic} should be mentioned.
Its interest lies in the attempt to describe all available experimental
data on exclusive $n$-particle cross sections, $\sigma_n{(E_0)}$, 
for producing  $n$ charged pions generated  by a primary
nucleon at different  energies $E_0$ (from $25$~GeV up to $5\cdot10^4$ GeV)
by using a NB (Pascal) MD, i.e.
\begin{equation}
	\sigma_n(E_0) = \frac{\nbar^n}{n!}  
	   \left[ 1+ \frac{\nbar}{k(E_0)}\right] ^{-n - k(E_0)}
  \prod_{j=1}^{n-1} \left( 1 + \frac{j}{k(E_0)} \right)  ,
\end{equation}
with $n=n_s - 1$, $n_s$ being the number of shower tracks. According to the fits
shown  in Figure \ref{fig:wolfendale}
 one can conclude that particles in cosmic ray are not 
independently produced but are  highly correlated. 
It should be noticed  that the Authors of \cite{Cosmic}
determined also the energy 
dependence of the two standard parameters of the proposed  phenomenological 
multiplicity distribution.
They found
\begin{equation}
	k^{-1}(E_0) = 0.4 (  1 - \exp [ - 1.8 \cdot 10^{-4} E_0] )
	\label{eq:mac1}
\end{equation}
and
\begin{equation}
	\nbar = 1.8 E_0^{1/4} . \label{eq:mac2}
\end{equation}
Particle correlations are controlled by $1/k$ and  become stronger
as the  primary energy becomes  larger. 
To the growth of $\nbar$  from 4.02 at primary
energy $E_0 = 25$ GeV  to 26.92 at $E_0 = 5\cdot 10^4$  
it corresponds in the same
energy range the decrease of $k$  parameter from 557 to 2.5 .
This   result is even more stimulating  when interpreted in the framework of 
clan structure analysis: the average number of clans,
$\Nbar$,  is growing from $\approx 4$ to 6.16 
and the average number of particles 
per clan, $\nbar_c$, from $\approx 1.0$ to 4.37, 
indicating a strong  clustering effect 
as the primary energy becomes larger (Table \ref{tab:mckeown}.)

\begin{table}
  \caption[NBD in cosmic rays data]{NB (Pascal) MD parameters
  calculated at different primary 
  nucleon energies $E_0$, using Eqs~(\ref{eq:mac1}) and
  (\ref{eq:mac2}) \cite{Cosmic}.}\label{tab:mckeown}
	\begin{center}
  \begin{tabular}{|r|cccc|}
		\hline
		$E_0$~~~~~ & $\nbar$ & $k$ & $\Nbar$ & $\nc$ \\
		\hline
		25 GeV &   4.02  &  557  &  4.01  &  1.0\\
		30 GeV &   4.21  &  464  &  4.19  &  1.0\\
		300 GeV &  7.49  &  47.6 &  6.96  & 1.08\\
		1000 GeV & 10.12 &  15.18 & 7.75  & 1.31\\
		5000 GeV & 15.14 &  4.21 &  6.42  &  2.36\\
		$10^4$ GeV& 18   & 3     &  5.84  &  3.08\\
		$1.5\cdot 10^4$ GeV & 19.92 &2.68 &5.71 & 3.49\\
		$5\cdot 10^4$ GeV & 26.92 & 2.5 & 6.16 & 4.37\\
		\hline
  \end{tabular}
	\end{center}
  \end{table}

\subsection{The occurrence of the NB (Pascal) regularity  in full phase-space 
in the accelerator region and the generalised multiperipheral
bootstrap model.}\label{sec:III.bootstrap}

In 1972 the  KNO scaling violation and the widening of the $n$ charged 
particle MD with the increase of $p_\text{lab}$ are confirmed in hadron-hadron
collisions in  the accelerator region \cite{ISR:1}. It is found that the
 NB (Pascal) MD describes $n$ charged particle MD's, $P_n$, 
when plotted vs.\ $n$; 
to the increase with $p_\text{lab}$ of 
the average charged multiplicity $\nbar$ one 
notices the decrease of the parameter $k$, i.e., the dispersion 
$D = \sqrt{\avg{\nbar^2} - \avg{\nbar}^2}$  becomes larger in contrast with
 multiperipheral model predictions and in agreement with
cosmic rays experimental findings. The inverse of $k$  NB  parameter,
$g^2 \equiv k^{-1}$, 
is interpreted in the generalised multiperipheral bootstrap
model \cite{AGCim:1+AGCim:6} 
as the ratio  between the pomeron-reggeon-particle 
vertex controlling diffraction phenomena ($g^2_P $) and the reggeon-reggeon 
particle vertex ($g^2_M $). $g^2$  is evaluated to vary between 0.01  
at 30 GeV/c and 0.67 in the above mentioned extreme  cosmic ray experiments.
The occurrence of the  NB distribution in the generalised multiperipheral 
bootstrap model  is understood as  the effect of the weighted superposition 
of $r$  multiperipheral diagrams  with $\nbar_r = {r \nbar}$  and $\nbar$ the 
average multiplicity of the first Poissonian  multiperipheral diagram, i.e., 
$\nbar = g^2_M \ln (s / m_a m_b) $ ($ m_a$ and $m_b$ are 
the masses of the initial 
particles and $\sqrt{s}$ the c.m.energy). 
The distribution function  which  weights the average multiplicity
in the set of multiperipheral diagrams is a function of $g^2$ and  $r$ and is
assumed to be  a gamma distribution 
\begin{equation}
  f(r, g^2 ) =  \frac{(r/ g^2 )^{(1 - g^2 ) / g^2 }}{
		{(1 - g^2 ) / g^2 }!}  \exp (- r/ g^2 )  .
\end{equation}
It follows
\begin{equation}
	\sigma_n = \sigma_{\text{tot}} \int_0^\infty  
	\frac{(r\nbar)^n}{n!}  \exp[-r\nbar]
	\frac{(r/ g^2 )^{(1 - g^2 ) / g^2 }}{
		{(1 - g^2 ) / g^2 }!}  \exp (- r/ g^2 )  d(r/g^2) ,
	\label{eq:III.13}
\end{equation}
that is
\begin{equation}
	\sigma_n  = \sigma_{\text{tot}}  \left( \frac{ \nbar }{ 1+g^2\nbar}
	                            \right)^n 
				\frac{ (1+g^2)\cdots\left(1+(n+1)g^2\right) }{ n! }
				(1 + g^2\nbar)^{-1/g^2}  .
 	\label{eq:III.14}
\end{equation}

\begin{figure}
  \begin{center}
  \mbox{\includegraphics[width=\textwidth]{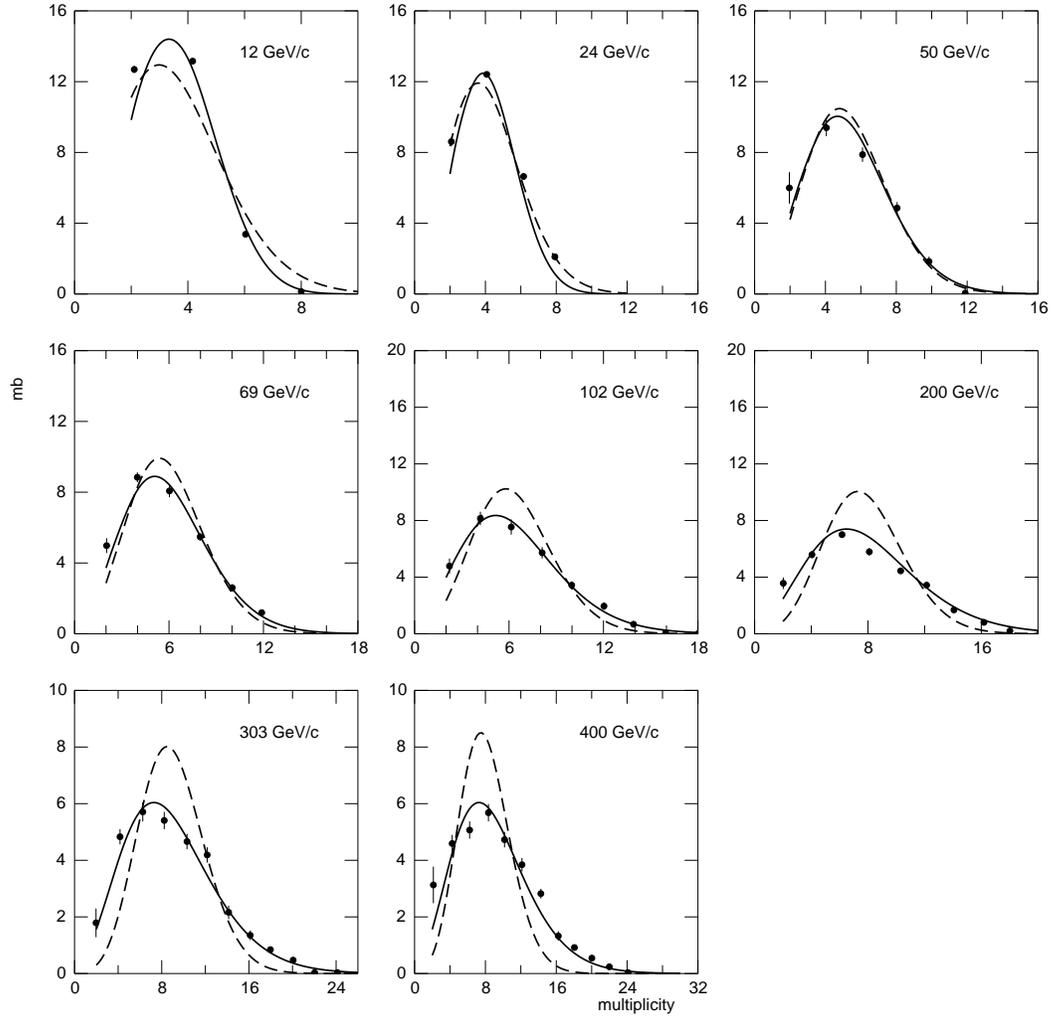}}
  \end{center}
  \caption[low energy NBD fits]{NB (Pascal) MD (solid line) and Poisson 
	(dashed line) formulae plotted
  vs.\ the number of charged particles $n$
  compared with data on \pp\ collisions at different 
	projectile momenta in the laboratory 
  frame \cite{AGCim:3+AGCim:4}.}\label{fig:antich}
  \end{figure}

\begin{figure}
  \begin{center}
  \mbox{\includegraphics[width=\textwidth]{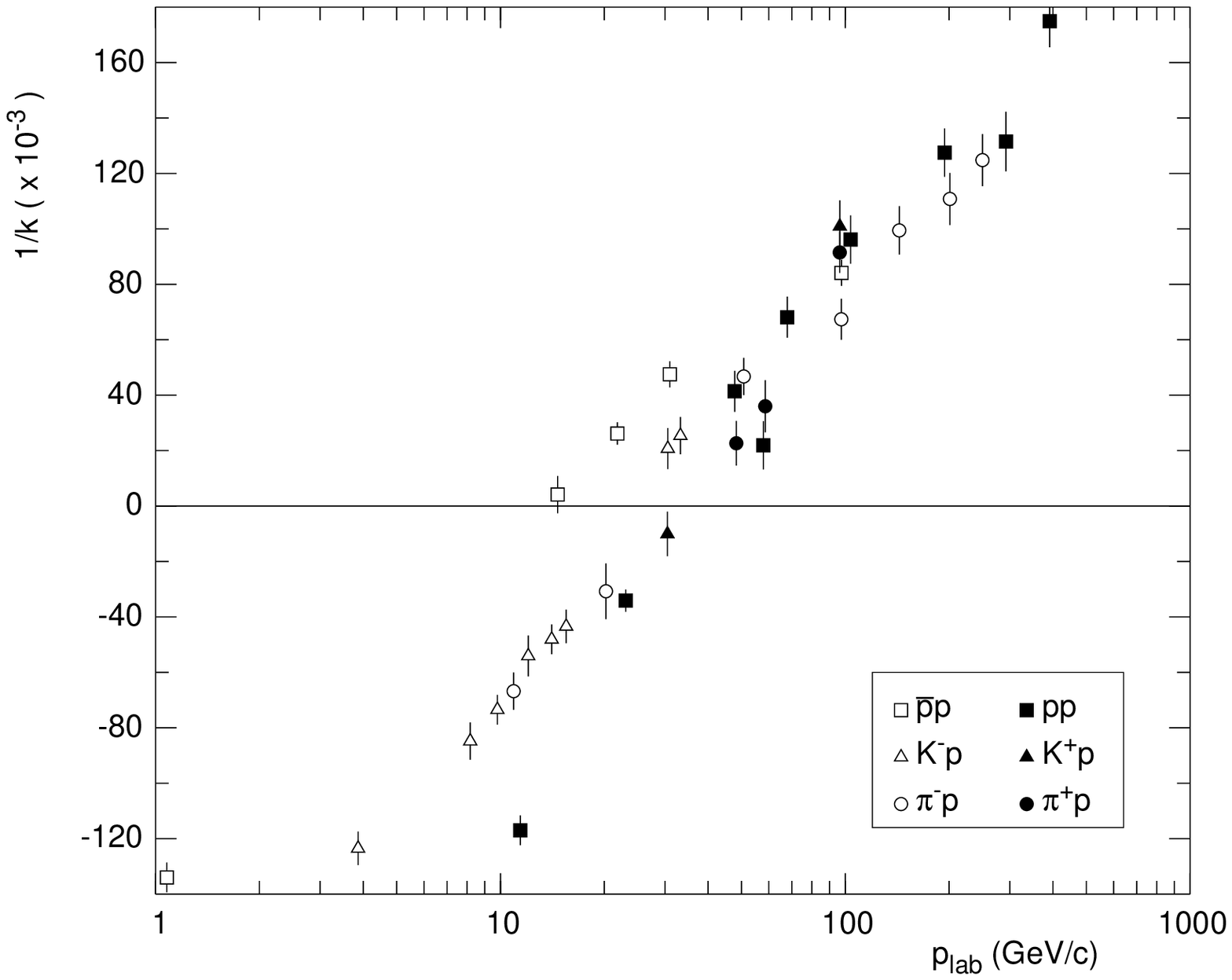}}
  \end{center}
  \caption[Energy behaviour of 1/k]{Overall $1/k$ data as a function of 
	projectile momentum in the laboratory frame $p_{\text{lab}}$ for the
  reactions listed \cite{AGCim:3+AGCim:4}.}\label{fig:garetto}
  \end{figure}

For $g^2_P = 0$,  $\nbar \approx g^2_M  \ln (s / m_a m_b)$ and
\begin{equation}
  \sigma_n \approx \sigma_{\text{tot}}  \nbar^n \frac{ e^{-\nbar} }{n!}
\end{equation}
and in general  $1/k(s/s_0)  = g_P^2/ g_M^2 h(s/s_0)$.
From a purely  phenomenological point of view, it should be pointed out that
multiplicity distributions of available  hadron-hadron collision 
experiments in the accelerator region  were all  successfully  
described by  the NB (Pascal) MD (see for example Fig.~\ref{fig:antich})
in a series of  papers which were overlooked
in the field. In addition,  deviations from 
Poisson MD were  interpreted as the onset of pomeron-reggeon vertex. Of 
particular interest was the fact that $k^{-1}$ parameter
was increasing  with c.m.\ energy  in the accelerator region as in cosmic rays
in all hadron-hadron reactions and  crossing the zero point  in the low energy 
domain (few GeV/c), as shown in Figure \ref{fig:garetto}, in 
agreement with later findings \cite{KittelNB,Kittel:Fest}.

An alternative interpretation of the regularity in terms of a stochastic
cell model was also suggested by one of the present Authors 
in 1973 \cite{AGCim:2+AGCim:5} and then extensively discussed  by P. Carruthers 
\cite{CarrShih}. $k^{-1}$ parameter is understood  in 
this framework as a stimulated emission factor, i.e., the fraction of particles 
already present stimulating the emission of an additional particle.

\subsection{The advent of the regularity in different classes of collisions 
   and in rapidity intervals, and  its interpretation in terms of
   clans.}\label{sec:III.clans}
 
\begin{figure}
  \begin{center}
  \mbox{\includegraphics[width=0.8\textwidth]{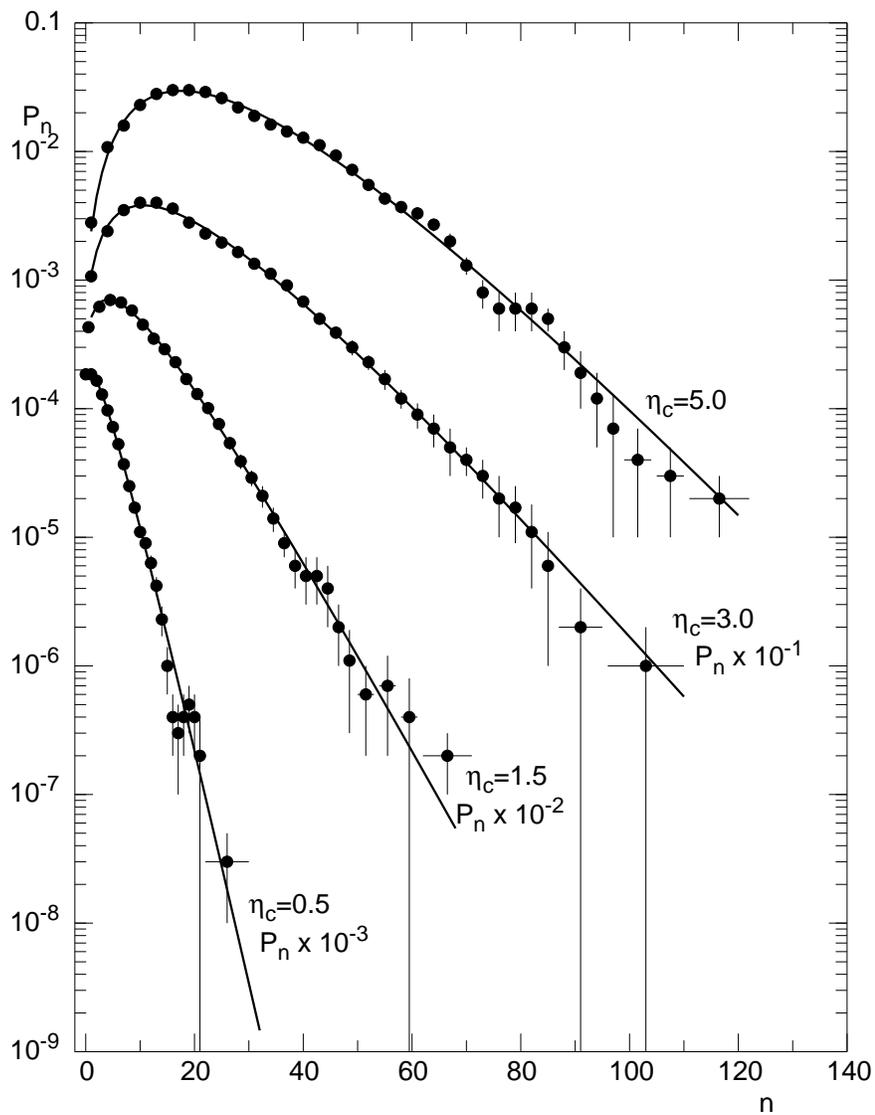}}
  \end{center}
  \caption[NBD fits in ppbar at 546 GeV]{MD $P_n$ vs.\ $n$
  in \ppbar\ collisions at c.m.\ energy 546 GeV in different
	pseudo-rapidity intervals are shown together with fits using the
	NB (Pascal) MD (solid lines) \cite{UA5:rep}.}\label{fig:alner}
  \end{figure}

In 1985, the UA5 collaboration rediscovered NB (Pascal) MD for describing
CERN  \ppbar\ collider MD data in full phase-space and in symmetric
rapidity intervals \cite{UA5:5}
as shown in Fig.~\ref{fig:alner}.

\begin{figure}
  \begin{center}
  \mbox{\includegraphics[width=0.8\textwidth]{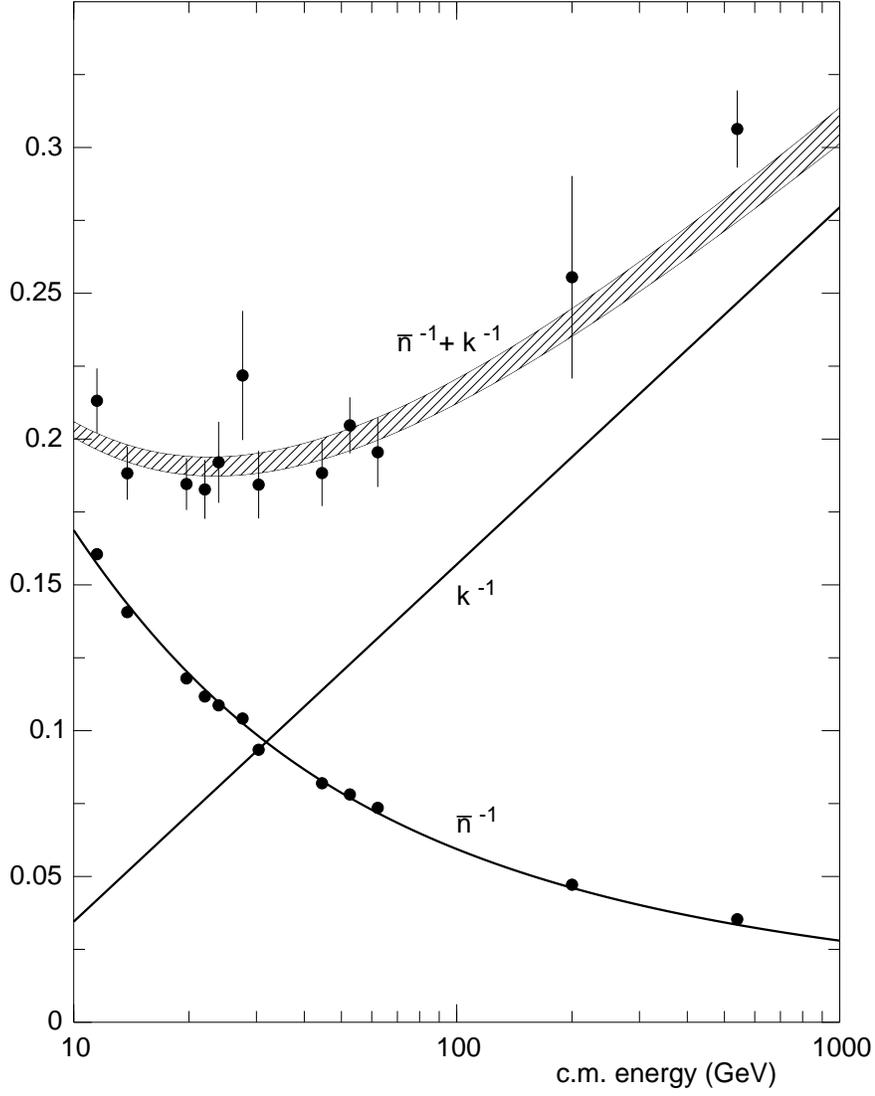}}
  \end{center}
  \caption[D2/n = 1/k + 1/n]{Compilation of measurements of the
  average charged multiplicity 
  $\nbar$ and of the normalised variance $D^2/\nbar^2$ in \pp\ and p\=p
  collisions. Superimposed are interpolations to the NB (Pascal) MD
  parameters $\nbar$ and $k$ and their sum (the band takes into
  account interpolation errors) according to the NB relation
	$D^2/\nbar^2 = 1/k+1/\nbar$ \cite{UA5:4}.}\label{fig:isr}
  \end{figure}

\begin{figure}
  \begin{center}
  \mbox{\includegraphics[width=\textwidth]{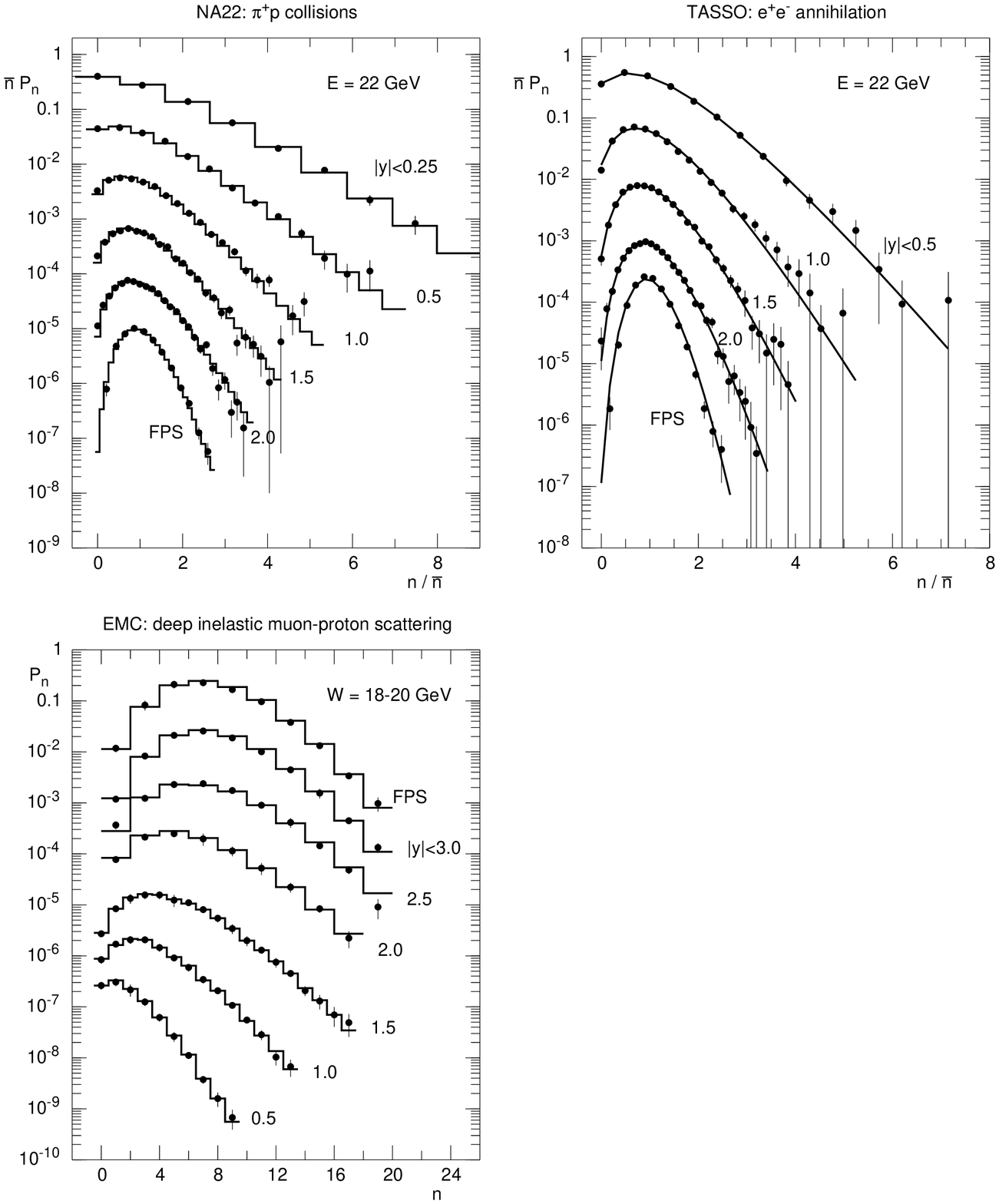}}
  \end{center}
  \caption[NA22, EMC, TASSO NBD's]{Multiplicity distributions for
  final charged particles 
  in various reactions but at similar c.m.\ energies; 
  data are taken in central symmetric rapidity intervals 
	$|y| < y_c$ and in full phase-space as shown by the labels. 
  The lines are fits with NB (Pascal) MD \cite{Singapore}.}\label{fig:emc}
  \end{figure}

In going from ISR and SERPUKHOV up to CERN \ppbar\ Collider data
\cite{UA5:4,Giacomelli},
$\nbar$ becomes  larger  and $k$ parameter smaller as the c.m.\ energy of the 
collision becomes larger,  as shown in Figure \ref{fig:isr},
confirming the general trends of NB parameters  
already observed in the accelerator and in the cosmic ray regions. 
In addition,  parameter $k$ is smaller in central pseudo-rapidity intervals,
suggesting the onset of a significant dynamical effect in central regions 
not hidden by conservation laws, and  indicating 
stronger particle correlations  with respect to the larger intervals where 
conservation laws are the dominant effect 
(see \cite{Schmitz} for a detailed review).

The next question concerned the possible extension of the regularity  
to other  classes of collisions, both  in the available energy domain and 
in symmetric (pseudo)-rapidity intervals.   
The answer came from NA22,  EMC  and   TASSO collaborations  
as shown in Figure \ref{fig:emc}. 
$P_n$ vs.\ $n$ data were very well fitted by NB (Pascal) MD's 
in hadron-hadron \cite{NA22}, lepton-hadron \cite{EMC} 
and lepton-lepton \cite{HRS:1} collisions at nearly the same 
c.m.\ energy both in FPS\ and in  (pseudo)-rapidity intervals. 
The success of 
the regularity in describing $n$ charged particle MD's 
in all classes of collisions
and in symmetric rapidity intervals, i.e., its universality, demanded
an interpretation in more physical terms. Previous phenomenological models, 
although leading to the experimentally observed $n$ charged particle
multiplicity distribution, 
were too poor and in a certain sense too naive 
for the new wide  domain of validity  of the regularity, especially in view also
 of the almost simultaneous  achievements of QCD as the theory of strong 
interactions.
This  search was developed in two moments.

Firstly the characterisation of the observed  regularity at hadron level in 
more physical terms was studied. Cascading and stimulated emission appeared as 
the natural candidates for the dynamical mechanisms leading to the 
observed distribution. It is interesting to point out that, according
to the  interpretation of $k^{-1}$, in the  stimulated emission framework   
$k^{-1}$ should be always larger than  or equal to 1. 
Since stimulating emission was  
excluded by experiments (i.e., the fraction of particles already present
stimulating the emission of a new particle  was larger in the total
sample of positive and negative particles than in the separate samples with
only positive or negative particles \cite{NA22:b}), the attention
was concentrated on the  cascading  mechanism.  Attention is paid to the 
possibility of particles already produced to emit additional particles and 
leads to  grouping of particles into clusters (which later were called
clans) \cite{AGLVH:0}.

\begin{figure}
  \begin{center}
  \mbox{\includegraphics[width=\textwidth]{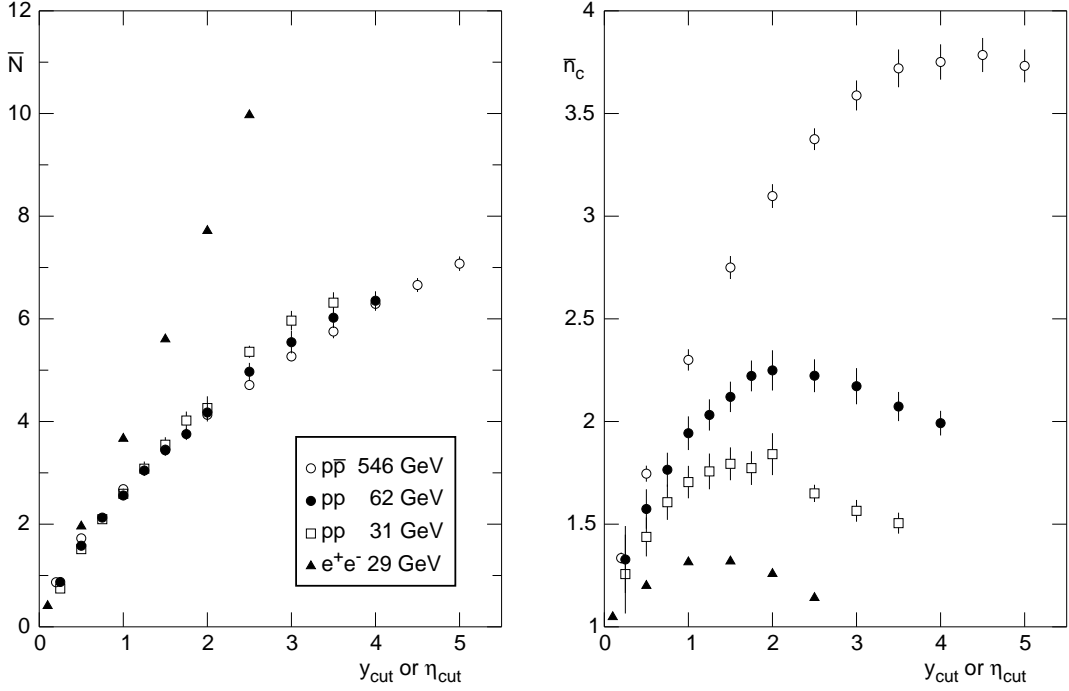}}
  \end{center}
  \caption[Clans parameters at ISR,TASSO]{Average number of clans,
  $\Nbar$ (left panel), and average 
  number of particles per clan, $\nc$ (right panel) vs.\ the
  half-width of the pseudo-rapidity (for 546 GeV data) or rapidity
  (for the other energies) interval \cite{ISR:1}.}\label{fig:clan}
  \end{figure}

The mechanism underlined by the NB (Pascal) regularity  goes as
follows (recall Section \ref{sec:I.3} for the underlying mathematical
structure).
Clans are by definition independently produced, each clan contains
at least one particle by assumption and all correlations are exhausted 
within each clan. Clan ancestors, after their production, generate additional
particles via cascading  according to a logarithmic multiplicity distribution. 
Two new parameters are  the average number of clans, $\Nbar$, and the average 
number of particles per  clan,  $\nc$. They are linked to the standard
NB parameters $\nbar$ and $k$ by the following non trivial relations
\begin{equation}
	\Nbar = k \ln ( 1 + \nbar/ k )  \qquad\text{and}\qquad 
	\nc = \nbar / \Nbar .
\end{equation}
Clan   structure analysis reveals new interesting properties  when  applied 
to above mentioned  collisions as shown in Figure \ref{fig:clan}.
In particular it is shown that  $\Nbar$(\ee) is much larger than 
$\Nbar$(\pp),   whereas $\nc$(\ee) is much smaller than
$\nc$(\pp), in addition clans in central rapidity intervals are larger 
than in more  peripheral intervals.
The deep inelastic case is intermediate among the previous two:
clans are less numerous than that in \ee\ but the average number
of particles per clan is much larger. 

\begin{figure}
  \begin{center}
  \mbox{\includegraphics[width=\textwidth]{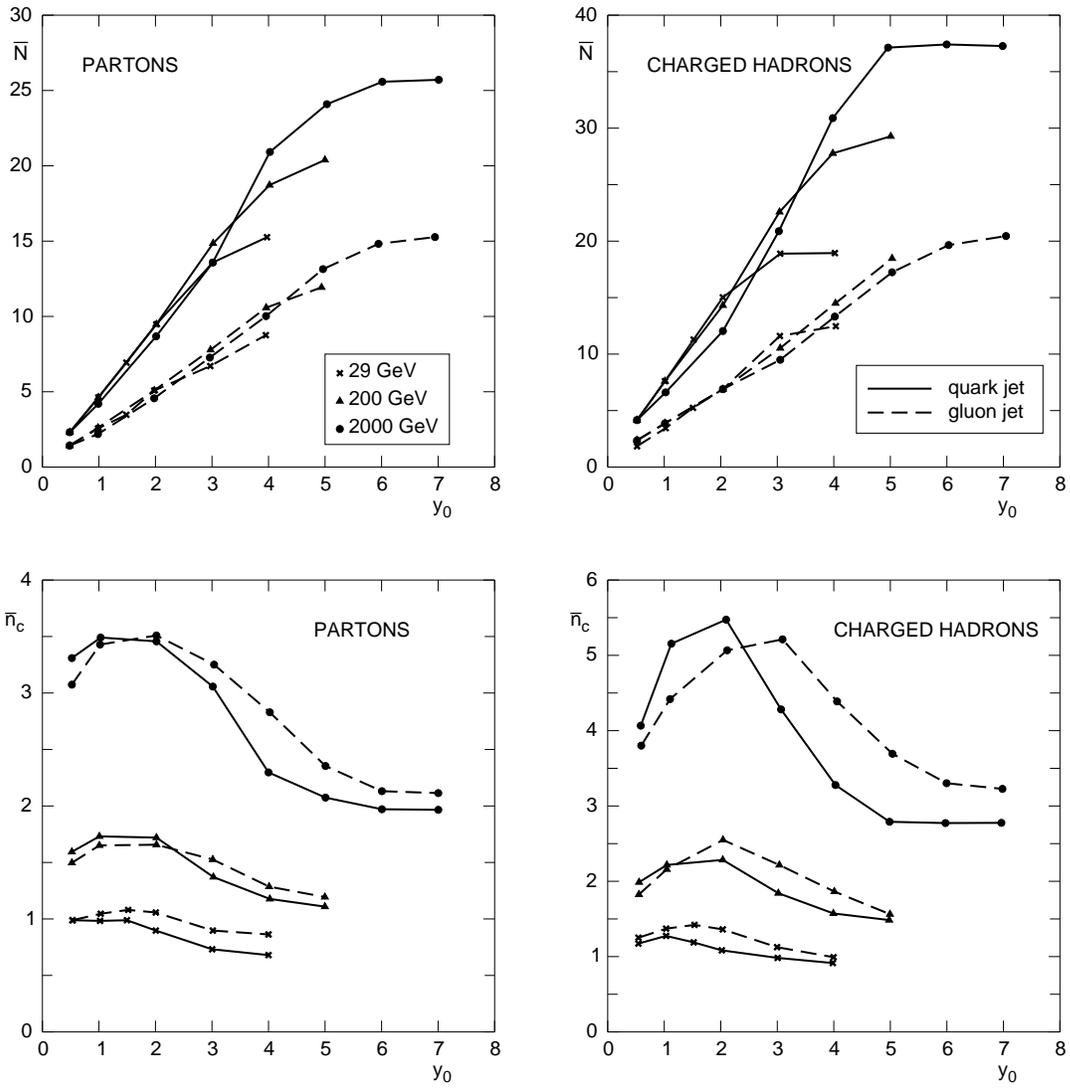}}
  \end{center}
  \caption[Clan parameters in qq and gg in Jetset]{Clan parameters at
  partonic and hadronic levels in the Jetset 
  7.2 Monte Carlo in the $q\bar q$ and $gg$ systems at different
  energies.
	Lines ares drawn to group together points pertaining to the same
  initial state ($q\bar q$ or $gg$) \cite{AGLVH:2}.}\label{fig:jetset}
  \end{figure}

Secondly, it was compulsory to estimate final $n$-parton MD's 
in a region were QCD had poor predictions. The idea
was to rely on Monte Carlo calculations. The choice of Jetset 7.2 seemed
particularly appropriate \cite{Jetset}. 
Jetset is based indeed  on DGLAP equations at
parton level and has Lund string fragmentation  as hadronization
prescription. Extending previous work by W. Kittel \cite{KittelNB},  
$q\bar q$ and $gg$ systems were studied as shown in Figure
\ref{fig:jetset} \cite{AGLVH:2}.  
Quite satisfactory
NB fits for final parton and charged hadron multiplicities in full phase-space 
and in symmetric rapidity windows  were found.
In addition, $k^{-1}$ was decreasing for growing symmetric rapidity intervals
at fixed c.m.\ energy and increasing with c.m.\ energy in fixed rapidity
intervals. Finally $\Nbar$ was approximately energy independent in 
a fixed rapidity interval and  linearly growing with rapidity variable
at fixed c.m.\ energy. 
These results, together with the previous finding
that  $q$- and $g$-jets can be interpreted as QCD Markov branching
processes, led to the conclusion that partonic clans are 
quite similar to bremsstrahlung gluon jets (BGJ's).  
At parton level, the claim is that clan ancestors are independent 
intermediate  gluon sources. 
This finding, together with the discovery that final  partonic and
hadronic MD's  in Jetset 7.2 were not independent but linked by the GLPHD 
(i.e., $\nbar_{\text{hadron}}  \approx  {\rho} \nbar_{\text{parton}} $
and  $k_{\text{parton}} \approx k_{\text{hadron}}$, see 
Section~\ref{sec:II.hadronization}), led 
to  a convincing interpretation of the differences seen in clan
structure analysis of lepton-lepton and hadron-hadron collisions:
in \pp\ collisions,
the independent intermediate gluon  sources  become active at very high 
virtuality (a remark which would suggest a larger population per clan
and strong colour exchange processes); a situation which is to be contrasted
with what is seen in \ee\ annihilation, where  gluon sources
are active quite late at relatively low virtuality, thus generating
a large number of BGJ with a small amount of particles per clan.  
New results from Tasso \cite{TASSO} and  
ISR \cite{Giacomelli} on charged particle MD's confirmed the
just examined properties. 

\begin{figure}
  \begin{center}
  \mbox{\includegraphics[width=0.8\textwidth]{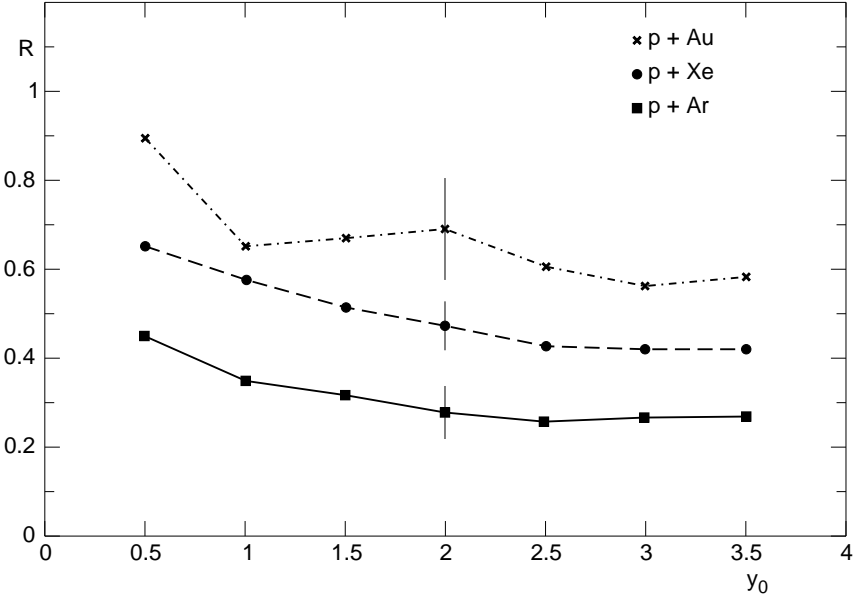}}
  \end{center}
  \caption[R in pA]{Ratio of the average number of partonic clans in p+A
  collisions to average number of partonic clans in \pp\ collisions as a
  function of the half-width of the rapidity interval. The error bar
  shown at $y_0=2$ is indicative of the error size at all $y_0$'s
	\cite{AGLVH:3}.}\label{fig:erre}
  \end{figure}

Interestingly, data on multiplicity distributions  
in proton-nucleus collisions \cite{Dengler+Bailly:1988bx} revealed 
new features when studied in terms of clan structure analysis
\cite{AGLVH:3}. In particular, 
it was found that the regularity is satisfied  at final hadron level in p+Au, 
p+Xe and p+Ar collisions. By applying then  GLPHD to the hadronic sector,
 partonic level properties in proton-nucleus  were determined. It was found 
that the partonic level was much simpler than the hadronic one. The mean number
of clan (BGJ) is controlled by the central region in rapidity (it is equal 
within errors for the forward and backward hemispheres) and it shows a weak
 A-dependence. The increase of the backward hemisphere multiplicity with
A is caused by the multiplicity increase  of the backward BGJ's, which might
be due to larger virtualities of their initial gluons and/or a more rapid
shower development. The relative increase with A of the average number of clans
in $pA$ with respect to \pp\  collisions is interpreted as  due
to multiple collisions in the target nucleus. 
Using Poisson statistics for the successive emission of the projectile, the
mean number of downstream collisions suffered by the projectile with a
target nucleon can be defined. It has been found that
\begin{equation}
	\nu'=[ \nu / (1 - e^{ -\bar\nu})] -1 ,
\end{equation}
with $\bar\nu = {A \sigma_{\text{inel}}(pp)}/{\sigma_{\text{inel}}(pA)}$
the mean number of inelastic projectile-nucleus collisions (they are 2.4,
3.3 and 3.9 for argon, xenon and gold respectively). 
By measuring the relative
increase of $\Nbar_{\text{parton}}$ 
due to the multiple collisions of the target nucleus A, i.e.,
\begin{equation}
	R = \Nbar_{\text{parton}} (pA) / \Nbar_{\text{parton}} (pp),
\end{equation}
(see Figure \ref{fig:erre}),
it should be pointed out that its comparison with 
the mean number of downstream collisions suffered by the projectile after 
its first collision  with a target nucleon  decreases from  
$\approx 0.27$ at $y_0=0.5$ to $\approx 0.17$ at $y_0=3.5$. 
It is remarkable that the above mentioned nuclear
effect leads to a strongly reduced capability of the wounded projectile
for further emission of BGJ's. 

In low and intermediate energy nucleus-nucleus collisions, 
it was also found that the NB (Pascal) regularity is obeyed \cite{NA35}, 
especially in small phase-space intervals,
and in particular that it leads to a very short correlation length
\cite{Tannenbaum}.

The lesson we learn is that
the occurrence of the regularity in all reactions suggests also here a two-step
production process:  to the independent intermediate gluon  sources
(the bremsstrahlung gluons), it follows their cascading according to
parton showers (the BGJ's). In ultrarelativistic nucleus-nucleus collisions,
one should expect more randomness and higher parton densities 
and, accordingly,  more time for  parton shower and clan formation.   
Reduced clan production is therefore,  when it occurs,  
an important signal which  should be studied with great attention (see
Section \ref{sec:III.thirdclass}.)

\subsection{The violation of the regularity and its occurrence at a 
more fundamental level of investigation, i.e., in different  classes of
events contributing to the total sample  in \ee\ annihilation  and
in \pp\ collisions}\label{sec:III.violation}

New experimental  facts were around the corner.
Collective variable properties in \ee\ annihilation and in \pp\ collisions
show impressive similarities and  quite intriguing differences. Both classes
of reactions are characterised by a shoulder structure 
\cite{DEL:1,DEL:2,UA5:3} visible in the 
intermediate multiplicity range in $n$ charged particle multiplicity
distributions $P_n$ when plotted vs.\ $n$  at LEP and at top \ppbar\
collider energies respectively. 
In addition, the ratio of factorial cumulants
$K_n$ to factorial moments $F_n$, i.e., $H_n$, when plotted
as a function of its order $n$, decreases sharply to a negative minimum and
follows then a quasi-oscillatory behaviour \cite{Gianini:1}. 
In order to explain both facts
in \ee\ annihilation,  properties of multi-parton final states can be
computed in the framework of perturbative QCD by exploiting its branching
structure and then extended to multi-hadron final states by comparing
partonic and hadronic distributions under the assumption of LPHD. Results
along this line are not fully  satisfactory.
An alternative phenomenological
approach consists in thinking that the mentioned effects are due to the
weighted superposition mechanism  of two classes of events, the first one 
with two jets and the second one with three or more jets, as identified by a
suitable jet finding algorithm \cite{DEL:single}. 
The phenomenological  guesswork is that both 
classes are described by a NB (Pascal) MD with characteristic, different
NB  parameters.  It turns out that in  this approach  the shoulder structure
in $P_n$ vs.\ $n$ and $H_n$ vs.\ $n$ oscillations are correctly
described \cite{hqlett:2}. 

In addition, a convincing description of the two effects occurring
also in \pp\ collisions can be obtained by the same mechanism assuming
that the superposition occurs between soft (without mini-jets) and
semi-hard (with mini-jets) events (whose counterpart at parton level
are single- and double-parton scattering) and that again each class of events
is described in terms of a NB (Pascal) MD. The intriguing difference
between the two collisions appears when one tries to describe the
energy dependence of FB multiplicity correlation strengths.

It is  found that at LEP \cite{OPAL:FB} 
the two-jet sample of events and the multi-jet sample
of events,  separately considered, do not show FB MC, whereas it has  to
be pointed out that FB MC occurs in the total sample of events. By knowing the 
average charged multiplicities and the dispersion of each class of events and 
the relative weight of two classes, FB MC of the total sample are
correctly reproduced  within experimental errors  thanks to 
a formula based on the weighted superposition mechanism of the two classes
of events. This situation should be contrasted with \pp\ collisions.
Here   FB MC are growing  in the separate samples of
events with and without mini-jets, and of course in the total sample of
events. It is shown, in addition, that FB MC cannot be explained in this case 
without introducing clans and particle leakage from clans in one hemisphere
to the opposite one, i.e., without assuming the form of the MD in each
sample of events \cite{RU:FB}. 
Clans of particles  become essential in this context.

In conclusion the regularity is not a property of the full sample of events
in the examined  collisions but a  characteristic  property of the substructures
of each collision, i.e., of the separate sample of events contributing to the
total sample. In the following, the  mentioned features of the 
substructures expected in two classes of collisions,
as they are guessed  today,  will be explored  
with the warning  to be jet-finding algorithm dependent in \ee\
annihilation and mini-jet definition dependent in \pp\ collision. A universally 
accepted  definition of the mentioned substructures is indeed per se  a  
fascinating search for future experimental work!

\begin{figure}
  \begin{center}
  \mbox{\includegraphics[width=0.8\textwidth]{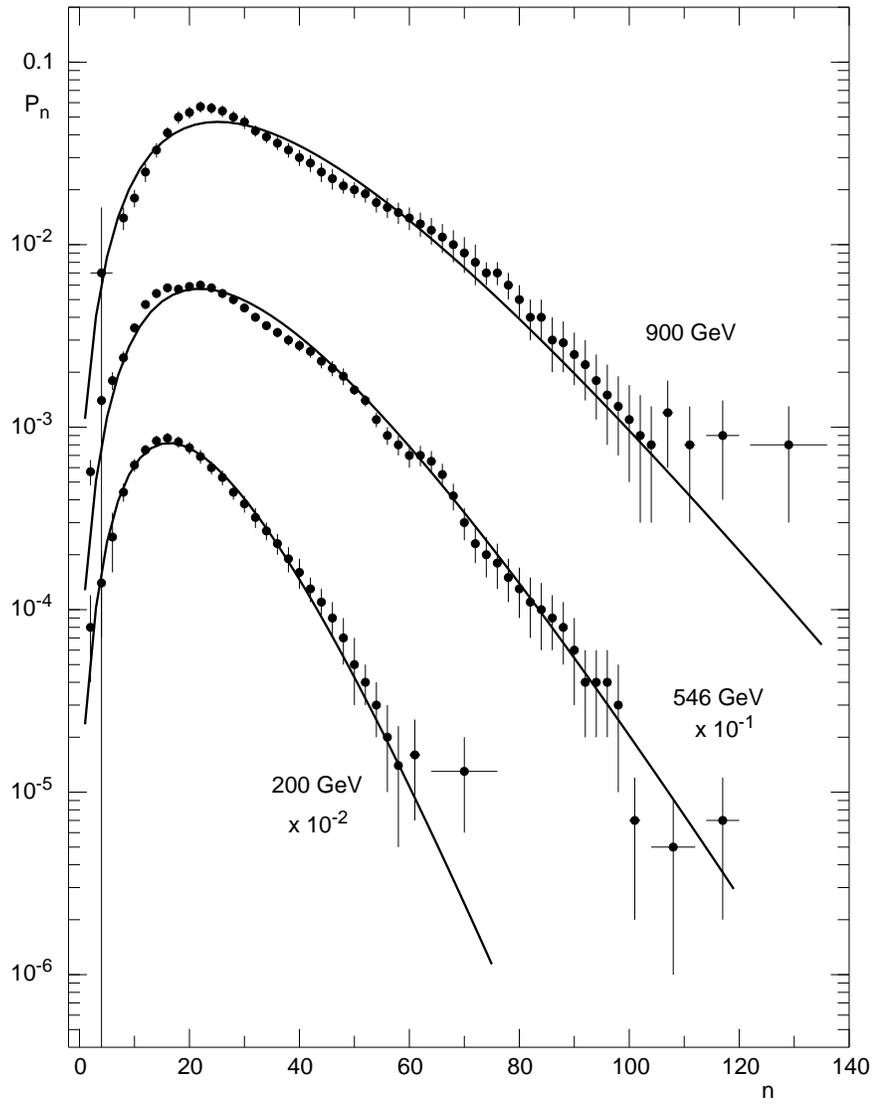}}
  \end{center}
  \caption[Shoulder in pp at 900 GeV]{MD's in full phase-space in \pp\
  collisions  
  compared with the
  NB (Pascal) fits. The shoulder structure is clearly visible,
  especially at 900 GeV \cite{Fug}.}\label{fig:fuglesanguno}
  \end{figure}

\begin{figure}
  \begin{center}
  \mbox{\includegraphics[width=0.8\textwidth]{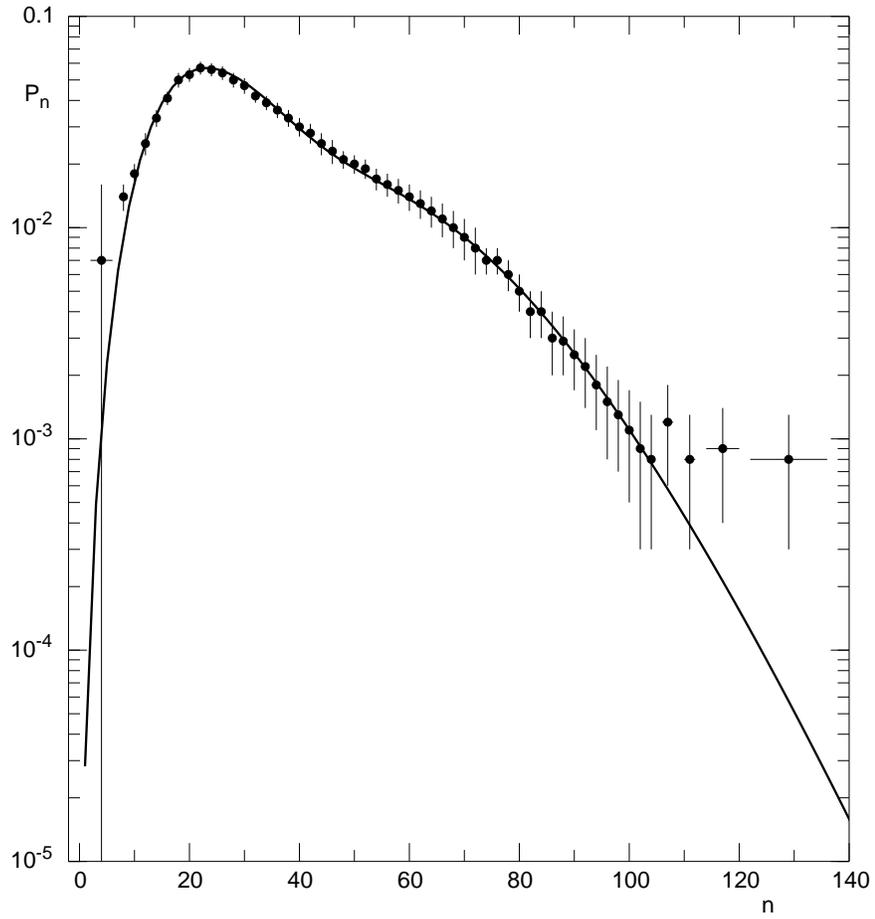}}
  \end{center}
  \caption[fit with two NBD's]{MD's in full phase-space at 900 GeV (as
  in the previous figure)   compared with the fit
  with the weighted superposition of two NB (Pascal) MD's, which now
  reproduces the data perfectly \cite{Fug}.}\label{fig:fuglesangdue}
  \end{figure}

\begin{figure}
  \begin{center}
  \mbox{\includegraphics{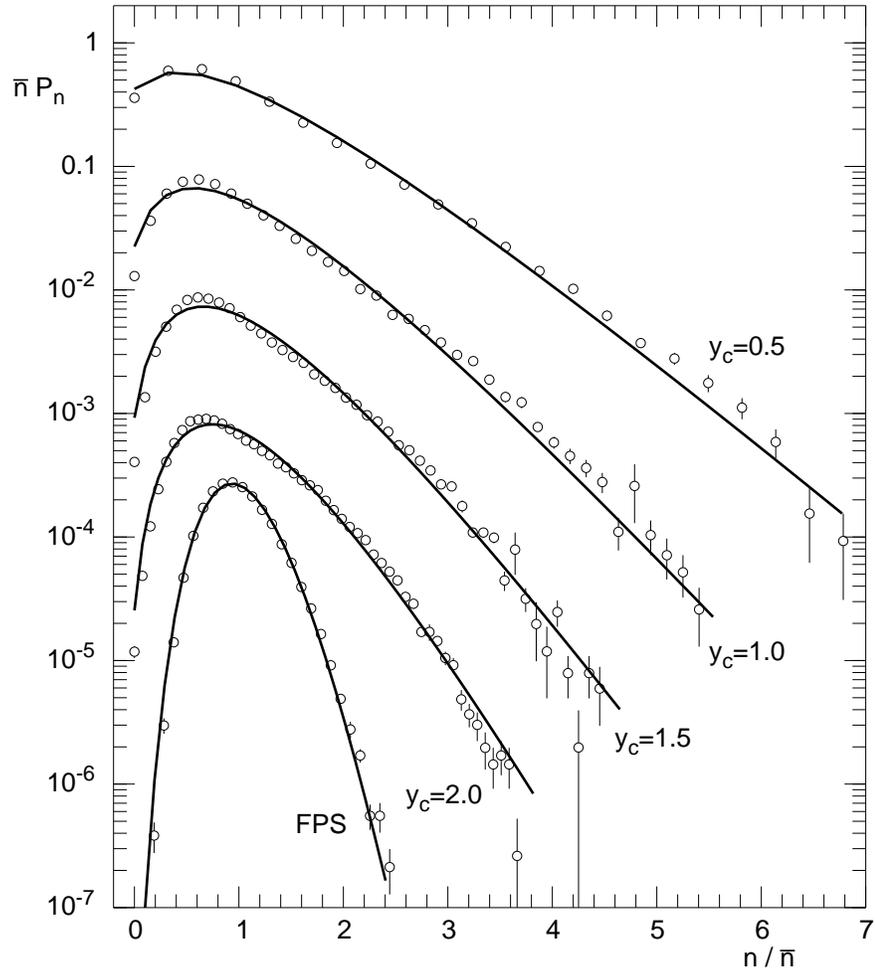}}
  \end{center}
  \caption[shoulder in e+e-]{Multiplicity distributions in different
  rapidity intervals 
  $|y| < y_c$ and in FPS in \ee\ annihilation at the $Z^0$ peak
		shown in KNO variables. 
  The lines show best fits with the NB (Pascal) MD, which
  does not reproduce the shoulder structure \cite{DEL:2}.}\label{fig:uvarov}
  \end{figure}

\begin{figure}
  \begin{center}
  \mbox{\includegraphics[width=0.85\textwidth]{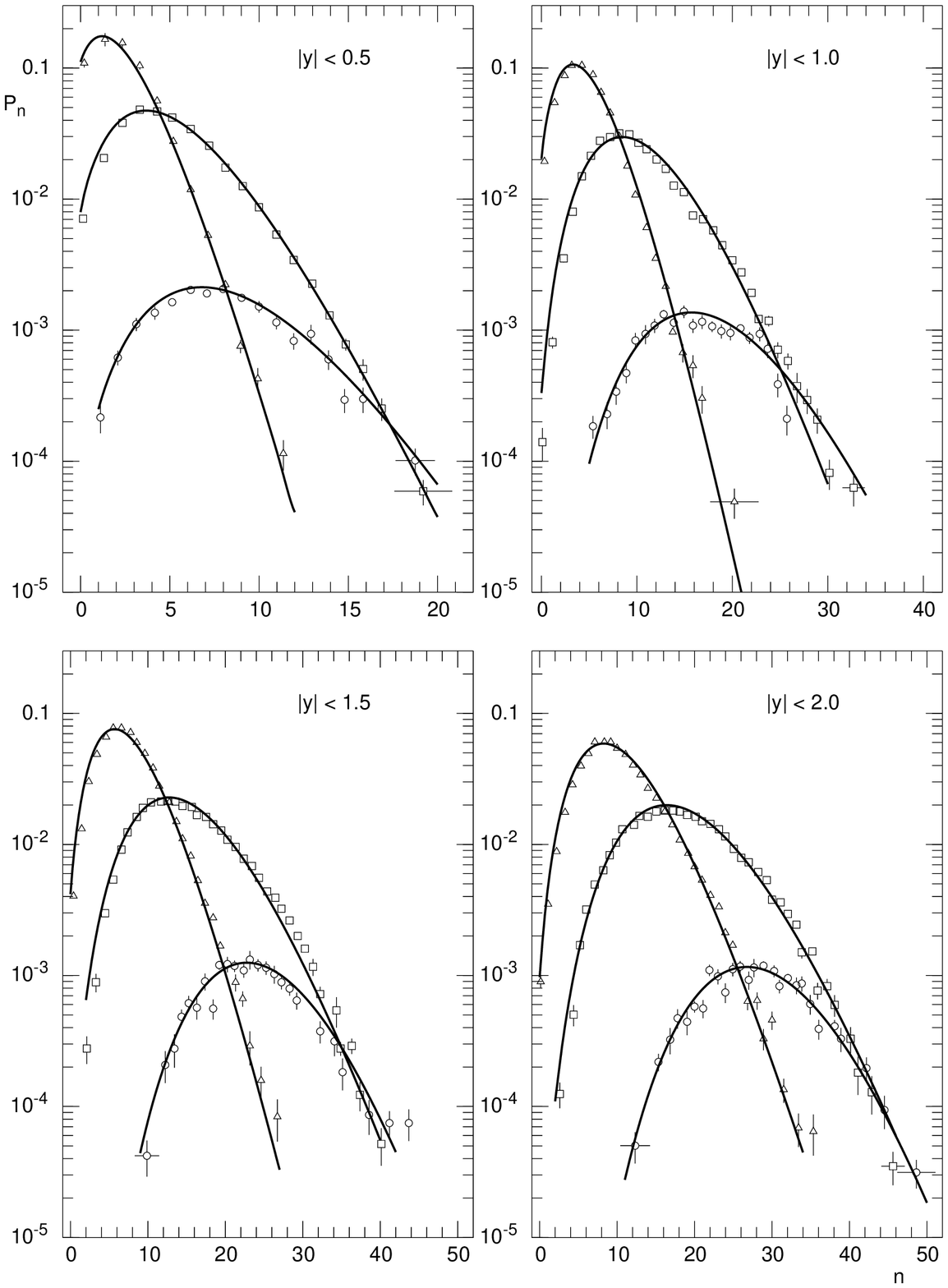}}
%% 		\begin{picture}(100,100)
%% 			\framebox(90,90){fig:delphi}
%% 		\end{picture}
  \end{center}
  \caption[e+e-: sum of 2, 3, 4 jet events]{Multiplicity distributions
  in different rapidity intervals 
  $|y| < y_c$ for 2-jet events (triangles), 3-jet events (squares)
		and 4-jet events (circles) in \ee\ annihilation at LEP. Each
		class is normalised to its fraction. Jets were
  counted using the JADE algorithm with parameter
  $Y_{\text{min}}=0.4$. The lines show best fits with the NB (Pascal)
  MD for each subsample \cite{DEL:single}.}\label{fig:delphi}
  \end{figure}

\subsection{Shoulder effect in $P_n$ vs.\ $n$  at top UA5 energy and in
	\ee\ annihilation 
and its removal by means of the weighted superposition mechanism
of different classes of events}\label{sec:III.shoulder}

The mentioned shoulder structure in $P_n$ vs.\ $n$
is well visible in UA5 data, especially at the top energy of 900 GeV,
as shown in Fig.~\ref{fig:fuglesanguno} by the comparison of data to
fits with one NB (Pascal) MD.
The same data were best fitted \cite{Fug} by the weighted
superposition of two NB (Pascal) MD's, as follows:
\begin{equation}
	P_n = \alpha P_n^{\text{NBD}}(\nbar_1,k_1) +
	  (1-\alpha) P_n^{\text{NBD}}(\nbar_2,k_2) .
	\label{eq:superpos}
\end{equation}
This structure corresponds to the superpositions of two classes of
events, here called generically 1 and 2, the MD of each class 
being described by a
NB (Pascal) MD, but with parameters which are different in each class.
The fraction of events of class 1 is given in Eq.~(\ref{eq:superpos})
by $\alpha$.
An example of the fit is shown in Fig.~\ref{fig:fuglesangdue}.

It was founds that the fitted values of $\alpha$ corresponded
to the fraction of event without mini-jets in analyses
performed by the UA1 collaboration  at the same energies 
\cite{UA1}, and the
relation between $\nbar_1$ and $\nbar_2$ in the fit also coincided
with UA1 findings ($\nbar_2 \approx 2\nbar_1$) \cite{UA1:minijets}.
Because the UA5 detector could not identify mini-jets, the just
mentioned analysis could not be verified completely; further
insight is now available from Tevatron (Sec.~\ref{sec:III.cdf}).

A very similar shoulder structure was found in \ee\ annihilation at
LEP: the precise data of DELPHI \cite{DEL:single}
are shown in Fig.~\ref{fig:uvarov}.
In this case it was possible to carry out the analysis by selecting
events with a fixed number of jets; this was done by DELPHI using the
JADE algorithm, and the stability of the fit was verified by varying
the resolution of the jet finder algorithm. The MD of each class
(i.e., 2-, 3- and 4-jet events) was successfully fitted by a NB (Pascal)
MD; an example is shown in Figure \ref{fig:delphi}.

%% \begin{figure}
%%   \begin{center}
%%    \mbox{\includegraphics[width=\textwidth]{plot_arsuno.ps}}
%%   \end{center}
%%   \caption{Charged particles' MD in full phase-space, 
%% $P_n$, at the $Z_0$ peak  from different \ee\ annihilation experiments
%% 		are
%% compared with the weighted superposition of two NB (Pascal) MD's;
%% dotted lines indicate the two components.
%% The lower part of the figure shows the residuals, $R_n$, 
%% i.e., the difference between
%% data and theoretical predictions, expressed in units of 
%% standard deviations.}\label{fig:arsuno}
%%   \end{figure}

\begin{figure}
  \begin{center}
   \mbox{\includegraphics[width=\textwidth]{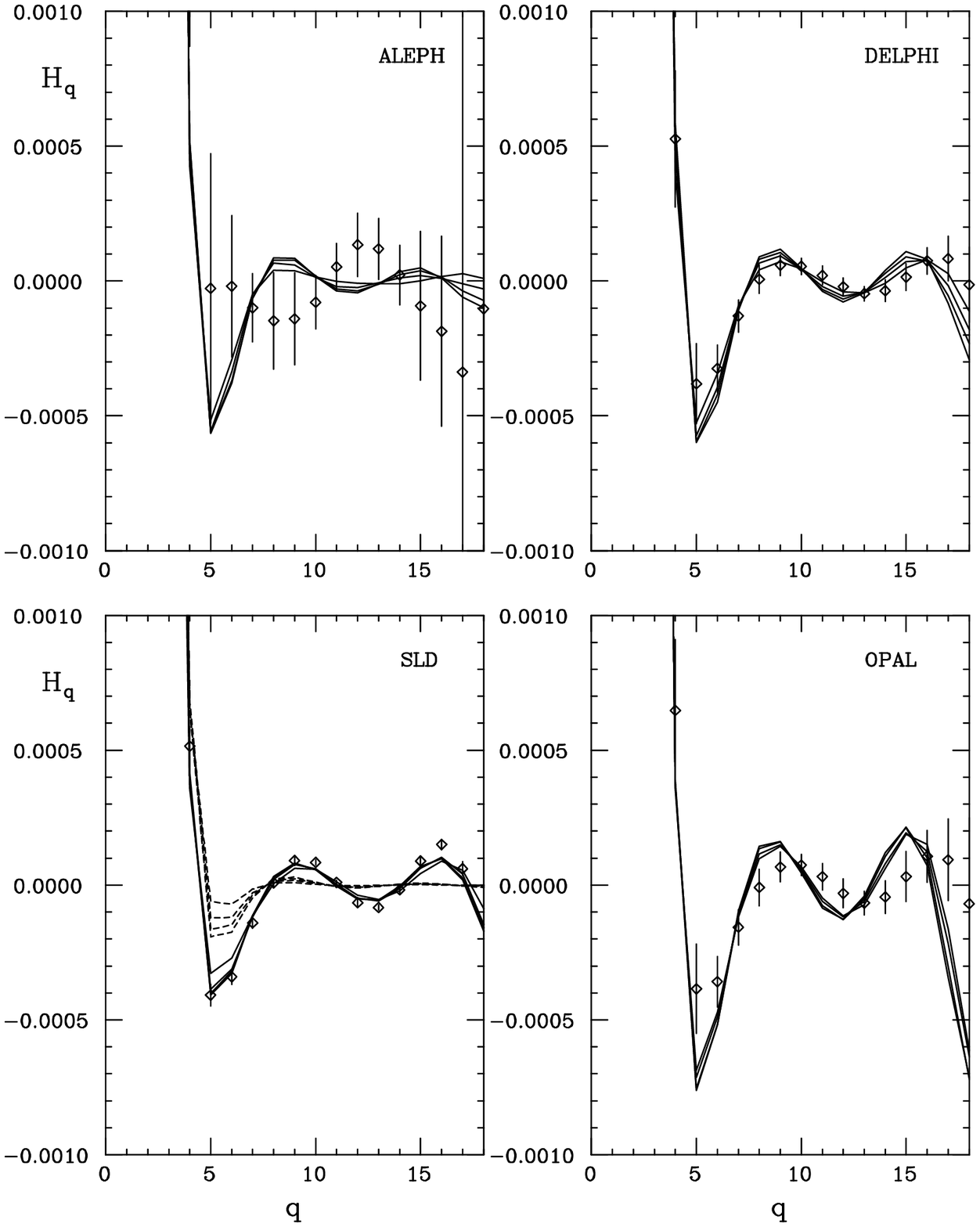}}
  \end{center}
  \caption[Hq at LEP]{The ratio of factorial cumulants over factorial moments,
$H_q$, as a function of $q$;
experimental data (diamonds) are compared to $H_q$
moments computed from the weighted superposition of two NB (Pascal) MD
(solid lines), truncated at the same multiplicity as the data.
For each experiment, different superposition appear corresponding to
different values of the resolution parameter of jet-finder algorithm.
The dashed lines in the SLD plot show predictions of
the same parametrisation as the solid lines
but without taking into account the effect of truncation.
In the figure only statistical errors are 
shown \cite{hqlett:2}.}\label{fig:arsuno}
  \end{figure}

\begin{figure}
  \begin{center}
  \mbox{\includegraphics[height=0.75\textheight]{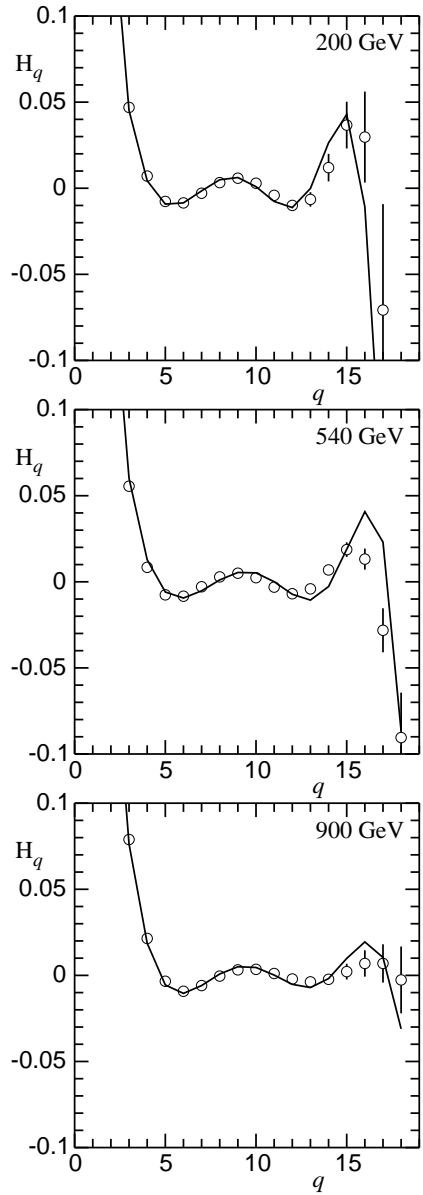}}
  \end{center}
  \caption[Hq in pp]{Ratio of factorial cumulant moments over factorial moments,
$H_q$, vs.\ the order $q$ in \ppbar\ collisions at c.m.\ energies
200 GeV, 546 GeV and 900 GeV. The solid line is the prediction
of the fit with the weighted superposition of two NB(Pascal) MD's
truncated at the same multiplicity as the MD 
data \cite{hqlett}.}\label{fig:arsdue}
  \end{figure}

\begin{figure}
  \begin{center}
  \mbox{\includegraphics[width=0.8\textwidth]{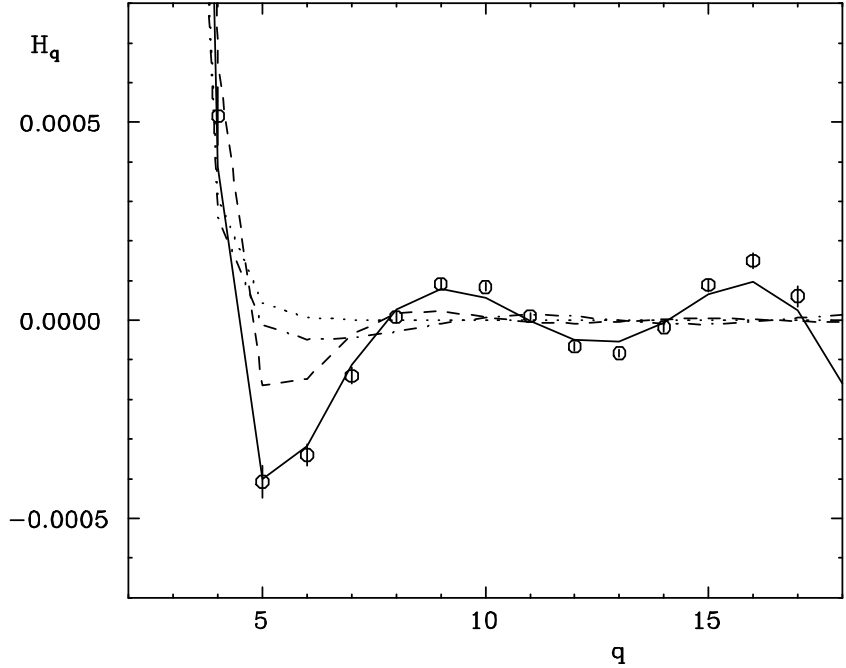}}
  \end{center}
  \caption[Hq at SLD]{Ratio of factorial cumulant moments over
		factorial moments, $H_q$ as a function of $q$;
\ee\ experimental data (circles) from the SLD Collaboration at 
at $\sqrt{s} = 91$ GeV
are compared with the predictions of several parameterisations, 
with parameters fitted to the data on MD's: a full NBD (dotted line); 
a truncated NBD (dot-dashed line); sum of two full NBD's (dashed line); 
sum of two truncated NBD's (solid line)
\cite{NijmegenAG}.}\label{fig:sld}
  \end{figure}

\subsection{$H_q$ vs $q$ oscillations  and the
	weighted superposition mechanism}\label{sec:III.oscillations}

After the prediction of sign oscillations of $H_q$ moments of the MD,
when plotted against the order $q$, at parton level in analytical QCD
calculations (See Sec.~\ref{sec:II.kinem.dla}), such oscillations
where looked for and found also in data \cite{Gianini:1}, 
again in all classes of collisions (see, e.g., Figs.~\ref{fig:arsuno} 
and \ref{fig:arsdue}).  
However, QCD predictions are only qualitatively in agreement with \ee\ data,
but quantitatively rather far, in particular in the computation of the
magnitude of $H_q$ moments.  Indeed, the fact that experimental data
are truncated ---due to the finiteness of the data sample--- plays an
important role in the magnitude of the oscillations \cite{hqlett}.

It was shown previously (Sec.~\ref{sec:I.5}) that one NB (Pascal) MD
also shows such sign oscillations when truncated ---a complete NB does
not, see Eq.~(\ref{eq:I.35}). That a single truncated NB can reproduce
\pp\ collisions data is shown by the solid lines in
Fig.~\ref{fig:arsdue}: notice that this happens even if the fits to
the MD are not satisfactory (recall Fig.~\ref{fig:fuglesanguno}.)

The same is not true in the case of \ee\ annihilation.  An accurate
analysis was performed on SLD data at 91 GeV c.m.\ energy, which are
badly fitted by one NB but well fitted by the weighted superposition
of two such distributions. The $H_q$ moments computed by SLD are shown
in Fig.~\ref{fig:sld}.  
The single NB (Pascal) MD, even when truncated, cannot
describe the $H_q$ moments (dot-dashed line in Fig.~\ref{fig:sld}.)
The weighted superposition of two complete NB (Pascal) MD's 
comes closer, but is not enough (dashed line in Fig.~\ref{fig:sld}.)  
It is only after
truncating, as in the data, the weighted superposition of two 
NB (Pascal) MD's that a successful description can be achieved (solid line in
Fig.~\ref{fig:sld}.)

\subsection{Towards the TeV energy domain in  \pp\ 	
collisions.}\label{sec:III.scenarios} 

The main scope of this Section is to explore possible scenarios 
for  multiparticle production in \pp\ collisions in the TeV energy domain 
following  our knowledge  of  the GeV energy region.

The main conclusion of the  phenomenological study on multiparticle
production  in \pp\ collisions in the GeV region  is that 
there are two classes of events, the class of  soft (without mini-jets)
events  and the class of  semi-hard (with mini-jets) events,
whose $n$ charged particle multiplicity distributions can be described 
in terms of NB (Pascal) MD's with characteristic $\nbar_i$ and $k_i$
($i$ = soft, semi-hard) parameters,  i.e., 
the NB (Pascal) MD regularity is still applicable 
but at a deeper level of investigation than initially thought. 
The  weighted superposition of the two components, each described by
a NB (Pascal) MD,  explains indeed at least three important experimental facts 
of multiparticle production in \pp\ collisions, i.e.,

\emph{a.}  the shoulder effect in  $P_n$ vs.\ $n$;

\emph{b.}  the oscillations in $H_q$ vs.\ $q$;

\emph{c.}  forward-backward multiplicity correlations (these will be
described in detail in Sec.~\ref{sec:III.fbmc}).

\begin{figure}
  \begin{center}
  \raisebox{5.9cm}{$\alpha$}\mbox{\includegraphics[width=0.6\textwidth]{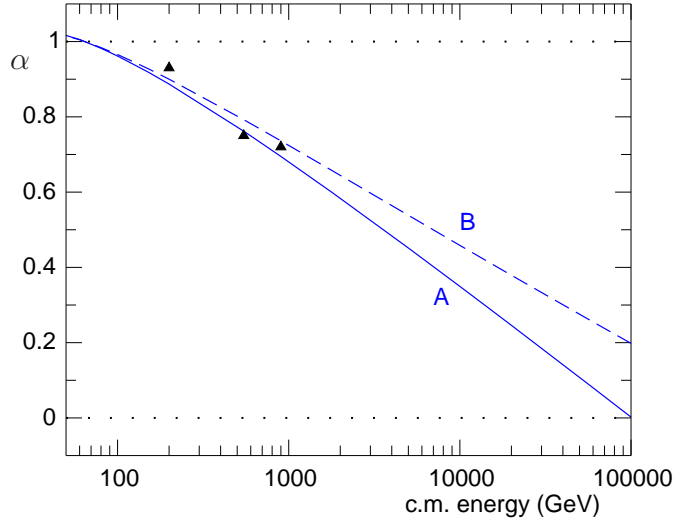}}
  \end{center}
  \caption[Extrapolations to TeV: alpha]{Energy dependence of 
the superposition parameter $\alpha$ (fraction of soft events)
in the two cases of a linear (solid line, Eq.~(\ref{eq:1A})) and 
quadratic (dashed line, Eq.~(\ref{eq:2B})) dependence of the
average multiplicity of the semi-hard component on c.m.\ energy.
The triangles are the result of the UA5 analysis
\cite{combo:prd}.}\label{fig:albertouno}
  \end{figure}

The success of the phenomenological analysis in  terms of NB (Pascal) MD's
of experimental data or available fits  on collective variable properties, 
together with the QCD roots of the MD's itself,  suggest to consider  
these results as  a sound  basis for the description of possible multiparticle
production scenarios   in  the TeV energy domain  accessible to future 
experiments at CERN LHC.
The approach we decided to  follow depends  in a crucial way  on  the 
NB (Pascal) MD parameters behaviour  and the problem one would 
like to solve  first  is  how to extrapolate these parameters
($\nbar_i$ and $k_i$   
($i= $ soft, semi-hard)) from the GeV  to the TeV energy region
\cite{combo:prd,combo:eta}.

\begin{figure}
  \begin{center}
  \mbox{\includegraphics[width=0.7\textwidth]{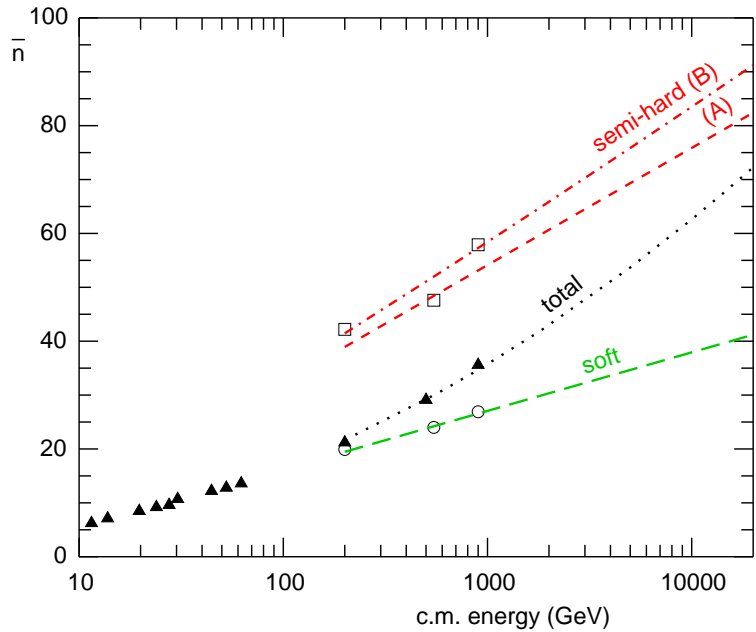}}
  \end{center}
  \caption[Extrapolations to TeV: Average multiplicity]{Average
		multiplicity $\nbar$ vs.\ c.m.\ 
energy. The figures shows experimental data (filled triangles) from
ISR and SPS colliders, the UA5 analysis with two NB (Pascal) MD's of SPS data
(circles: soft component; squares: semi-hard component), together with
our extrapolations (lines: dotted: total distribution; dashed: soft
component; short-dashed: semi-hard component, Eq.~(\ref{eq:1A});
dot-dashed: semi-hard component, Eq.~(\ref{eq:2B})
\cite{combo:prd}.}\label{fig:albertounoA}
  \end{figure}

\subsubsection{The average charged  multiplicities.} 
Extrapolations from UA5  collaboration results  on soft events
suggest
\begin{equation}
	\nbar_{\text{soft}} (\sqrt{s}) = - 5.54 + 4.72 \ln \sqrt{s} ,
\end{equation}
and estimates  for  semi-hard events from UA1 collaboration
\cite{UA1:minijets} give
\begin{equation}
	\nbar_{\text{semi-hard}} (\sqrt{s})  \approx  2 \nbar_{\text{soft}}
	(\sqrt{s}) \qquad\text{(case A)} \label{eq:1A}  % (A)
\end{equation}
or
\begin{equation}
	\nbar_{\text{semi-hard}} (\sqrt{s})  =  2 \nbar_{\text{soft}} (\sqrt{s})
  + 0.1 \ln^2 \sqrt{s} \qquad\text{(case B)}  \label{eq:2B} .  % (B)
\end{equation}
Since the average charged multiplicity is well parametrised by
$\nbar_{\text{total}}(\sqrt{s}) = 3.01 - 0.474 \ln \sqrt{s}
                                     + 0.75 \ln^2  \sqrt{s} $
and  the total charged average multiplicity $\nbar_{\text{total}}$ in terms
of the superposition of two components with weight $\alpha$  turns out to be
\begin{equation}
	\nbar_{\text{total}}  (\sqrt{s}) = \alpha \nbar_{\text{soft}}
  + (1-\alpha) \nbar_{\text{semi-hard}} ,
\end{equation}
it follows that
\begin{equation}
	\alpha_0 = 2 - \nbar_{\text{total}} / \nbar_{\text{soft}}
\end{equation}
in the case (A) and 
\begin{equation}
	\alpha = 1+(\nbar_{\text{soft}} -
	\nbar_{\text{total}})/(\nbar_{\text{soft}} - 0.1 \ln^2 \sqrt{s})
\end{equation}
in the case (B).
We expect therefore for cases (A) and (B) in the TeV region  the general trends 
of $\alpha$ and the average charged multiplicities shown in  Figures
\ref{fig:albertouno} and \ref{fig:albertounoA}.

\begin{figure}
  \begin{center}
		{\small $\Nbar$\hspace*{6cm}$\nc$}\\
  \mbox{\includegraphics[width=\textwidth]{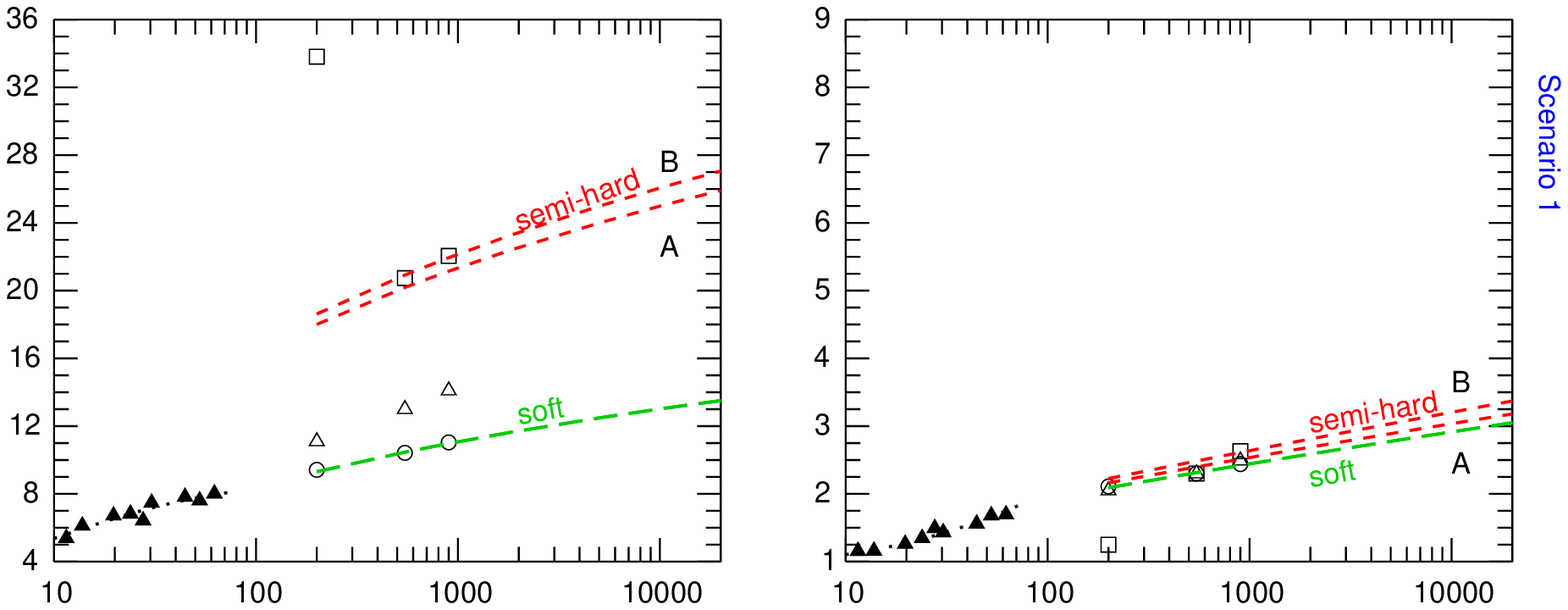}}
  \mbox{\includegraphics[width=\textwidth]{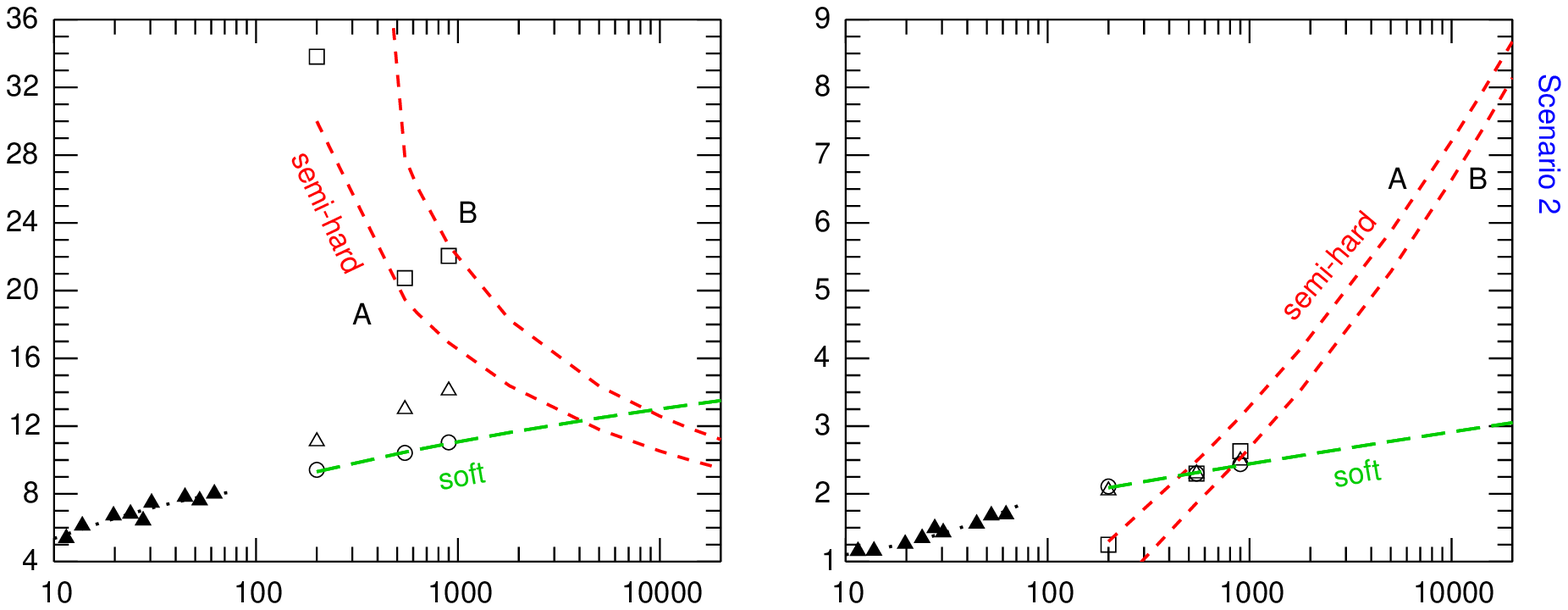}}
  \mbox{\includegraphics[width=\textwidth]{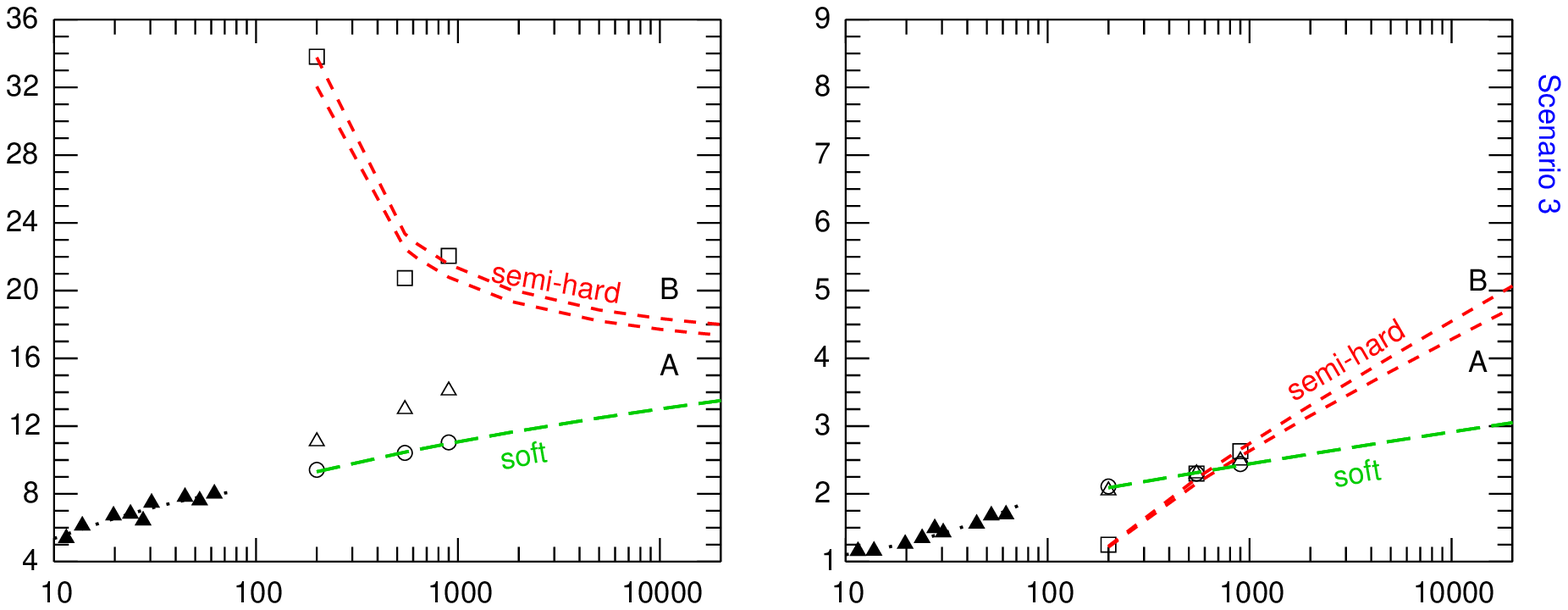}}
  \end{center}
  \caption[Extrapolations to TeV: clan parameters]{Clan parameters $\Nbar$
(panels in the left columns) and $\nc$ (panels in the right column)
are plotted for the scenarios described in the text
(from top to bottom: first row: scenario 1; second row: scenario 2;
third row: scenario 3).
The figures shows experimental data (filled triangles) from
ISR and SPS colliders, the UA5 analysis with two NB(Pascal) MD's of SPS data
(circles: soft component; squares: semi-hard component), together with
our extrapolations (lines: dotted: total distribution; dashed: soft
component; short-dashed: semi-hard component)
\cite{combo:prd}.}\label{fig:albertodue}
  \end{figure}

\subsubsection{The $k$ and $k_i$ ($i$=soft, semi-hard)  parameters.}

For the soft component  it is not too daring  to assume that KNO scaling
is not violated in the new energy domain, and accordingly to decide
that  $D^2_{\text{soft}} /\nbar^2{_{\text{soft}}}$ remains approximately
constant ($\approx 0.143$) 
 with $k_{\text{soft}}$ above 200 GeV c.m.\  energy  $\approx  7$. 
Assumptions for the semi-hard component are more delicate and at least 
three possibilities which could  characterise different scenarios
should be examined.

In scenario 1, being $k_{\text{semi-hard}}$ in  Fuglesang's fit
\cite{Fug} approximately 13 
at 900 GeV  and $\nbar_{\text{semi-hard}}$ even larger than
$\nbar_{\text{soft}}$  it is assumed 
that $k_{\text{semi-hard}}$ remains approximately constant with
$D^2_{\text{semi-hard}} /\nbar^2{_{\text{semi-hard}}} \approx 0.09$, 
i.e., in this scenario KNO scaling is not violated in the two separate 
components contributing to the total sample of events.

In scenario 2,  a strong KNO violation is assumed  with  
$D^2_{\text{semi-hard}} /\nbar^2_{\text{semi-hard}}$ growing logarithmically
with c.m.\ energy above 200 GeV  and  $k_{\text{semi-hard}}$ falling
from 79 at 200 GeV to $\approx 3$ at 14 TeV ($k_{\text{semi-hard}}
\approx {1/\ln\sqrt{s}}$).

Scenario 3 is the  QCD-inspired one, i.e.,
\begin{equation}
	k_{\text{semi-hard}}^{-1} = a + b \sqrt{\alpha_{\text{strong}}}
\end{equation}
with $\alpha_{\text{strong}}  \approx (\ln Q/Q_s)^{-1}$, $Q$ and
$Q_s$   are respectively the initial virtuality 
and the cut-off of the parton shower. $Q_s$, $a$ and  $b$ are determined
by a least square fit to the values of $k_{\text{semi-hard}}$ given by
UA5 collaboration 
at 200 GeV, 546 GeV and 900 GeV. It follows
\begin{equation}
	k_{\text{semi-hard}}^{-1} = 0.38  -  \sqrt{ 
		\frac{0.42 }{ {\ln(\sqrt{s} / 10)}} }
\end{equation}
The decrease of $k_{\text{semi-hard}}$  is milder than in scenario 2, it goes
from  13 at 200 GeV to  7 at 14 TeV; interestingly, the  QCD 
inspired behaviour of $k_{\text{semi-hard}}$   is intermediate among
the previous two, 
with no KNO  scaling and strong KNO scaling  violation respectively,
which describe probably quite extreme situations.

\subsubsection{Clan structure analysis of \pp\ collisions in the TeV region}
Following the above general discussion on the possible behaviours of
NB (Pascal) MD's parameters in the TeV  region,  their interpretation in terms 
of the average number of clans, $\Nbar$, and of the average number of 
particles per clan, $\nc$, reveals unsuspected new features which deserve 
particular attention (See Figure \ref{fig:albertodue}).

Of course, in scenario 1, $\Nbar_{\text{soft}}$ and
$\Nbar_{\text{semi-hard}}$ are expected to continue 
to increase ---although $k_{\text{soft}}$ and $k_{\text{semi-hard}}$
remain constant--- in view of the 
increase of the average charged multiplicity in each component.
Clans are quite numerous but their  size (i.e., $\bar n_{c,\text{soft}}$
and $\bar n_{c,\text{semi-hard}}$)
  although still growing with c.m.\ energy remains quite small. Notice that the 
behaviour of clans of the soft component is the  same  by assumption in all 
scenarios. The situation is similar to what we know from the GeV
region. What makes this analysis striking is the remark that
$\Nbar_{\text{semi-hard}}$ 
in the 2nd and 3rd scenarios  becomes smaller with the increase of the  c.m.\
energy, and the average number of particles per clan very large. 
The phenomenon is enhanced in particular in scenario 2 where 
$D^2_{\text{semi-hard}} /\nbar^2{_{\text{semi-hard}}}$  shows a
dramatic increase with energy 
as requested by strong KNO scaling violation.
The reduction of $\Nbar_{\text{semi-hard}}$  and the simultaneous
quick  increase of  
$\nbar_{\text{semi-hard}}$  suggests that in the scenarios 2 and 3
the available c.m.\ energy  
goes more in particle production within  a clan rather than in new clan 
production, differently from what was found in scenario 1. 

\begin{figure}
  \begin{center}
  \mbox{\includegraphics[width=0.8\textwidth]{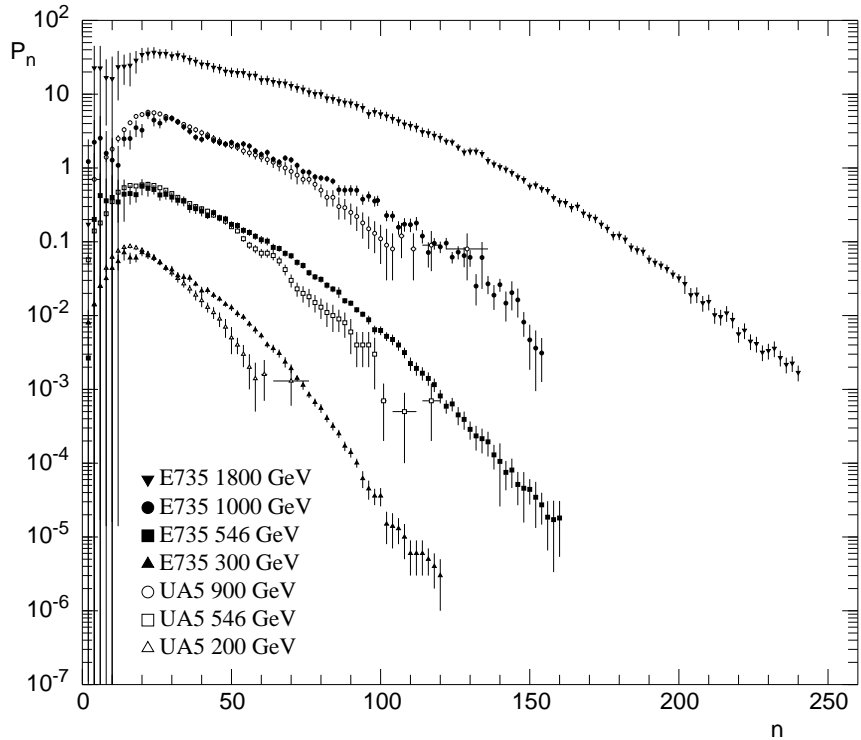}}
  \end{center}
  \caption[E735 results vs UA5]{E735 results on charged particle
  multiplicity distributions 
  in \pp\ collisions at various energies
  in full phase-space compared with UA5 results at similar energies. 
  Data from the two experiments which were taken at nearly the same
  energy are rescaled by the same factor \cite{walker}.
  }\label{fig:UA5E735}
  \end{figure}

Data on MD's at 1.8 TeV c.m.\ energy (from E735 experiment
\cite{walker}), when compared with our predictions 
are closer to scenario 2 but with an even wider MD.
It is to be stressed that E735 results
on full phase-space multiplicity distributions
do not completely agree with those obtained at comparable energies
at the S{\ppbar}S collider by the UA5 Collaboration
\cite{UA5:rep,UA5:3}, see Fig.~\ref{fig:UA5E735}. 
Tevatron data are more precise than
S{\ppbar}S data at larger multiplicities (they have larger
statistics and extend to larger multiplicities than UA5 data), but
much less precise at low multiplicity.
Both sets of data show a shoulder structure, but the Tevatron MD is
somewhat wider. It should be noticed that E735 data are measured
only in $|\eta|<3.25$ and $\pt > 0.2$ GeV/$c$
then extended to full phase-space via a Monte Carlo program.
Notwithstanding the mentioned discrepancy, there is consensus in
accepting the presence of (at least) two substructures in the data,
either related to the varying impact parameter \cite{two-comp} or to
multiple parton scattering \cite{Walker:p,DelFabbro:2001rs}.
In terms of the scenarios discussed in this Section,
one can reasonably argue that scenario 1 is the less probable to occur and that
therefore the above mentioned finding on clan reduction 
in scenarios 2 and 3  raises some interesting questions on the properties
of  clans themselves,  i.e., how far the number of clans of the semi-hard 
component could be reduced in our framework? is  
the reduction to one or very few clans  compatible with the structure
of the semi-hard component? should we expect, with the increase of the 
c.m.\ energy of the collision, the onset of a new component to be added to
the previous two? The attempt to  answer  these questions is postponed to  
the next sections, where also we would like to address the related question on 
the real nature of clans, which seems quite natural at this stage of our
search, i.e., are clans  observable (massive) objects?

\subsubsection{Analysis in pseudo-rapidity intervals}
We would like to  point  out that our study 
on the possible three scenarios in the TeV region based on extrapolations
of the experimental knowledge of the GeV energy domain can be extended from 
full phase-space  to  rapidity intervals.
 
Since only after the classification of events (soft and semi-hard) has been 
carried out do we look at phase-space intervals, it seems quite reasonable 
to assume that the weight factor $\alpha$ depends on the c.m.\ energy 
(as in full phase-space) and not on the pseudo-rapidity interval $\eta_c$.

The $\eta_c$ dependence comes therefore from $\nbar_i$ and $k_i$
parameters only.

\begin{table}
	\caption[Values of 1/k in the three scenarios]{Values of 
$1/k_{\text{total}}$, $1/k_{\text{soft}}$
and $1/k_{\text{semi-hard}}$ in our extrapolations 
for different rapidity intervals and for full phase-space (FPS).
$1/k_{\text{soft}}$  and, in scenario 1,  $1/k_{\text{semi-hard}}$,
are energy independent and their value is given in the table; in the
other cases we give the parameters and the form of the energy dependence
\cite{combo:eta}.}\label{table:kcombo}
\newcommand{\rs}{\sqrt{s}}
\begin{center}
\begin{tabular}{@{\extracolsep{1pt}}ccccc}
\hline
interval  & soft comp. & scenario 1 & scenario 2 & scenario 3 \\[2mm]
$|\eta| \le \eta_c$ & $k_{\text{soft}}^{-1}$ & $k_{\text{semi-hard}}^{-1}$ &   
   $k_{\text{total}}^{-1}(\eta_c,\rs) =$ &
   $k_{\text{semi-hard}}^{-1}(\eta_c,\rs) =$ \\[1mm]
 &&&  $a + b\ln\rs$ & 
   $C + D/{\sqrt{\ln(\rs/10)}}$\\[3mm]
\hline
$\eta_c = 1$ & 0.294 & 0.217 & 
        \parbox[c][1.5\height][c]{2.5cm}{$a = 0.02$\\$b = 0.08$} & 
        \parbox{3cm}{$C = 0.97$ \\$D = -1.6$} \\
\hline
$\eta_c = 2$ & 0.286 & 0.172 & 
        \parbox[c][1.5\height][c]{2.5cm}{$a = -0.06$\\$b = 0.08$} & 
        \parbox{3cm}{$C = 0.88$ \\$D = -1.5$} \\
\hline
$\eta_c = 3$ & 0.250 & 0.156 & 
        \parbox[c][1.5\height][c]{2.5cm}{$a = -0.12$\\$b = 0.08$} & 
        \parbox{3cm}{$C = 0.72$ \\$D = -1.2$} \\
\hline
FPS &        0.143 & 0.077 & 
        \parbox[c][1.5\height][c]{2.5cm}{$a = -0.082$\\$b =0.0512$} & 
        \parbox{3cm}{$C = 0.38$ \\$D = -0.65$}\\
\hline
\end{tabular}
\end{center}
\end{table}

%\subsubsection{average multiplicity in pseudo-rapidity intervals.} 
In order to be consistent with our assumptions in full phase-space,
the single particle density must  show an energy independent plateau which
extends in pseudo-rapidity for some units around $\eta = 0$ in each direction.
Accordingly, the height of the plateau ($\nbar_0$) of the soft and semi-hard
component is fixed equal to 
\begin{equation}
	\nbar_{0,\text{soft}} \approx 2.45  
\end{equation}
(a value compatible with low energy data \cite{Giacomelli} where only the soft
component is present) and
 \begin{equation}
	 \nbar_{0,\text{semi-hard}} \approx 6.4
 \end{equation}
(notice that  the assumption of an energy independent plateau for the
semi-hard component is not compelling  and that  the following  increase
with c.m.\ energy of   $\nbar_{0,\text{semi-hard}}$ is again
compatible with all our 
previous discussion: $\nbar_{0,\text{semi-hard}} \approx 6.3 + 0.07 \ln
\sqrt{s}$).  

Finally
\begin{equation}
	\nbar_i (\eta_c) = 2 \nbar_{0,i} \eta_c   (i=\text{soft},\text{semi-hard})
\end{equation}
and 
\begin{equation}
	\nbar_{\text{total}} (\eta_c,  \sqrt{s}) = \alpha_{\text{soft}} (\sqrt{s})
  \nbar_{\text{soft}} (\eta_c) + [ 1 - \alpha_{\text{soft}}
  (\sqrt{s})] \nbar_{\text{semi-hard}} 
  (\eta_c).
\end{equation}

\begin{figure}
  \begin{center}
  \mbox{\includegraphics[width=0.9\textwidth,height=9cm]{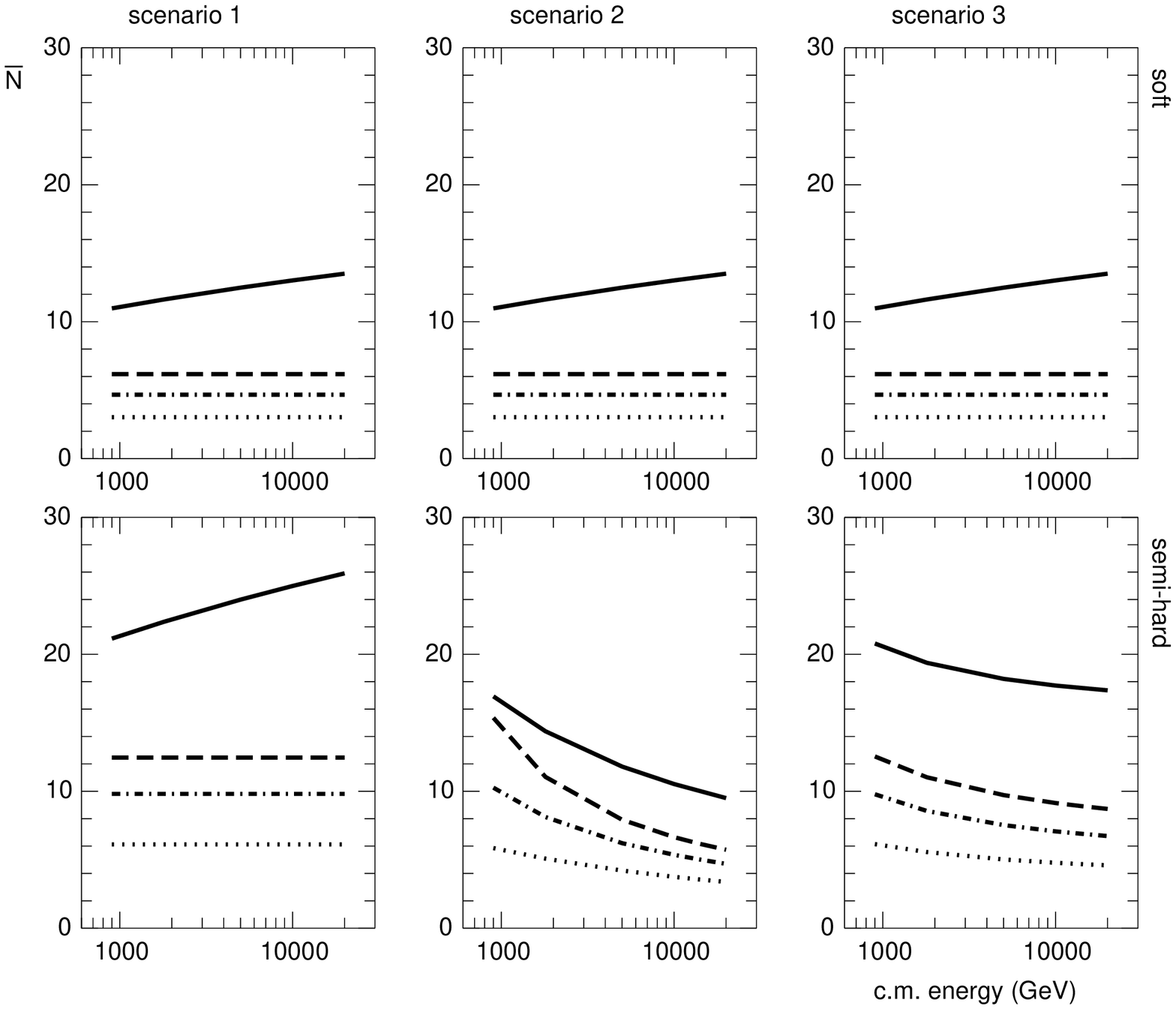}}
  \mbox{\includegraphics[width=0.9\textwidth,height=9cm]{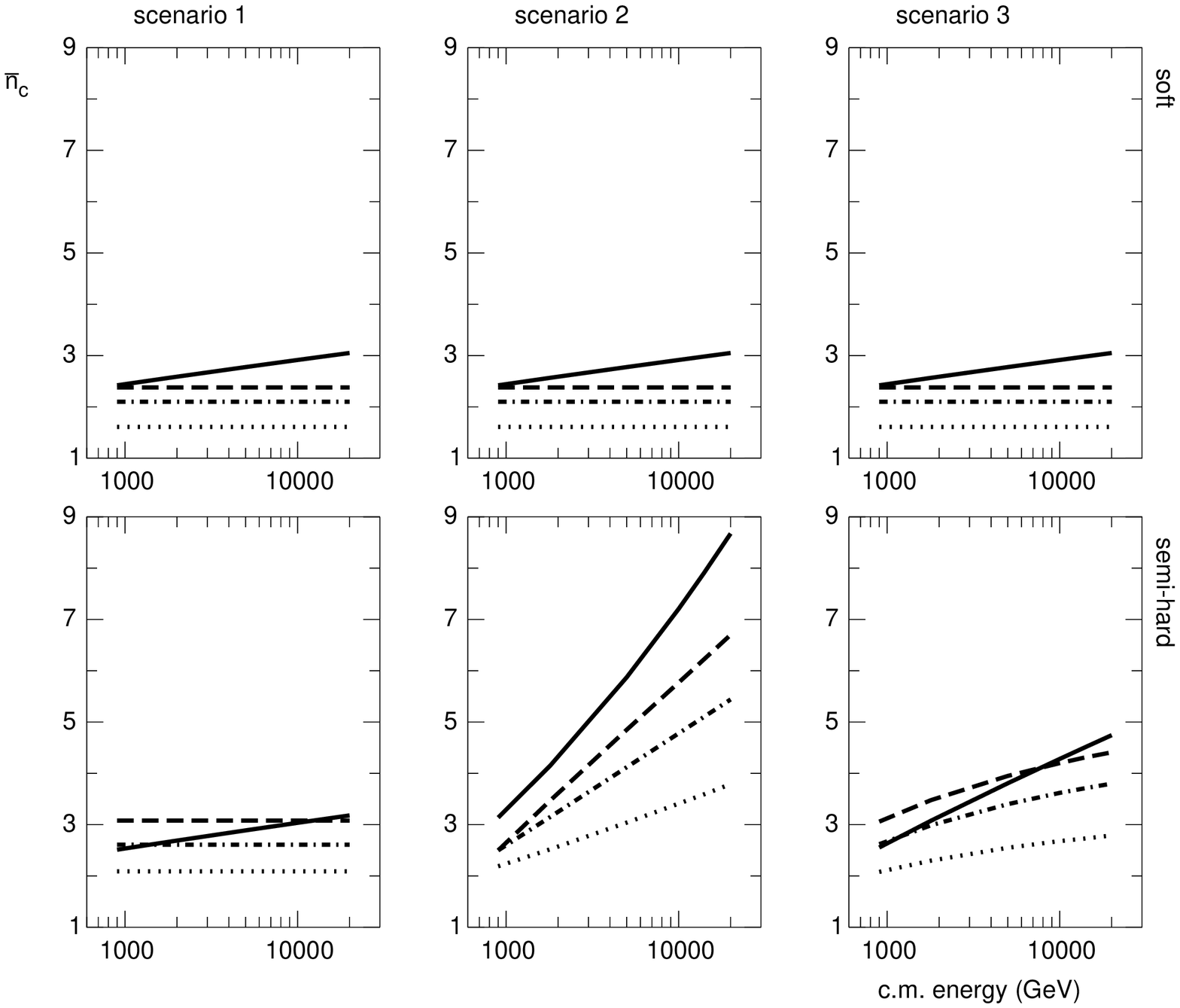}}
  \end{center}
  \caption[Extrapolations to TeV in rapidity intervals]{The average
		number of clans, $\bar N$, (top two rows) and the  
average number of particles per clan, $\nc$, (bottom two rows)
are plotted against the c.m.\ energy for three rapidity intervals
(dotted line: $\eta_c=1$; dash-dotted line: $\eta_c=2$;
dashed line: $\eta_c=3$) and for f.p.s. (solid line), 
for each scenario (in columns, from left
to right: scenario 1, 2 and 3) and for each component (in rows, from top to
bottom: soft and semi-hard) \cite{combo:eta}.}\label{fig:albertotre}
  \end{figure}

%\subsubsection{Dispersion in pseudo-rapidity intervals}
By recalling that
\begin{multline}
			\nbar^2_{\text{total}}(\eta_c,\sqrt{s}) 
			\left( 1 + \frac{ 1 }{k_{\text{total}}(\eta_c,\sqrt{s}) } \right)
			=
			\alpha(\sqrt{s}) \nbar^2_{\text{soft}}(\eta_c,\sqrt{s})  \left(1 +
			\frac{1}{k_{\text{soft}}(\eta_c,\sqrt{s}) } \right)  \\
			+ 
			(1-\alpha(\sqrt{s}))
			\nbar^2_{\text{semi-hard}}(\eta_c,\sqrt{s})  \left(1 +
			\frac{1}{k_{\text{semi-hard}}(\eta_c,\sqrt{s}) } \right)  
\end{multline}
results contained in table  \ref{table:kcombo}  are obtained.
Clan structure analysis confirms the general trends found in full phase-space,
favouring a lower number of clans of smaller size in more restricted rapidity
intervals; of  particular interest is again   the semi-hard component.
The results are illustrated in Fig.~\ref{fig:albertotre}.
In scenario 1, as the energy increases  one notices  
a copious production of clans of nearly equal size
in all rapidity intervals. In scenario 2, in all
pseudo-rapidity intervals,  to the higher aggregation of newly created
particles into existing clans it follows the aggregation of clans into 
super-clans favouring stronger long range correlations.
Scenario 3 (the QCD-inspired one) leads to predictions which are ---as 
usual---  intermediate between the previous two.

\subsection{Hints from CDF}\label{sec:III.cdf}
At the Tevatron, 
the subsample of events with no energy clusters above 1.1 GeV  and
the subsample of remaining events  were separated in the total sample
of \ppbar\ events collected in the pseudo-rapidity window $|\eta|<1$  
with CDF minimum bias trigger at $\sqrt s = 630$ GeV
and at $\sqrt s = 1800$ GeV \cite{CDF:soft-hard}.

\begin{figure}
  \begin{center}
  \mbox{\includegraphics[width=0.45\textwidth]{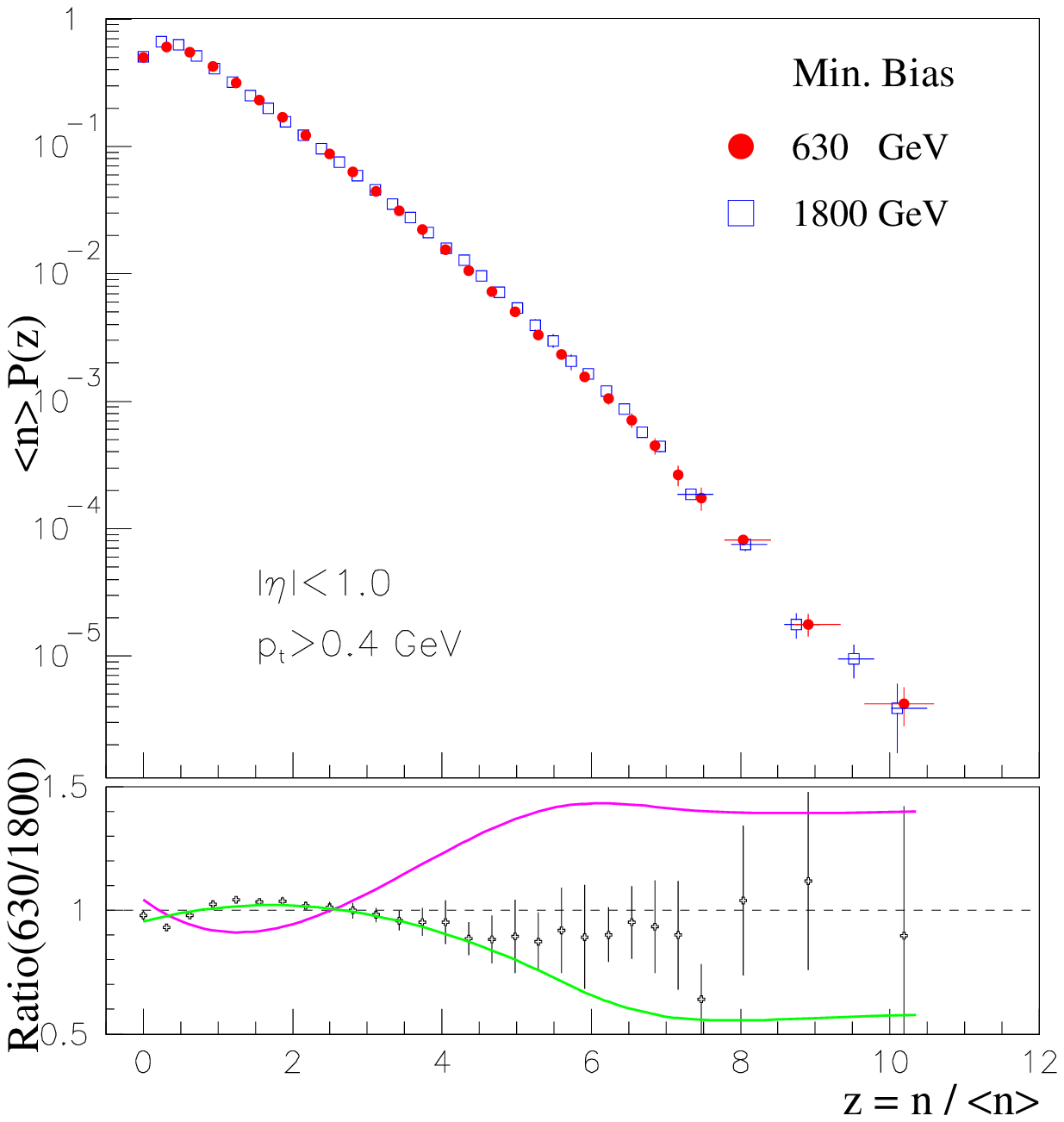}}\\
  \mbox{\includegraphics[width=0.45\textwidth]{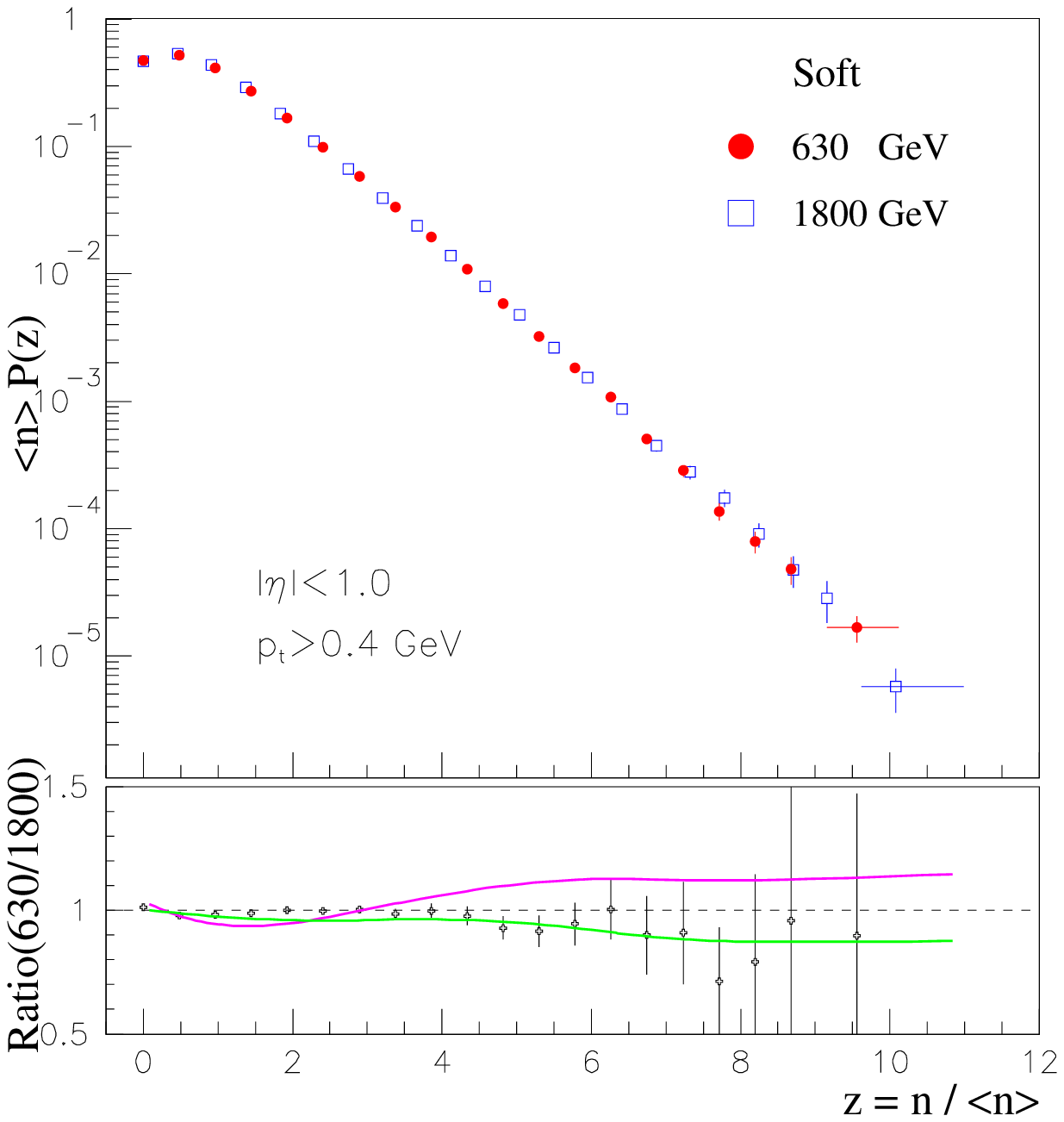}}
  \mbox{\includegraphics[width=0.45\textwidth]{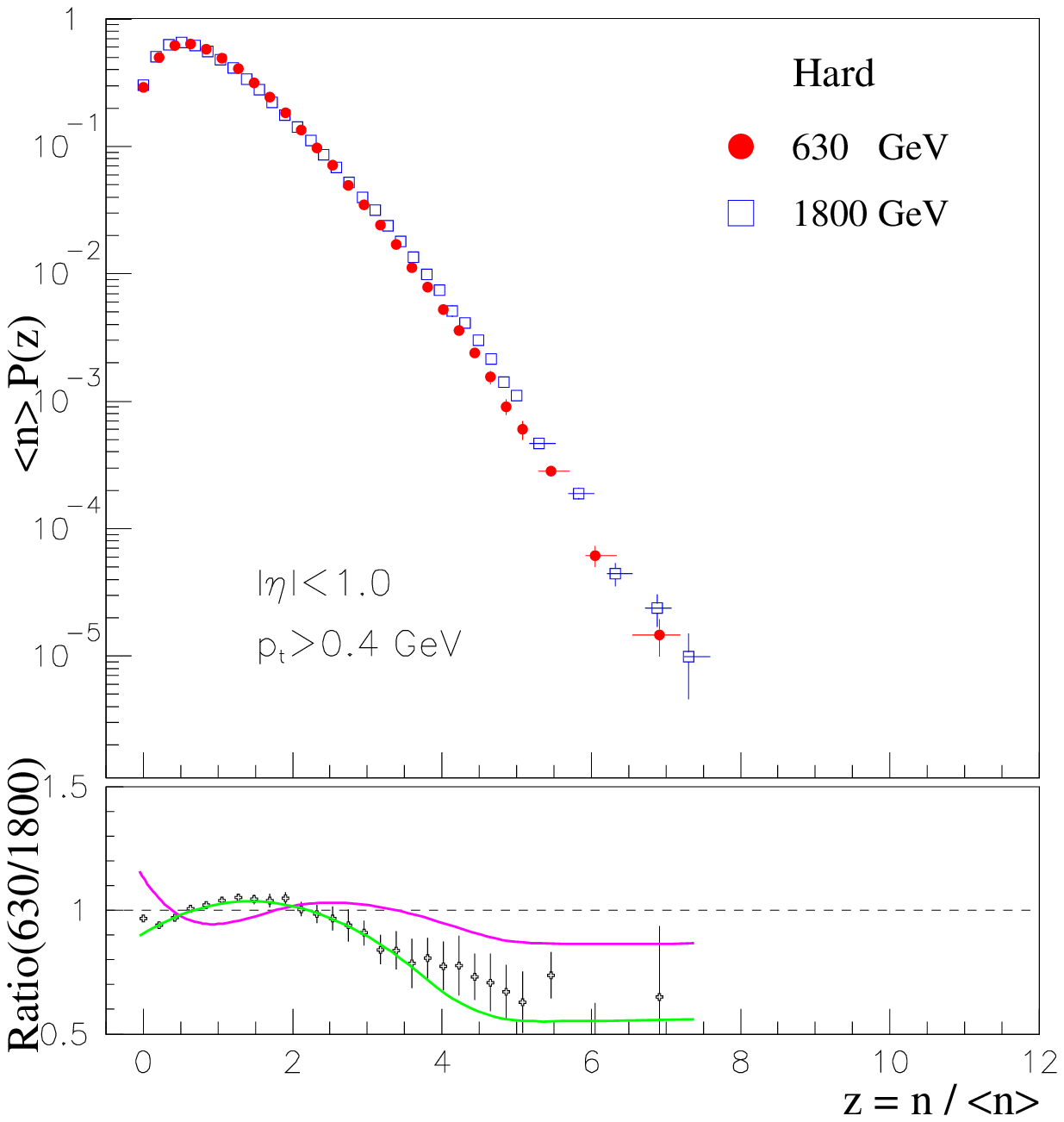}}
  \end{center}
  \caption[CDF data on MD's]{Multiplicity distributions for the full
		minimum bias, 
		the `soft' and the `hard' samples at 1800
  and 630 GeV from CDF; data are plotted in KNO variables. In the
  bottom panel of each figure
  the ratio of the two above distributions is shown. The two
  continuous lines delimit the band of all systematic 
	uncertainties \cite{CDF:soft-hard}.}\label{fig:alberto11}
  \end{figure}

\begin{figure}
  \begin{center}
  \mbox{\includegraphics[width=0.45\textwidth]{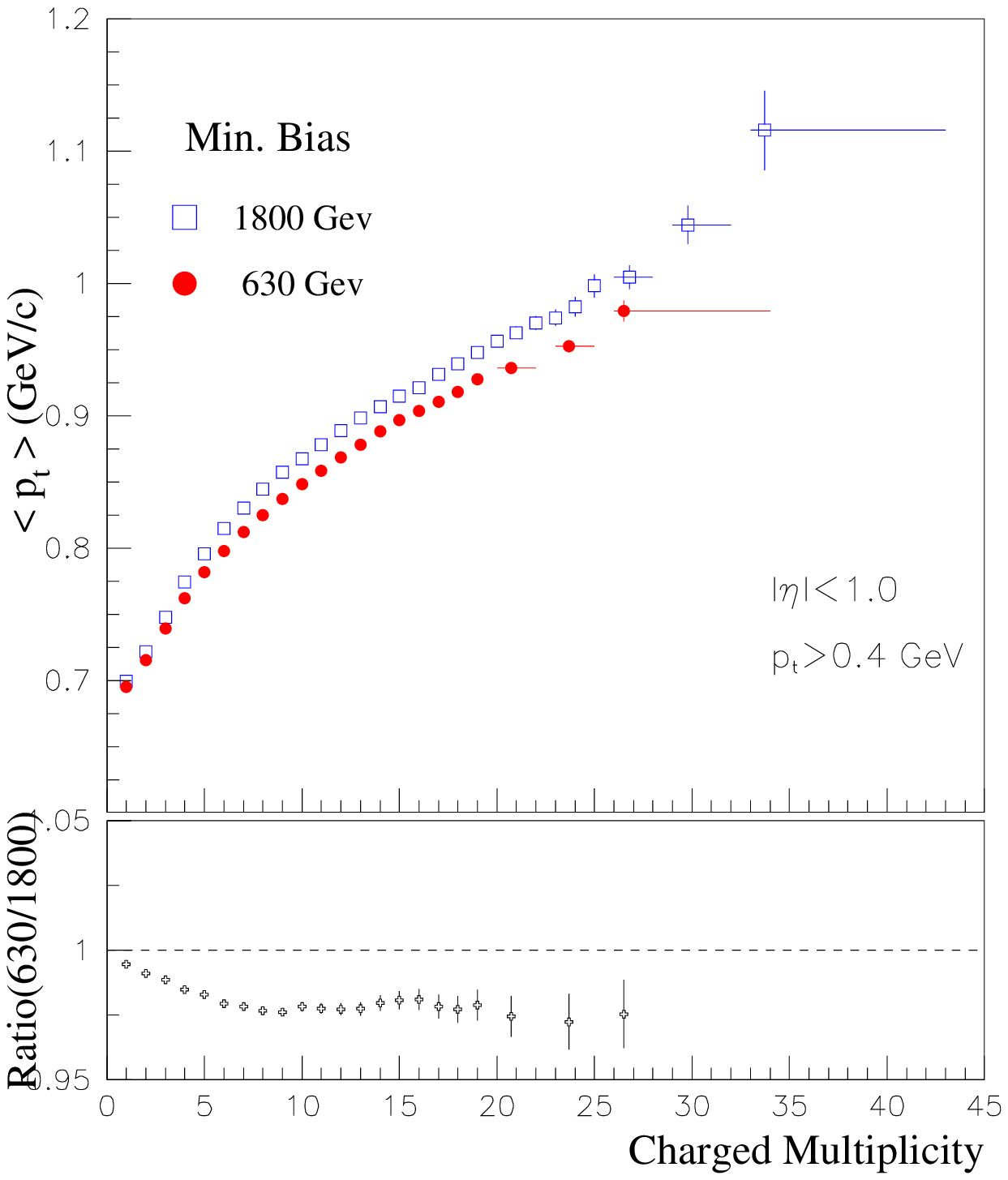}}\\
  \mbox{\includegraphics[width=0.45\textwidth]{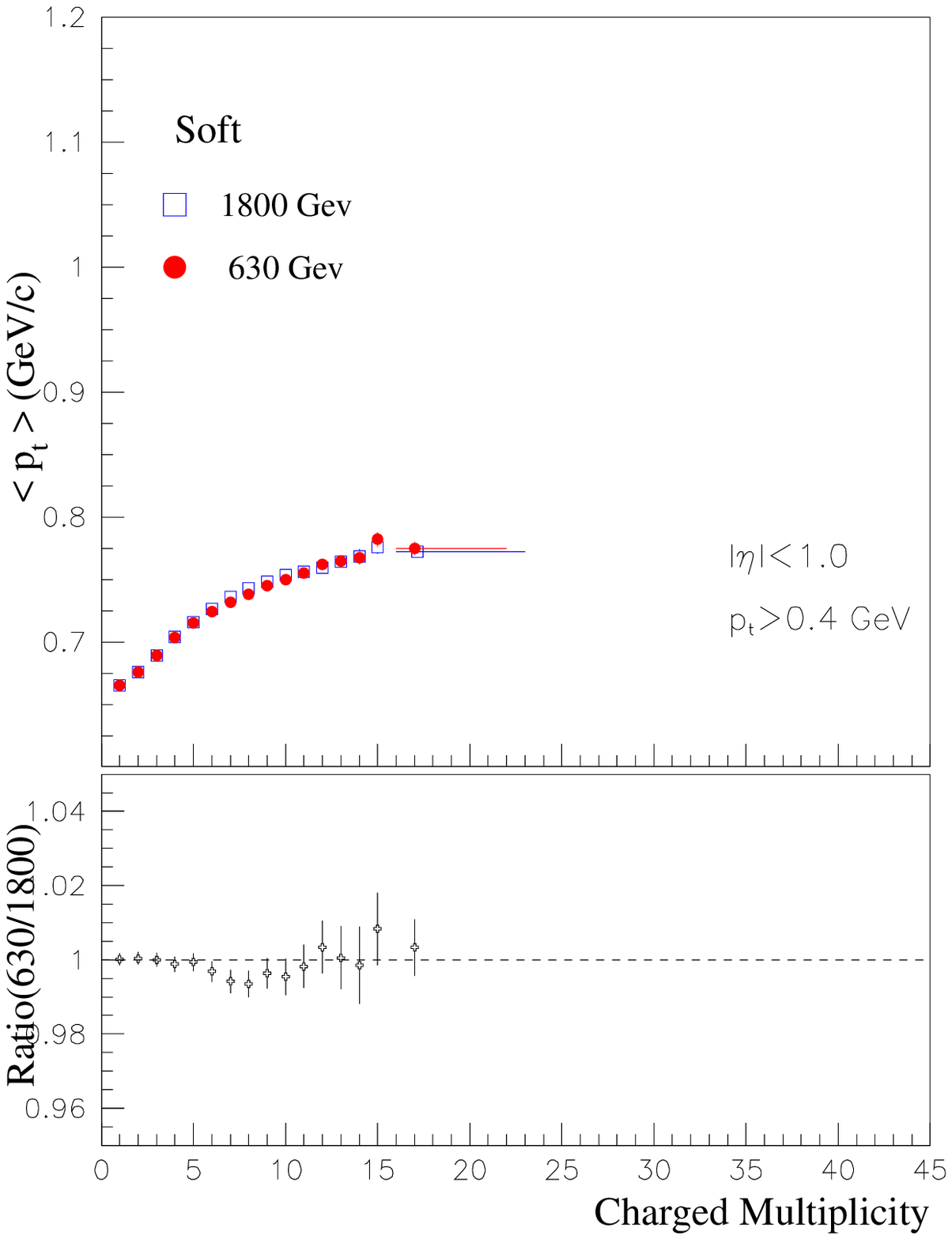}}
  \mbox{\includegraphics[width=0.45\textwidth]{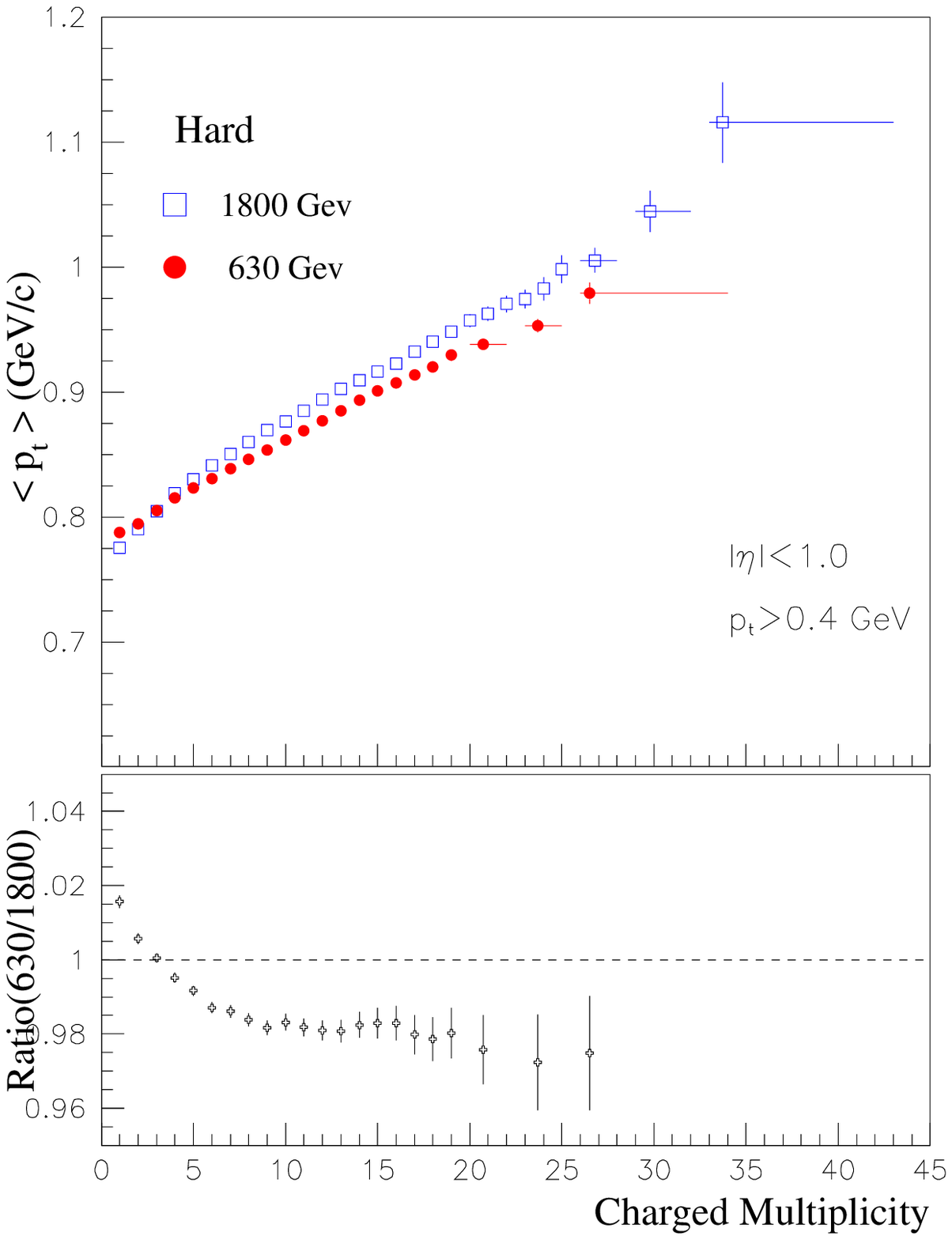}}
  \end{center}
  \caption[CDF data on mean pt]{Mean transverse momentum vs
		multiplicity for the full  
		minimum bias, the `soft' and the `hard' 
		samples at 1800 and 630 GeV from CDF. 
    On the bottom panel of each figure
    the ratio of the two curves is shown 
    \cite{CDF:soft-hard}.}\label{fig:alberto12}
  \end{figure}

Being the two samples of events highly
enriched in soft and hard interactions respectively and the Collaboration 
quite aware of the difficulty of a correct   identification and separation of 
jets with transverse energy lower than 5 GeV,  events with no clusters
were called `soft' and those with at least one cluster, `hard'.
$n$ charged particle multiplicity distributions for the
minimum bias  samples at 
the two energies were plotted in the KNO form and showed, in the interval
$|\eta | < 1$,  KNO scaling violations. 
When the two samples of events are
separately plotted the soft component satisfies KNO scaling, whereas the
hard component clearly violates KNO scaling, suggesting that the behaviour
of the second component is the main  cause of KNO scaling violations in 
the full sample, as shown in Fig.~\ref{fig:alberto11}.

Interestingly the mean transverse momentum  $\avg{p_T}$ when plotted versus 
$n$ charged multiplicity for the full minimum bias sample is larger at
1800  GeV than at 630 GeV, whereas it does not grow with the c.m.\ energy
in the soft sample (See Fig.~\ref{fig:alberto12}.) 
The different behaviour at the two
energies in the full minimum bias sample  is therefore entirely due to the 
hard sample   and it is probably the effect  of high transverse energy 
interactions (mini-jets in the literature).

These results, although limited to a single pseudo-rapidity interval
only, provide an interesting experimental support to the two main
assumptions which justified our approach to the TeV energy domain in
\pp\ collisions, i.e.:
 
\emph{a.} that there are at least two components in the total sample
of events at threshold of the TeV energy domain (they are called
`soft' and `semi-hard' in Ref.\ \cite{Fug} and by us (Sections
\ref{sec:III.shoulder} and \ref{sec:III.scenarios}), and `soft' and
`hard' in the paper by the CDF Collaboration);

\emph{b.} that the soft component satisfies KNO scaling. 

In addition, as far as the second component (the semi-hard one) is
concerned, the scenario with strong KNO scaling violation or the one
based on QCD inspired behaviour seem favoured with respect to the
scenario with KNO scaling.

\subsection{Forward-backward multiplicity correlations.  
The demand for  the existence of clans.}\label{sec:III.fbmc}
 
The experimental problem consists in studying, in each event, the number
of particles falling  in the forward  hemisphere (F) and in the backward
hemisphere (B) \cite{RU:FB}. 
Usually, the average number of particles
in the B hemisphere, $\nbar_B$, is 
successfully parametrised in terms of the 
number of particles in the F hemisphere, $n_F$, 
according to the following linear relation:
\begin{equation}
  \nbar_B(n_F) = a + \bFB n_F ,
\end{equation}
where $\bFB$ is the FB multiplicity correlation strength:
\begin{equation}
	\begin{split}
		\bFB & \equiv \frac{ \text{Cov}[n_F, n_B]}{(\text{Var}[n_F] 
                                  \text{Var}[n_B])^{1/2}}\\
		     & = \frac{ \avg{(n_F -\nbar_F) (n_B - \nbar_B)} }{ 
			         \left[\avg{(n_F -\nbar_F)^2}  \avg{(n_B - \nbar_B)^2}
		           \right]^{1/2}} .  \label{eq:III.FB.define}
	\end{split}
\end{equation}

The existence of strong correlations between particles 
in the two hemispheres is an important effect and could be a
signal of large  colour exchange among partons   at  parton level. It is 
instructive with this aim  to examine the experimental situation  on FB 
multiplicity correlations  in \pp\ collisions and \ee\  annihilation.

In \pp\ collisions the F hemisphere coincides with the region of
the outgoing proton, the B hemisphere is the symmetric region in the opposite
direction. 
In \pp\ collisions, $\bFB$ is growing with c.m.\ energy and is rather large.
At 63 GeV  (ISR)  $\bFB = 0.156 \pm 0.013$ and at 546 GeV (UA5) 
$\bFB = 0.43 \pm 0.01$  in $1< |\eta | < 4$  and  $\bFB = 0.58 \pm 0.01$ in
$0< |\eta| < 4$. Its general trend with c.m.\ energy  is given by
\begin{equation}
  \bFB = - 0.019 + 0.061 \ln s    \label{eq:linearbFB}
\end{equation}
In \ee\  annihilation the F hemisphere is chosen randomly  between the
two hemispheres defined by the plane perpendicular to the thrust axis.
OPAL collaboration has found $\bFB = 0.103 \pm 0.007$ for the total
sample of events and  $\bFB \approx 0$ 
in the separate two- and three-jets  sample
of events. TASSO Collaboration finds $\bFB = 0.080 \pm 0.016 $ in the
total sample of events, we do  not have estimates in this case of $\bFB$ in the 
separate samples of events. In conclusion in \ee\  annihilation
$\bFB$ grows with energy but is rather small.

Accordingly our interest on the theory side  was limited to symmetric reactions 
only \cite{RU:FB}, although a generalisation to asymmetric collisions  can be 
worked out \cite{RU:FBasymm}.

The number of F and B particles, $n_F$ and $n_B$, are of course random
variables with  $n_F + n_B =  n$   and 
\begin{equation}
	P_n = \sum_{n=n_F+n_B}  P_{\text{total}}(n_F, n_B).
\end{equation}
$P_{\text{total}}(n_F, n_B)$ is the joint distribution for the
weighted superposition 
of different classes of events and is equal to $\alpha P_1 (n_F, n_B) +
(1 - \alpha)   P_2 (n_F, n_B)$; $\alpha$ is the weight of class 1 of events
with respect to the total sample of events. 
The calculation of $\bFB$ according to Eq.~(\ref{eq:III.FB.define})
leads then to the following result:
\begin{equation}
        \bFB = \frac{\alpha b_1 {D^2_{n,1}}/(1+b_1) +
                  (1-\alpha) b_2 {D^2_{n,2}}/(1+b_2) +
                          \frac{1}{2}\alpha(1-\alpha)(\nbar_{2} - 
                                \nbar_{1})^2}
                   {\alpha  {D^2_{n,1}}/(1+b_1) +
                          (1-\alpha)  {D^2_{n,2}}/(1+b_2) +
                                 \frac{1}{2}\alpha(1-\alpha)(\nbar_{2} -
                        \nbar_{1})^2}  .
\label{eq:b_total}  %   (A)
\end{equation}
Here $b_1$ and $b_2$ are the correlation strengths of events of class 1 and 2,
$D_1$ and $D_2$ the corresponding dispersions and $\nbar_1$ and $\nbar_2$ the
corresponding average multiplicities.

For $b_1 = b_2 = 0$, one has $\bFB \to b_{12}$  and the previous
formula is reduced to  the following one
\begin{equation}
        b_{12} = \frac{
                 \frac{1}{2}\alpha(1-\alpha)(\nbar_{2} - \nbar_{1})^2}
              {\alpha  {D^2_{n,1}} +
                                (1-\alpha)  {D^2_{n,2}} +
                   \frac{1}{2}\alpha(1-\alpha)(\nbar_{2} - \nbar_{1})^2} .
        \label{eq:b_12}    % (B)
\end{equation}

It is  a general property of the two formulae to be  independent of the 
specific form of the $n_1$ particles and $n_2$ particles  multiplicity 
distributions, $P_{n_1}$ and $P_{n_2}$,  and to depend on
$\alpha$, $\nbar_1$, $\nbar_2$, $D_1$ and $D_2$ parameters only.

OPAL collaboration  measured FB multiplicity correlation strengths in
\ee\  annihilation in the two separate samples of events
(2-jet sample and 3-jet sample) and in the total sample \cite{OPAL:FB}. 
It has been found
that $b_1 \approx b_2 \approx 0$ and $\bFB = b_{12} = 0.103 \pm 0.007$  
respectively. Being 
$\alpha = 0.463$, $\nbar_1 = 18.4$,  $D_1^{2} = 25.6$,  $\nbar_2 =24.0$
and $D_2^{2} = 44.6$, the theoretical estimate of  $b_{12}$  according
to formula (\ref{eq:b_12})
turns out to be 0.101  in perfect agreement with the  experimental 
value $0.103\pm0.007$. The success of our formula (\ref{eq:b_12})
in describing FB
multiplicity correlations  in \ee\  annihilation  should be contrasted
with its failure in describing  $\bFB$ in \pp\ collisions: 
$b_1$ and $b_2$ are
expected to be different from zero in the latter case  and the more general
formula (\ref{eq:b_total}) is needed.
The  parameters  of such general formula can be computed
under the assumptions that :
 
a) particles are independently produced;

b) they are binomially distributed in the F and B hemispheres;

c) the overall MD for each component (soft and semi-hard) is a
NB (Pascal) MD
with characteristic parameters $\nbar_i$ and $k_i$ ($i=1,2$).

\noindent It can then be
concluded that charged particles FB multiplicity  correlations are \emph{not 
compatible} with independent particles emission   and that charged particles FB 
multiplicity correlations are \emph{compatible} with the production of
particles in   clusters (i.e., clans in view of assumption c).
It follows that:

\emph{a}. the joint probability distribution $P(n_F, n_B)$ is written as the
convolution 
over the number of produced clans and over the partitions of forward and 
backward produced particles among clans;

\emph{b}. the symmetry argument should be used;

\emph{c}. the introduction of a leakage parameter $p$ is needed:
the new parameter $p$  controls the probability that a binomially distributed 
particle (and  generated by one clan lying in one hemisphere) has to leak 
in the opposite hemisphere ($p=1$ means no leakage, the variation domain of $p$ 
is $1/2 < p < 1$  and  $p < 0.5$ implies that the clan to which the emitted 
particle is belonging  is classified in the wrong hemisphere);

\emph{d}. a covariance parameter, $\gamma$,
between F and B particles within a clan  is also introduced;

\emph{e}. within this framework clans are binomially produced in the F and B
hemispheres 
with the same probabilities and particles belonging to a clan are
independently produced in the two hemispheres.

From these assumptions one gets that:

\emph{i}. clan structure analysis is applicable for each component,

\emph{ii}. clans in each component are independently emitted (their
distribution is  Poissonian),

\emph{iii}. clans are binomially distributed in the two hemispheres,

\emph{iv}. logarithmically  produced particles from each clan are also
binomially  
distributed in  the F and B hemispheres but with different probabilities
(the corresponding  leakage parameters are different).

The parameters of formula (\ref{eq:b_total}) can then be calculated.
We find indeed  for the single  component strength
\begin{equation}\label{eq:C}
        \begin{split}
        \bFB^{(i)} &= 
				     \frac{D^2_N - 4\avg{d^2_{N_F}(N)}(p-q)^2 + 4\Nbar\gamma/\nc^2}
                   {D^2_N + 4\avg{d^2_{N_F}(N)}(p-q)^2 - 4\Nbar\gamma/\nc^2
 + 
                                        2\Nbar D^2_c/\nc^2}\\
                &= \frac{D^2_n/\nbar - D^2_c/\nc -
        4\avg{d^2_{N_F}(N)}(p-q)^2\nc/\Nbar + 4\gamma/\nc }{
        D^2_n/\nbar  + D^2_c/\nc +
        4\avg{d^2_{N_F}(N)}(p-q)^2\nc/\Nbar - 4\gamma/\nc }  .
        \end{split}
\end{equation}
Formula (\ref{eq:C}) is valid for the class of CPMD's. 
Assuming in addition,
according to \emph{iv}, that clans in each component are of NB (Pascal) MD
type  with $k_i$ and $\nbar_i$  parameters,  one has
\begin{equation}	\label{eq:D}    %  (D)
	\begin{split}
		\bFB^{(i)} & = 2 \nbar_i p_i (1- p_i)/
          (\nbar_i + k_i - 2 \nbar_i p_i (1- p_i)) \\
		& = 2 \beta_i p_i (1-p_i) /(1 -  2 \beta_i p_i (1-p_i)) ,
	\end{split}
\end{equation}
with  $\beta_i = \nbar_i / (\nbar_i + k_i)$.
It should be remembered that the leakage parameter $p_i$ is the fraction of 
particles within one clan which fall in the same hemisphere where
the clan was produced. 
Assuming that the semi-hard component is negligible at $\sqrt{s} = 63$ GeV
and knowing $\bFB (\sqrt{s}= 63~\text{GeV}) = 0.156 \pm 0.013$ from experiments,
and of course $\nbar_{\text{soft}}$, $k_{\text{soft}} $
leakage parameter for the soft component can be determined. 
We find $p_{\text{soft}} = 0.78$, i.e., 22\% of the particles
are expected to leak from clans in one hemisphere to the opposite one.

\begin{figure}
  \begin{center}
  \mbox{\includegraphics[width=\textwidth]{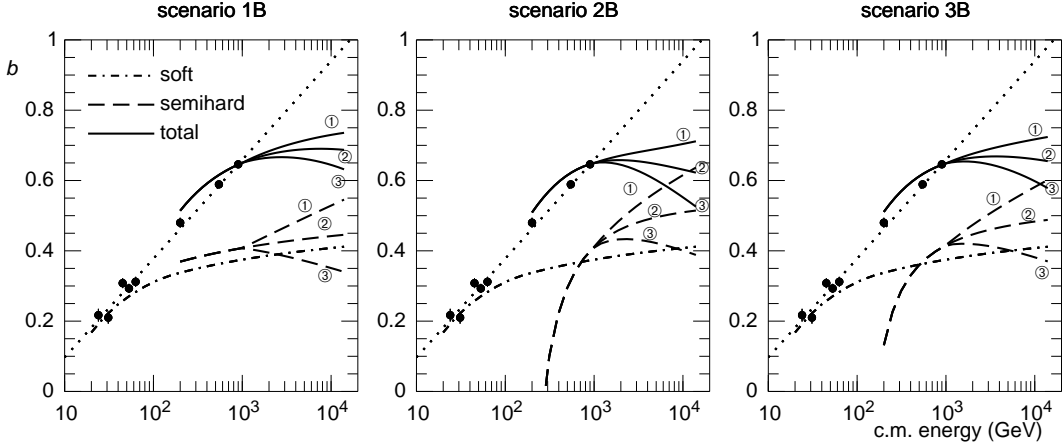}}
  \end{center}
  \caption[FB coefficient vs c.m. energy]{Predictions for the
  correlation coefficients  
  for each component (soft and semi-hard) 
  and for the total distribution in $p\bar p$ collisions.
  For each scenario, three
  cases are illustrated, corresponding to the three numbered branches: 
  leakage increasing with $\sqrt{s}$ (upper branch, \zapf{\char'300}),
  constant leakage (middle branch, \zapf{\char'301}) and
  leakage decreasing with $\sqrt{s}$ (lower branch, \zapf{\char'302}).
  Leakage for the soft component is assumed constant at all energies.
  The dotted line is a fit ---see Eq.~(\ref{eq:linearbFB})--- to 
  experimental values \cite{RU:FB}.}\label{fig:ugodottedline}
  \end{figure}

\begin{figure}
  \begin{center}
  \mbox{\includegraphics[width=\textwidth]{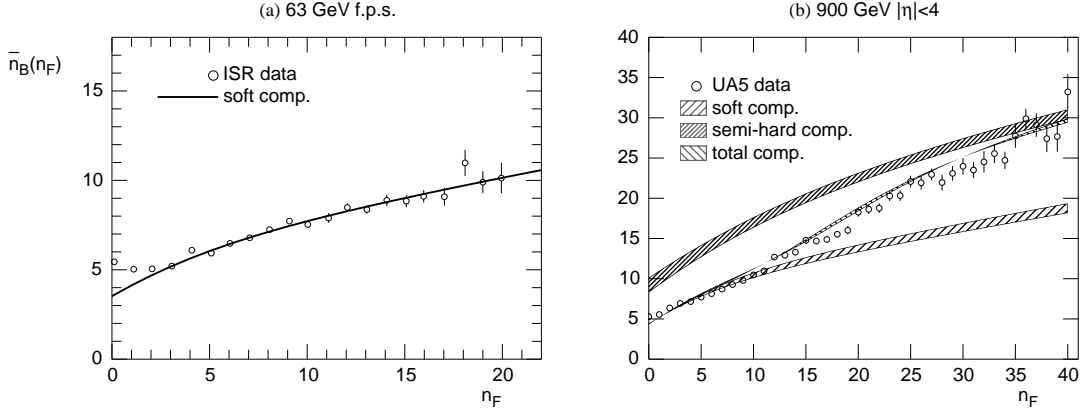}}
  \end{center}
  \caption[FB coefficient: weighted superposition]{Results of the
  weighted superposition 
  model for $\nbar_B(n_F)$ vs.\ $n_F$ compared 
  to experimental data 
  in full phase-space at 63 GeV (a) and in the
  pseudo-rapidity interval $|\eta|<4$ at 900 GeV (b)
	\cite{RU:FB}.}\label{fig:cioni}
  \end{figure}

The relatively small increase of the average number of  particles per clan 
from 63 GeV to 900 GeV for the soft component ($\nc$ goes from  $\approx
2$ to $\approx 2.44$)  
suggests to consider $p_{\text{soft}}$ constant throughout the GeV
energy region. 
Accordingly, being  $\nbar_{\text{soft}}$ and $k_{\text{soft}} $ 
known at 546 GeV,   $\bFBxx{soft}$ for
the soft component can be determined at such energy; its value, inserted in
Eq.~(\ref{eq:b_total}), 
allows the determination of $\bFBxx{semi-hard}$ in view of the fact that
$\bFBxx{total} = 0.58$.
It is found that $p_{\text{semi-hard}}$($\sqrt{s} = 546$ GeV) = 0.77.
The relatively small increase of the average number of particles per clan
also for the semi-hard component, from 200 GeV up to 900 GeV ($\nc$ goes
from 1.64 to 2.63), suggests to take also $p_{\text{semi-hard}}$
constant in the GeV region.

Under the just mentioned assumptions:

a) the correlation strength c.m.\ energy dependence is correctly reproduced
in the GeV energy range from ISR up UA5 top c.m.\ energy  and follows
the phenomenological formula  $\bFB = - 0.019 + 0.061 \ln s$
(Figure \ref{fig:ugodottedline}).

b) $\nbar_B(n_F)$ vs $n_F$  behaviour at 63 GeV c.m.\ energy (ISR data) is
quite well reproduced in terms of the soft component (a single
NB (Pascal) MD) only 
and at 900 GeV c.m.\ energy (UA5 data) in terms of the weighted superposition
of soft and semi-hard components, i.e., of the superposition of two
NB (Pascal) MD's.  (Figure \ref{fig:cioni}).

Concerning the general trends of  
$\bFB$ vs $\sqrt{s}$ and  of $\nbar_B(n_F)$ vs $n_F$
in the TeV energy domain, the situation is not unanimous.
In some approaches one expects a continuous increase towards the value 1:
either reasoning on the use of a soft and a hard component
within the eikonal model for mini-jets production \cite{jdd:FB}
or (in a purely statistical analysis) on the independent 
production of pairs of particles \cite{Carruthers:FB},
rather than of individual particles or clusters/clans.
The three scenarios  extrapolated from our
knowledge on \pp\ collisions in the GeV region and discussed in 
Section~\ref{sec:III.scenarios}, provide different predictions.
Being for instance in all scenarios at 14 TeV  
$\nbar_{c,\text{soft}} \approx 2.98$
($\nbar_{c,\text{soft}}$ is  $\approx 2$ 
at 63 GeV and $\approx 2.63$ at 900 GeV) to
assume $p_{\text{soft}}$ constant  ($\approx 0.78$) also in the new
energy domain  seems 
quite reasonable. The situation for $p_{\text{semi-hard}}$ is different. 
With the exception of the first scenario for which the validity of KNO
scaling is assumed also for the semi-hard component and 
$\nbar_{c,\text{semi-hard}}$
goes from  $\approx  2.63$ at 900 GeV to  $\approx 3.28$ at 14 TeV
after a very weak 
increase in the GeV energy region, $p_{\text{semi-hard}}$ constant is
an unrealistic  
assumption;  $\nbar_{c,\text{semi-hard}}$ in the second scenario where strong 
KNO scaling violation is allowed is indeed  $\approx 2.63$ at 900 GeV  
but  $\approx 7.96$   at 14 TeV   and in the third scenario (the QCD
inspired one) the   $\nbar_{c,\text{semi-hard}}$  
values at the two energies are $\approx 2.5$ and
$\approx 5$ at the two energies.

What should be noticed is that in all above mentioned scenarios $\bFB$
is bending in the new region and that $\bFB$ bending is more pronounced if
$p_{\text{semi-hard}}$ increases (less particle leakage from clans) and
it is less 
pronounced when $p_{\text{semi-hard}}$  decreases (more particle
leakage from clans).   
$\bFBxx{total}$ bending can be therefore  reduced by allowing more
particle leakage  
from individual  clans, i.e., by allowing a larger particle flow from one
hemisphere to the opposite one.  The real question is how far in energy the 
empirical  fit on $\bFBxx{total}$ 
c.m.\ energy dependence  will continue to find 
experimental support  without spoiling the main assumptions of our approach 
to FBMD correlations.

\begin{figure}
  \begin{center}
  \mbox{\includegraphics[width=\textwidth]{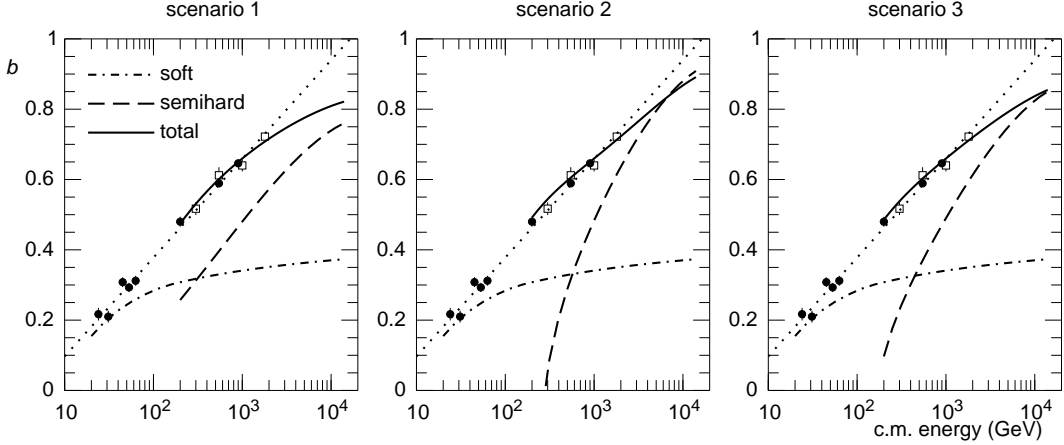}}
  \end{center}
  \caption[FB coefficients Energy dependence]{Energy dependence of the
        correlation coefficients  
        for each component (soft and semi-hard) 
        and for the total distribution in $p\bar p$ collisions.
        The dotted line is a fit ---see
        Eq.~(\ref{eq:linearbFB})--- to experimental 
        values \cite{RU:FBproblems}.}\label{fig:alberto}
  \end{figure}

It is indeed instructive to reexamine the three scenarios extrapolated in
the TeV energy domain 
in order to include in our scheme the  $\bFBxx{total}$
points at 1000 GeV and 1800 GeV  measured by E735 Collaboration at Fermilab
to sit on the same straight line in Figure \ref{fig:alberto}  of the
other points at lower c.m.\ energies. 
It is remarkable that the inclusion of the
new points demands that  $p_{\text{semi-hard}}$ must decrease and
accordingly particle  
leakage from one hemisphere to the opposite one must increase in all scenarios.

Satisfactory results for $\bFBxx{total}$ energy dependence are obtained 
in all three scenarios in Figure \ref{fig:alberto} by taking 
$p_{\text{soft}} \approx 0.8$ and 
$p_{\text{semi-hard}} \approx 0.84 - 0.07 \ln (\sqrt{s} / 200)$
for $\sqrt{s} \ge 200$ GeV. 

It should be pointed out that in the
scenario characterised by a semi-hard component with strong KNO scaling
violation (scenario 2) FBMC strength has a less steep behaviour with
the increase of the c.m.\ energy of the collision and its saturation towards
1 ---as that of $\bFBxx{total}$--- is quicker than in the other scenarios.
It turns out that  the linear behaviour of $\bFBxx{total}$ with the increase of
c.m.\ energy shown in Figure \ref{fig:alberto} is incompatible  with
our approach to FBMD correlations above 2.5 TeV in scenario 1, above 3.5 TeV
in scenario 2  and above 5 TeV in scenario 3. By incompatible we mean that
leakage parameter $p_{\text{semi-hard}}$ cannot be adjusted in order to
reproduce the  
eventual linear behaviour of $\bFBxx{total}$ above the mentioned c.m.\ energy
extreme values in the different scenarios without spoiling the model itself. 

On the contrary the experimental finding of the
mentioned linear fit of $\bFBxx{total}$ at higher c.m.\ energy, say at LHC,
would demand in our approach  the existence of a third class of events  to be 
added to the soft and semi-hard ones with even  larger leakage of particles 
from clans in one hemisphere to the opposite one than that given by the
a logarithmic decrease with c.m.\ energy of $p_{\text{semi-hard}}$. 

It seems that 
in the new class  clan production is therefore disfavoured with respect to the
production of more  particles within clans. The main characteristic 
of such events is to be composed of few high particles density clans
generated by necessarily  high virtuality ancestors.
This fact if confirmed would have an interesting  counterpart at
parton level and allow a suggestive interpretation of the completely different 
behaviour of the FB multiplicity correlations in \pp\ collisions and
\ee\  annihilation.

As already seen, in \ee\  annihilation, FB multiplicity  correlations are
almost inexistent in the two component and very weak in the total sample of
events (an effect entirely due   to the superposition of the two classes 
of events). Here clans are numerous but quite small in size, i.e., the
independent intermediate gluon sources (the BGJ's) ---the clan  ancestors
at parton level---  are expected to be generated  quite late in the
production process  at relatively low virtualities, implying small
colour exchange  among partons and then very weak FB multiplicity correlations
among hadrons.

In \pp\ collisions clans are less numerous than in \ee\  annihilation
but their size is larger, i.e., more particles are contained within
each clan. At parton level one should expect that BGJ's are generated
quite early in the production process at relatively high virtualities
with enough room for many parton splittings  and consequently
stronger  colour exchanges. 
All that at hadron level will lead to 
quite remarkable FBMC's. In this framework the eventual third class
of events whose existence would be determined  by the observation of
a linear   increase of $\bFBxx{total}$ with c.m.\ energy in  the proper  TeV
energy domain will have some peculiar characteristics, which make its finding 
particularly appealing. 
First of all one should see in these events 
a reduction of the average number of clans to few units in order to guarantee 
a high particle population per clan   and a quite consistent particle leakage
from one hemisphere to the opposite one. 
At parton level the fewer BGJ's 
production of the new class of events with respect to the semi-hard one
should be originated by a very high virtuality process ---much higher  than 
for  semi-hard events--- leading to  high parton density regions characterised
by  huge colour exchanges.

\begin{figure}
  \begin{center}
  \mbox{\includegraphics[scale=0.9]{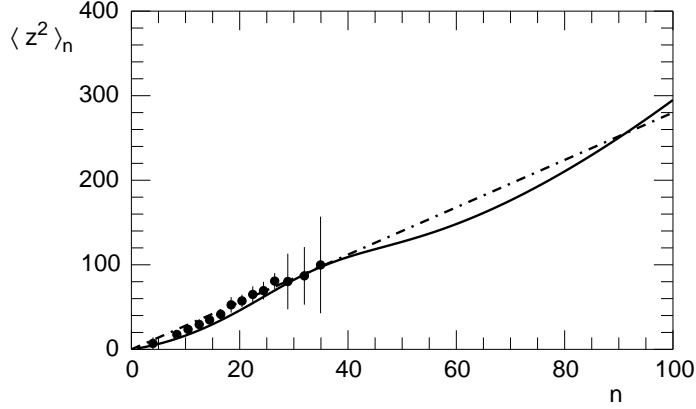}}
  \end{center}
  \caption[z2 vs n]{$\avg{z^2}_n$ vs $n$ at 900 GeV in the interval
           $1<|\eta|<4$. 
           Data points are from UA5 Collaboration,
           the solid line is the result of our model in $0<|\eta|<4$,
           the dash-dotted line is a linear fit \cite{RU:FBproblems}.
        }\label{fig:albertozetaquadrouno}
  \end{figure}

Another variable sometime used in the literature in addition to $\bFB$
should be mentioned: it  is the   average of the
forward backward particle multiplicity difference over all
events at fixed multiplicity $n$, i.e., $\avg{z^2}_n = 
\avg{n_F - n_B}_n$. The new 
variable works at a deeper level of investigation  than $\bFB$
(the latter is related 
to the average of  $\avg{n_F - n_B}_n$ over all multiplicities)  
and it is of
particular interest for global properties of the collisions related to 
average $n$. In our framework one has
\begin{equation}
     \avg{z^2}_n = \frac{4 d^2_{n_F,\text{soft}}(n) 
					             \alpha P_{\text{soft}}(n)}{
					P_{\text{total}}(n)} +
             \frac{4 d^2_{n_F,\text{semi-hard}}(n) (1-\alpha) 
							 P_{\text{semi-hard}}(n)}{P_{\text{total}}(n)} ,
\end{equation}
with $P_{\text{total}}(n) = \alpha P_{\text{soft}}(n) + 
(1 - \alpha) P_{\text{semi-hard}}(n)$.

In Figure \ref{fig:albertozetaquadrouno}, data points from UA5 collaboration 
in $1 < |\eta| < 4$   at 900 GeV   on $\avg{z^2}_n$ vs.\ $n$  
are compared with our 
model results   in full phase-space  (solid line)
as well as with the behaviour predicted in
the model of Chou and Yang \cite{Chou:FB+Chou:geometrical}
(dashed line). 
The latter model supposes a composition of `stochastic' (i.e., Poissonian)
fluctuations in $z$ with `non-stochastic' ones 
(in $n$, obeying KNO scaling), composition which is 
regulated by the collision geometry; 
in this scheme one obtains a linear 
dependence of $\avg{z^2}_n$ on $n$.
Below $n \approx 40$,
no differences between the two models are visible and both models are in
agreement with experiments. 
Above $n \approx 40$, our model shows a clear "hump"
structure. 
The situation  in such range is
not experimentally clear  and consequently no conclusions can be drawn 
on this point.  
Predictions on $\avg{z^2}_n$  vs.\  $n$  behaviour in
the possible scenarios expected in the TeV energy domain are discussed in 
reference \cite{RU:FBproblems}.
The behaviour of the variable $\avg{z^2}_n$ has also been used
as one possible evidence for a phase-transition \cite{Gutay:plasma}.

\subsection{Are clans massive objects?}\label{sec:III.clanmass}

The success of clan structure analysis  in interpreting multiparticle 
production phenomenology  raises,  as pointed out at the end of section  
\ref{sec:III.scenarios},
many intriguing  questions. Among them the problem of the clan mass
is of particular interest. The first search on the subject goes back to
A. Bialas and A. Szczerba \cite{Bialas:clanmass}. 
The attempt is based on  a  generalisation
of  standard clan Poissonian production mechanism with two additional  
assumptions  on  clans distribution in rapidity variable  and  on the
angular distribution of particle decay  in a clan.
Let us review briefly the main content of this search.
Clans, Poissonianly distributed and independently emitted in bremsstrahlung-like
fashion,  are characterised using energy and (longitudinal) momentum 
conservation by the following  single-clan (pseudo)-rapidity density 
\begin{equation}
	\frac{d N}{ d\eta} = \lambda (1- x^+)^\lambda  (1 - x^-)^\lambda ,
\end{equation}
with  $x^{\pm} =  (m_T/\sqrt s) e^{\pm \eta}$.

Notice that $\lambda$ is the plateau height (the average number of clans
per rapidity unit), $m_T$ the clan  transverse mass  ($m_T = \sqrt{
m^2 + p_T^2}$), $\eta$ is the clan (pseudo)-rapidity and
$\sqrt s$ the c.m.\ energy (clan emission is limited to the
interval  $|\eta| < \ln (\sqrt s / m_T)$.
Furthermore  particle  probability density function  inside a clan 
(assumed to be in $\eta_0$) on the hypothesis of  isotropic decay is given by
\begin{equation}
	\Phi(\eta, \eta_0) = \left[2 \omega \cosh^2 \left(
		\frac{\eta - \eta_0 }{\omega}\right)\right]^{-1} ,
\end{equation}
where $\omega$ is a free parameter; distribution amplitude is proportional to
$\omega$  and $\omega = 1$ corresponds to an isotropic decay.

By assuming next that   particles will be produced in each clan 
according to a logarithmic MD, whose generating function is  given by
$g_{\text{log}}(z) = \ln (1 - z\beta) / \ln (1 - \beta) = 
\sum_n  P_n^{\text{(log)}} z^n$
and the  average number of particles per clan  by 
\begin{equation}
	\nc^{\text{(log)}} = \frac{\beta }{(\beta - 1) \ln (1 -\beta)} ,
\end{equation}
with $\beta = \nbar /(\nbar + k)$, one finds that
the generating function for the MD in the  interval $\Delta \eta$  turns
out to be
\begin{equation}
	G (\lambda, \Delta \eta)
	= \exp \left\{ \int \frac{ dN }{d\eta }
	  \frac{ \ln\left[ 1 - \frac{ \beta }{1-\beta}
	p(\eta_0;\Delta\eta)(z-1)\right] }{ \ln(1-\beta) } d\eta_0 \right\}
		\label{eq:III.36}
\end{equation}
with 
\begin{equation}
	p(\eta_0, \Delta  \eta) = 
  \int_{\Delta  \eta } \Phi (\eta, \eta_0) d \eta_0 .
\end{equation}
$p(\eta_0, \Delta \eta)$ describes the fraction of particles generated by 
a clan  with (pseudo)-rapidity $\eta_0$ falling in the interval $\Delta\eta$.
It is assumed in addition that emitted particles by each  clan are independently
produced in rapidity. 
With  a convenient choice of the parameters ($\lambda = 0.855$,
$m_T = 3.15$ (GeV/c$^2$), $\omega = 1.45$, $\beta = 0.90$) 
UA5 data at 546 GeV c.m.\ 
energy are approximately reproduced by the generating function of the model
$G(z, \Delta \eta)$. 
The corresponding $n$ charged 
particle MD, $P_n$, is   not a NB (Pascal) MD except in full
phase-space and forward backward multiplicity correlations are not correctly
reproduced by the model. 
These considerations notwithstanding,
the above results  are in our opinion quite instructive, although not
satisfactory,   
in that they show  a possible way  to determine clan width and mass. 
Accordingly, we decided to extend the  search on clan masses   
performed in reference \cite{Bialas:clanmass} 
in the  full sample of events of the
collision to the individual components  contributing to the collision itself.
The idea is that, if clans are massive,  clan masses are very probably 
different in the different classes of events  in
a given collision and again presumably  they differ from one class of 
collisions to the other \cite{RU:clanmass}.

\begin{figure}
  \begin{center}
  \mbox{\includegraphics[width=\textwidth]{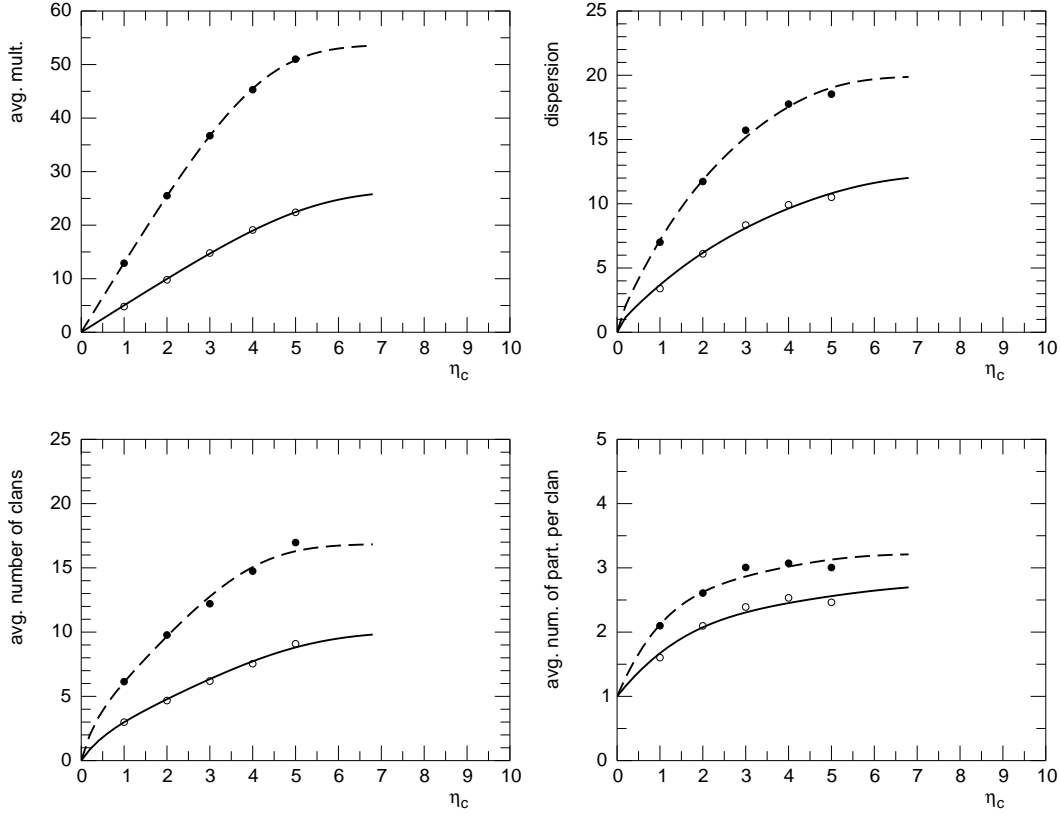}}
  \end{center}
  \caption[Clan mass: fit D and n]{Fit to the average multiplicity and
  dispersion  
  in different pseudo-rapidity intervals $[-\eta_c,\eta_c]$ for the two
  components of the MD in $p\bar p$ collisions at 900 GeV:
  soft component: open circles (data) and solid line (fit); 
  semi-hard component: filled circles (data) and dashed 
  line (fit) \cite{RU:clanmass}.}\label{fig:albertosette}
  \end{figure}

\begin{figure}
  \begin{center}
  \mbox{\includegraphics[width=\textwidth]{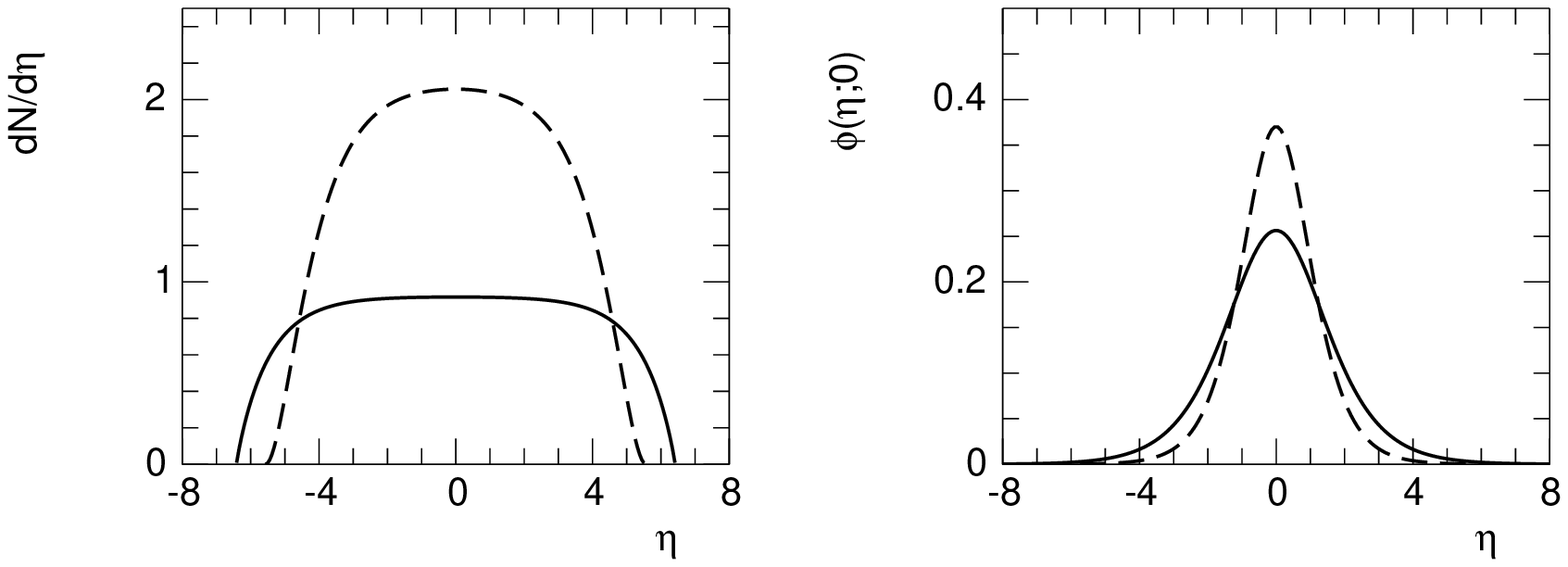}}
  \end{center}
  \caption[Clan mass: clan density]{Clan density $dN/d\eta$, and
        single particle  
        pseudo-rapidity probability density in a clan $\phi({\eta};0)$,
        for the soft (solid line) and semi-hard (dashed line)
        component at 900 GeV c.m.\ 
				energy \cite{RU:clanmass}.}\label{fig:albertootto}
  \end{figure}

Let us examine \pp\ collisions first, assuming, in agreement with the
results of   
our search, that we have two classes of events: with and without  mini-jets.
This distinction is, as we know, not unique  and at the present stage of our 
knowledge we should rely on
reasonable guesswork in order to strengthen our intuition in regions where
experimental data on a collision  are lacking or not performed in terms of
their components. 

Although it is not clear how much semi-hard component events contaminate
the soft sample, the assumption  that at 63 GeV  one has only
soft events is well supported by the fact that the shoulder effect in
$P_n$ vs $n$  
at such c.m.\ energy is negligible (one single NB (Pascal) MD
describes experimental data quite  well).
A NB fit to the data at 63 GeV  in $\eta_c < 2.5$  is indeed
correctly reproduced  
with the following choice of the parameters of the generating function
Eq.~(\ref{eq:III.36}):
$\lambda = 1.14$, $m_T = 1.80$ (GeV/c$^2$), $\omega = 0.84$,
$\beta=0.79$.  They are
obtained by  fitting with the least squares method  the average multiplicity  
$\avg{n}$ and   the quantity $D^2 / \avg{n}^2  -  1/\avg{n}$ of the distribution
given by the generating function   $G (\lambda, \Delta \eta) $ 
to the corresponding moments of the NB (Pascal) MD (i.e., $\nbar$ and $k^{-1}$)
in the pseudo-rapidity intervals $\Delta \eta = [- \eta_c, \eta_c]$ with
$\eta_c < 2.5$. 

The same method is then  applied separately  to soft and semi-hard components 
which, according to our experience, control the dynamics in \ppbar\ collisions 
at 900 GeV c.m.\ energy.
Above four parameters are fitted in the two components by using available 
data at 900 GeV collected in terms  of NB fits in pseudo-rapidity intervals
$|\eta_c| < 1,\dots, 5$.
The fits to the average charged multiplicity and dispersion as well as
to the average number of clans and to the average number of particles
per clan  in different (pseudo)-rapidity intervals for the two components  
contributing to the total MD turns out to be  quite  good as shown in 
Figure \ref{fig:albertosette}.
 Clan density  and single-particle
pseudo-rapidity probability density in  a clan  separately for the two
components are also shown in Fig.~\ref{fig:albertootto}.

Clan transverse masses and plateau heights at 900 GeV are much higher in the 
semi-hard than in the soft component 
($m_{T,\text{semi-hard}}$ = 3.43 (GeV/c$^2$) $\gg$
$m_{T,\text{soft}}$ = 1.47 
and  $\lambda_{\text{semi-hard}}= 2.09 \gg$ $\lambda_{\text{soft}}= 0.92$)
whereas the distribution width is much higher in the soft (1.95) than
in the semi-hard (1.35) component. 
This fact shows that heavier particles are
produced more in the semi-hard than in the soft component. In addition,
the average number of particles per clan is bending in larger rapidity
intervals in both components  suggesting that clans are larger in central
rapidity intervals than in the peripheral ones. 
Accordingly, leakage parameters
in FB multiplicity correlations should be larger when clans have larger
masses and their particle content is distributed  in more central rapidity
intervals. Indeed, in the present framework, it is possible to write
an approximate expression of the leakage parameter $p$ (controlling FB
correlations, see section \ref{sec:III.fbmc}):
\begin{equation}
	p = 1 + \frac{ \omega }{2 \Delta\eta }
	  \ln \left[ \frac12 \left( 1 + e^{-2\Delta\eta/\omega}\right)
			\right] ;
\end{equation}
this formula was obtained summing $\Phi(\eta,\eta_0)$ over $\eta$ and
averaging the result over $\eta_0 \in \Delta\eta$.
The extreme value of $p$, 1 and 1/2, are obtained respectively for
$\omega\to 0$ and $\omega\to\infty$.
%% The comparison with results at 63 GeV for the soft component
%% is quite awkward due to semi-hard event  contamination which was assumed
%% negligible in this case as discussed previously.

An interesting application of our approach to the two- and three-jets
components at LEP c.m.\ energy in \ee\ annihilation shows that
clan transverse masses $m_T $  are larger in the three-jets sample of events
(1.10 GeV/$c^2$) than in the two-jets sample of events  (0.62 GeV/$c^2$)
and that both masses are much lower that the masses expected at ISR
energy in \pp\ collisions. In conclusion clans could have masses which vary
with c.m.\ energy in a collision, they are different in the different
 components of a  collision and vary  from one class of collision to
another one. The word is again to experiments which should clarify
in addition to clan masses properties the existence of eventual other
clan quantum numbers.

\subsection{The reduction of the average number of clans and signals of new
physics in full phase-space and in restricted rapidity
intervals}\label{sec:III.thirdclass} 

It has been shown in previous Sections that the weighted superposition 
mechanism of two classes of events in high
energy collisions explains a series of experimental facts assuming that
the $n$ charged particle multiplicity distribution, for each class
of events, is described in terms of a NB (Pascal) MD  with characteristic
parameters $\nbar_i$ and $k_i$ (with $i$ = soft, semi-hard).
The experimental facts we refer to are those that were presented in
Section \ref{sec:III.scenarios} and successfully described in the
previous sections:
\emph{a}) the shoulder structure in the intermediate multiplicity range;
\emph{b}) the quasi oscillatory behaviour of the ratio of factorial
cumulants, $K_q$, 
to factorial moments, $F_q$, when plotted as a function of its order $q$
(after an initial decrease towards a negative minimum at  $q\approx5$);
\emph{c}) energy dependence of the strength of forward-backward multiplicity
correlations.
All these facts implied the relevance of NB (Pascal) regularity for
classifying  different  classes of events and of its interpretation
in terms of clan structure analysis.

Our attention is focused here on clan behaviour for the semi-hard component
in the two most realistic   scenarios (according to the results of CDF
Collaboration mentioned in Section~\ref{sec:III.cdf}),  
i.e., scenarios  2 and 3.
Clans general behaviour  at 900 GeV and 14 TeV  is summarised in the following 
table

\begin{center}
\begin{tabular}{|c|cc|cc|}
\hline
& \small $\Nbar$ (900 GeV) & \small $\Nbar$ (14 TeV) & 
    \small $\nc$ (900 GeV) & \small $\nc$   (14 TeV)\\[0.1cm]
\hline
 scenario 2 &&&&\\
 $k_{\text{sh}} \sim  (\log\sqrt{s})^{-1}$  & 23 & 11 &  2.5 &  7\\
\hline
 scenario 3  &&&&\\
 $k_{\text{sh}} \sim   (\sqrt{\log{s}})^{-1}$  & 22 & 18 & 2.6 & 5\\
\hline
\end{tabular}
\end{center}

In going from the GeV to the TeV energy domain
$\Nbar_{\text{semi-hard}}$ decreases and  
$\nbar_{c,\text{semi-hard}}$ increases as requested by a clan
aggregation process with  
higher particle population per clan.

Maximum clan aggregation would correspond of course to the reduction
of  $\Nbar_{\text{semi-hard}} \to 1$, i.e., 
remembering the definition
\begin{equation}
	\Nbar_{\text{semi-hard}} = k_{\text{semi-hard}} 
	\ln\left( 1 + \frac{ \nbar_{\text{semi-hard}} }{ 
		                  k_{\text{semi-hard}} } \right) ,
\end{equation}
to the following relation between
$\nbar_{\text{semi-hard}}$  and $k_{\text{semi-hard}}$
\begin{equation}
	\nbar_{\text{semi-hard}}  = k_{\text{semi-hard}} [ \exp (
	1/k_{\text{semi-hard}} ) - 1 ] ,
\end{equation}
which, being  $\nbar_{\text{semi-hard}}   > k_{\text{semi-hard}}$,
implies $k_{\text{semi-hard}} < 1$.

Natural  questions are: is it that just an asymptotic property of the
semi-hard component, or   the characteristic property of an effective
third class of events to be added to the soft and semi-hard ones?

\begin{figure}
  \begin{center}
  \mbox{\includegraphics[width=0.7\textwidth]{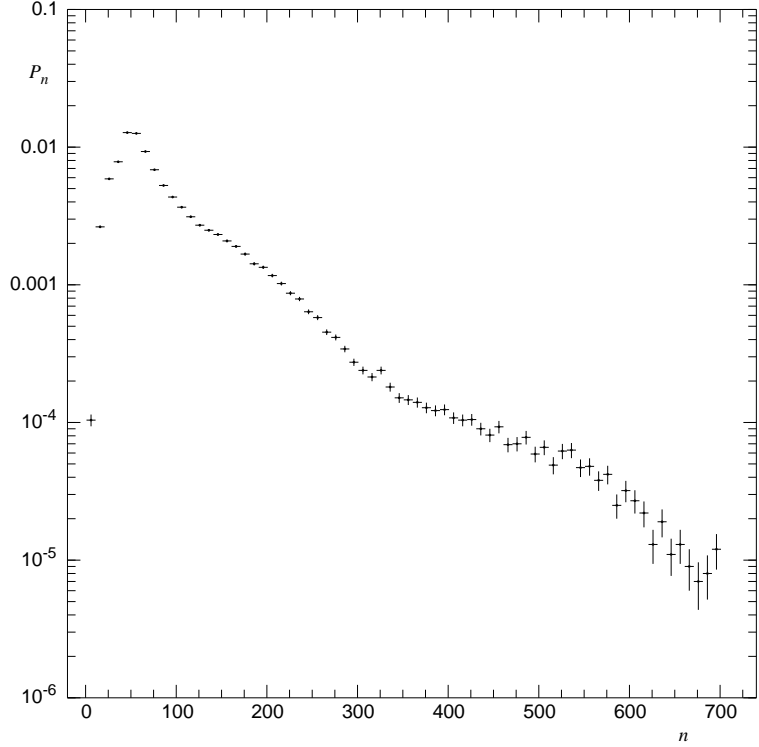}}
  \end{center}
  \caption[Pythia MD]{$n$ charged particle multiplicity distribution $P_n$
    predicted for minimum bias events in full phase space by 
    Pythia Monte Carlo (version 6.210, default parameters using
    model 4 with a double Gaussian matter distribution)
    at 14 TeV c.m.\ energy, showing two shoulder 
		structures \cite{RU:NewPhysics}.}\label{fig:alberto13}
  \end{figure}

\begin{figure}
  \begin{center}
  \mbox{\includegraphics[width=0.7\textwidth]{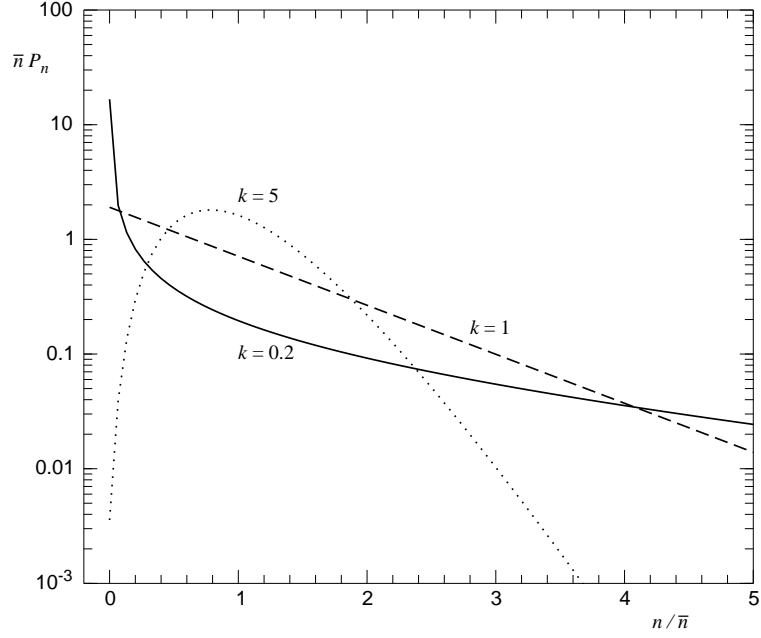}}
  \end{center}
  \caption[MD with different values of k]{Multiplicity distributions,
      in KNO form, with various  
      values of $k$ and the same average multiplicity $\nbar$, show
      different curvatures \cite{RU:NewPhysics}.}\label{fig:alberto14}
  \end{figure}

Being the asymptotia of the semi-hard component corresponding to 
$\Nbar_{\text{semi-hard}} \to 1$ a quite extreme c.m.\ energy region
in scenarios 2 and 3, 
completely outside the energy range available to future  experiments,
the first choice seems too far. 
On the contrary the suggestion to try an 
answer to the second question is of particular interest. 
In order to proceed along this line,
let us anticipate  at 14 TeV the new class of event with the above mentioned
property dictated by semi-hard clan reduction (i.e., $k_{\text{third}}
< 1$) \cite{RU:NewPhysics}. 

It is interesting to remark that
total    $n$-particle MD, $P_n^{total}$, for the  
minimum bias event sample in full phase-space from Pythia version 6.210, 
(default parameters,  with double Gaussian  matter distribution, model 4),
when plotted vs.\ multiplicity $n$, shows at 14 TeV a two-shoulders structure
(Fig.~\ref{fig:alberto13}.) 
The second shoulder is similar although lower than the
first one. 
In order to interpret the second shoulder one needs
a third class of events:
the second shoulder is the superposition 
of the events of the second class and those of the third class.

Therefore we decided to explore in our scenarios
the consequences of the existence of a third class of events
with $k_{\text{third}} < 1$ already at 14 TeV.

The fact that $k_{\text{third}}$ is less than one 
has  important consequences and fully characterises the
properties of the new class :

\emph{a.} Two particle correlations are stronger in the event of the
third class than in the event of the semi-hard component 
\begin{equation}
  \nbar^2_{\text{third}} / k_{\text{third}}  = \int C_2 (\eta_1
  ,\eta_2 )  d \eta_1 d\eta_2 
  > \nbar^2_{\text{semi-hard}} / k_{\text{semi-hard}} .
\end{equation}

\emph{b.} Since cumulants depend on $1/k_{\text{third}}$ which is much
higher than 
$1/k_{\text{semi-hard}}$ 
also cumulants of the third component are in general much larger than
  cumulants of the semi-hard component.

\emph{c.} Since $\Nbar_{\text{third}} =  1$ it follows that  leakage parameter
$p_{\text{third}} = 1/2  $, 
i.e., it reaches its maximum value, being also $b_{\text{third}} \approx 1$ from
\begin{equation}
	b_{\text{FB,third}} = \frac{ 2 b_{\text{third}} p_{\text{third}} }{ 1 -
	2b_{\text{third}} p_{\text{third}} (1- p_{\text{third}}) }  ,
\end{equation}
one has that $\bFBxx{third} \to 1$  and that also forward backward
multiplicity correlation strength of the third class of events is larger
than that of the semi-hard component.

It should be remembered that the new class of events is described by a
NB (Pascal) MD with $\nbar_{\text{third}} \gg k_{\text{third}}$ and
that $k_{\text{third}} \ll 1$, 
i.e.,  a log-convex gamma MD, well approximated for $k_{\text{third}} \to 0$ by
a logarithmic distribution (Fig.~\ref{fig:alberto14}.)

\begin{figure}
  \begin{center}
  \mbox{\includegraphics[width=0.7\textwidth]{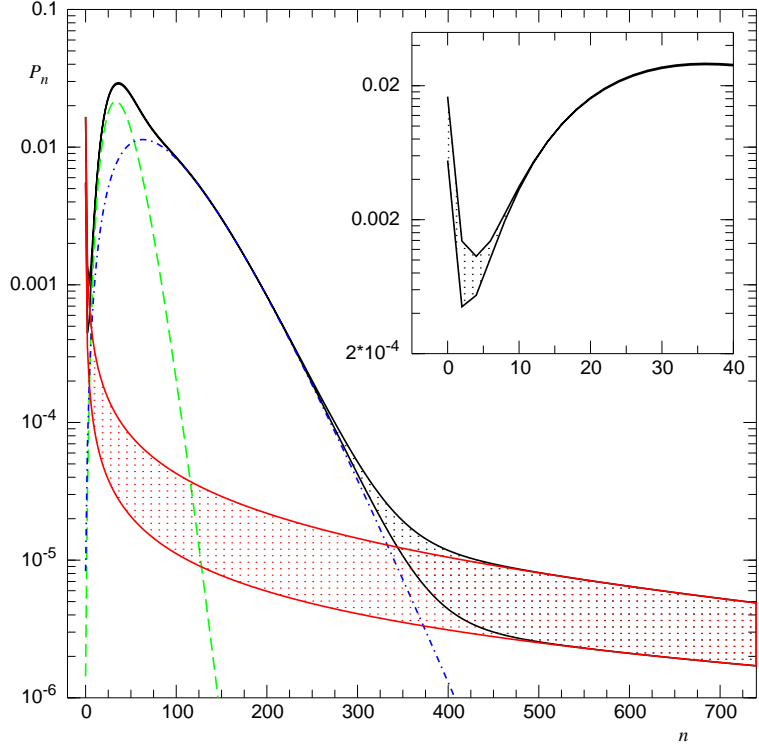}}
  \end{center}
  \caption[MD at 14 TeV with elbow (fps)]{$n$ charged particle
        multiplicity distribution $P_n$ expected at 14 TeV 
				in full phase-space in presence
        of a third (maybe hard) component with
        $\Nbar_{\text{third}}=1$, showing one 
        shoulder structure and one `elbow' structure. The band illustrates
        the range of values of parameters $\nbar_{\text{third}}$, 
				$k_{\text{third}}$ and $\alpha_{\text{third}}$
  discussed in the text \cite{RU:NewPhysics}.}\label{fig:alberto15}
  \end{figure}

\begin{figure}
  \begin{center}
  \mbox{\includegraphics[width=0.7\textwidth]{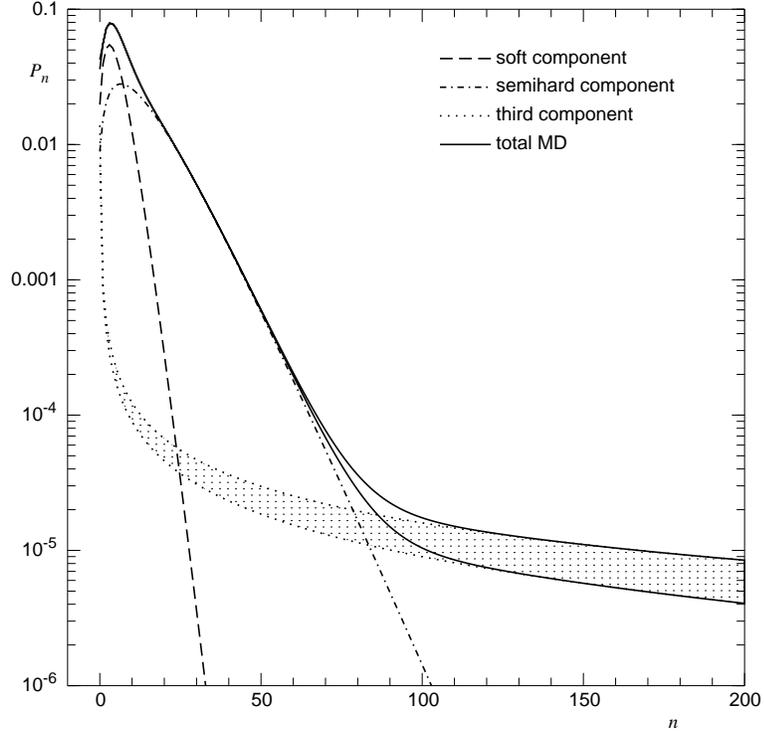}}
  \end{center}
  \caption[MD at 14 TeV with elbow (rap-int)]{Multiplicity
     distribution in $|\eta|<0.9$ for the      
     scenario described in the text (solid line); the three 
        components are also shown: soft (dashed line), semi-hard
    (dash-dotted line) and the third (dotted line) \cite{RU:NewPhysics}.}\label{fig:alberto16}
  \end{figure}

It is assumed therefore
that the classes of events contributing to the total sample 
at 14 TeV are the following:
 
\paragraph{Class I: soft events (no mini-jets).}
$\Nbar_{\text{soft}}$ is here large and growing with c.m.\ energy and
$\nbar_{c,\text{soft}}$ quite 
small; $P_{n,\text{soft}}$ obeys KNO scaling 
and $k_{\text{soft}}$ is constant.

\paragraph{Class II:  semi-hard events  (with mini-jets).}
$\Nbar_{\text{semi-hard}}$ is decreasing with c.m. energy whereas
$\nbar_{c,\text{semi-hard}}$ 
is increasing and it larger than $\nbar_{c,\text{soft}}$; KNO scaling
is violated 
and $k_{\text{semi-hard}}$ decreases.

\paragraph{Class III: the benchmark of the new class is
	$k_{\text{third}} < 1$.}
Being $\Nbar_{\text{third}}$ (reduced to few units and) approximately
equal to 1,  
quite large forward backward  multiplicity correlations are expected  and
---according  to the results of section~\ref{sec:III.fbmc} --- a
leakage parameter from one  
hemisphere to the opposite   one  close to its maximum. 
The fact that
hadronic clans are very few with a high particle density per clan
should have consequences also at parton level where one should find a
huge colour exchange process   from a relatively small number of high virtuality
ancestors presumably controlled by a mechanism harder than in the components
of the other two classes.

Generalising previous results we  expect that $P_n^{\text{total}}$ 
be obtained as 
the weighted superposition of the above mentioned three classes of events
each described by a NB (Pascal) MD with characteristic parameters $\nbar_i$ and
$k_i$ ($i$ = soft, semi-hard, third) i.e.
\begin{equation}
	\begin{split}
	P_n^{\text{total}} & = \alpha_{\text{soft}} P_n^{\text{soft}}
  (\nbar_{\text{soft}}, 
  k_{\text{soft}})\\
  &\quad + \alpha_{\text{semi-hard}} P_n^{\text{semi-hard}}
  (\nbar_{\text{semi-hard}},  k_{\text{semi-hard}})\\
  &\quad + \alpha_{\text{third}} P_n^{\text{third}} (\nbar_{\text{third}},
  k_{\text{third}}) ,
	\end{split}
\end{equation}
with  $\alpha_{\text{soft}} + \alpha_{\text{semi-hard}}  +
\alpha_{\text{third}} = 1$ 
and $\alpha_{\text{soft}}$,   $\alpha_{\text{semi-hard}}$ and
$\alpha_{\text{third}}$  the weight factors 
of each class of events with respect to the total sample 
(see Fig.~\ref{fig:alberto15}.)

$P_n^{\text{total}}$ has a characteristic elbow structure for large
$n$ and a narrow peak 
for $n$ close to zero. Both trends are consequences of the log-convex 
($k_{\text{third}} < 1$) gamma shape of the $n_{\text{third}}$-charged
multiplicity distribution 
of the new  component which shows a high peak at very low multiplicities
and a very slow decrease for large ones. 
The contrast with Pythia Monte
Carlo calculations predictions  is remarkable as can be seen just by 
inspection of Figures \ref{fig:alberto13} and \ref{fig:alberto15}. 

The search on global properties  can be extended from full phase-space to
rapidity intervals. Since Tevatron
data seem to favour, among our scenarios, the one based on strong  
KNO-scaling violation (i.e., scenario 2 with $k_{\text{total}}^{-1}$
growing linearly in $\ln s$),  we decided to discuss
in the rapidity interval $|\eta |< 0.9$  (which will be available at LHC with
Alice detector) this scenario only.
The weight factor of each 
component has been taken to be in the  rapidity interval considered the same 
as in full phase space.  Two extreme situations has been allowed for the third 
component 

(i)  the third component is distributed uniformly over the whole phase space

(ii) the third component has a very narrow plateau and falls entirely within the
interval $| \eta | <  0.9$ (see \cite{RU:NewPhysics} for details.)

Results are shown as a band  in Fig.~\ref{fig:alberto16}.
It should be pointed out that the general trend of 
$P_n^{\text{total}} (\eta_c , \sqrt s )$
vs.\ $n$  is quite similar to that already seen in full phase-space. 
Notice that
the narrow peak at very low multiplicity (again due to the third component)
is here hidden  by the standard peaks 
of the soft and semi-hard components which 
are shifted to lower multiplicities than in full phase-space.
In order to make the comparison easier,
in the following Table  are given  the parameters for the extrapolated
component multiplicity distributions at 14 TeV in full phase-space,
assuming that events of the third class comprise 2\% of the total:

	\begin{center}
  \vspace*{0.4cm}
  \begin{tabular}[t]{|r|ccccc|}
		\hline
		  FPS\vphantom{\LARGE H} &    \%   &  $\nbar$  &  $k$ &  $\Nbar$ &  $\nc$\\
      \hline
		  soft    &      41  &   40        &    7     &       13.3    &
		  3.0\\
			semi-hard  &    57  &   87      &     3.7     &      11.8&
		  7.4\\
			third      &    2  &  460       &    0.1212    &     1     &
      460\\
			\hline
  \end{tabular}
	\end{center}

\noindent and  the parameters for the extrapolated component 
multiplicity distributions at 14 TeV in the 
pseudo-rapidity interval $|\eta|< 0.9 $:

	\begin{center}
  \vspace*{0.4cm}
  \begin{tabular}[t]{|r|ccccc|}
      \hline
		  $|\eta|<0.9$\vphantom{\LARGE H} &    \%   &   $\nbar$   &   $k$
		  &  $\Nbar$ &  $\nc$\\ 
      \hline
		  soft    &41 &    4.9     &     3.4       &   3.0      &   1.6\\
			semi-hard  & 57 &    14  &         2.0   &       4.2  &   3.4\\
			third (i) & 2  &   40    &       0.056   &     0.368  &  109\\
			third (ii) & 2  &   460   &       0.1212  &      1     & 460\\
			\hline
  \end{tabular}
	\end{center}

\paragraph{(i) events evenly spread over the whole  rapidity range}:
in this case only  37\% of the clan is contained within the pseudo rapidity
interval $|\eta| < 0.9 $,   $k_{\text{third}}$ is even much less than 1 and
$\nbar_{c,\text{third}} \approx 40$.

\paragraph{(ii)  event concentrated in $|\eta|< 0.9$}:
the single clan is fully contained in $|\eta| < 0.9$, its parameters
are of course the same as those in full phase-space, but particle
density in rapidity is much higher than in (i).

The average number of particles per clan both in full phase-space
and in the rapidity interval is much larger in the semi-hard than in 
he soft component (semi-hard clans are larger than soft clans.)

Coming to forward-backward multiplicity correlations,  the overall
strength for three components is given by
\begin{equation}
    \bFB = \frac{
             \sum_{i=1}^3 \alpha_i \frac{b_i D^2_i}{1+b_i} + 
    \frac12 \sum_{i=1}^3 \sum_{j>i}^3 \alpha_i\alpha_j(\nbar_i - \nbar_j)^2
  }{
             \sum_{i=1}^3 \alpha_i \frac{D^2_i}{1+b_i} + 
    \frac12 \sum_{i=1}^3 \sum_{j>i}^3 \alpha_i\alpha_j(\nbar_i - \nbar_j)^2
  } .
        \label{eq:third.fb}
\end{equation}

By taking the quite reasonable assumption that
the leakage parameter in $|\eta|<0.9$
is the same as in full phase-space, it
has been found:

\begin{center}
  \begin{tabular}{|r|ccl|}
                \hline
                     &    FPS   &   $|\eta|<0.9$ & \\
    \hline
    first/soft       &    0.41   &     0.25 & \\
    second/semi-hard  &    0.51   &     0.45 & \\
    third            &    0.9995 &     0.997 & (i)\\
                     &           &     0.9995& (ii)\\
    total (weighted) &    0.98   &     0.92 & \\
                \hline
  \end{tabular}
        \end{center}

Notice that $b_{\text{FB,semi-hard}}$ is always larger than 
$b_{\text{FB,soft}}$.
Then $b_{\text{FB,third}}$ tends  to saturate in all cases its maximum
which is 1. 

The total FB multiplicity correlation strength resulting from the
weighted superposition of the contributions of all classes of events 
is larger in FPS than in the rapidity interval but closer to its asymptotic
value. 
As already pointed out, stronger FB correlations at hadron level
suggest  stronger colour exchange process at parton level and  this effect is
clearly enhanced in and by the third component.

Clan reduction is an important phenomenon which has already observed
in proton-nucleus collisions (see Section~\ref{sec:III.clans}). 
The attempt is to
explore a possible link and to compare  predictions in \pp\ collisions 
with those on  nucleus-nucleus collisions in terms of the energy density
variable, $\varepsilon$, by using 
Bjorken formula \cite{Bj:energy}
\begin{equation}
        \varepsilon = \frac32 \frac{\avg{E_T}}{V} 
                \left.\frac{dn}{dy}\right\vert_{y=0},
\end{equation}
where $\avg{E_T}$  is the average transverse energy per particle, $V$
the collision 
volume and $dn/dy$ the particle density at mid-rapidity.
The volume has been estimated  with proton radius $\approx 1$ fm and
formation time $\tau \approx 1$ fm. 
Lacking general expectations for $\avg{E_T}$,
for the soft component the value measured at ISR 
and for the other components the values measured by CDF (which should
be intended as a lower bound leading to lower bounds for energy densities as 
well) has been used. The results are summarised in the following Table:

\begin{center}
  \begin{tabular}[t]{|r|cccc|cc|}
                \hline
                 our scenarios& soft & semi-hard & 
                   \multicolumn{2}{c}{(i) third (ii)}\vline &
                   \multicolumn{2}{c}{(i) total (ii)}\vline\\
    \hline
             $dn/dy$ &  2.5 & 7 & 20 & 230 & 10.8 & 19.2 \\
             $\avg{E_T}$ (MeV)&  350 & 500 & 500 & 500 & 500 & 500 \\
             $\varepsilon$ (GeV/fm${}^3$)& 0.4 & 1.6 & 4.7 & 54 & 2.5 & 4.5\\
                \hline
  \end{tabular}
        \end{center}

The energy density for the semi-hard component in our scenario at 14 TeV
is of the same order of magnitude  found at AGS at 5.6 GeV in O+Cu collisions
($\varepsilon ~\approx 1.7$).
The energy density for the third component in the spread out scenario  is
comparable with its value recently measured at RHIC in Au+Au collisions
($\varepsilon ~\approx 4.6$)
The energy density for the third component in the other extreme scenario
with high concentration is $\approx 54$, even larger, being $dn/dy$ much larger,
than the LHC expectations for Pb+Pb collisions ($\varepsilon ~\approx 15$)

Of course the above estimate are only indicative, in view of the application
of Bjorken formula to \pp\ collisions, as well as for the choice of
the values of the parameters.
These considerations notwithstanding, it is possible that some
characteristic behaviour of observables seen at RHIC in AA
collisions could be reproduced at LHC in \pp\ collisions.

\newpage

\section{CONCLUSIONS}\label{sec:conclusions}
Clan concept has been originally introduced in multiparticle phenomenology
in order to interpret approximate regularities observed in $n$ charged particle
MD's in all classes of collisions. 
The occurrence of the same regularity
at parton level as a  QCD Markov branching process in the solution of
KUV differential equations in LLA with a fixed cut-off regularization
prescription 
led to consider partonic  clans as independent intermediate gluon sources 
similar to bremsstrahlung gluon jets. 

The sudden violation of the regularity at higher
c.m.\ energies and in larger rapidity intervals in \pp\ collisions and in 
\ee\ annihilation, with the appearance of a shoulder structure in
$n$ charged particle MD's,  opened a new horizon in the field  and suggested
a more accurate search at a deeper level of investigation. 
The attention on 
experimental data analysis moved from the full sample of events to the separate
samples of events of the single components or substructures of the 
collisions. 
It has been an important discovery to find that the same
regularity violated in the total sample of events was satisfied in the
single components, each described by characteristic and in general different 
parameters. 
Their weighted superposition 
provided a satisfactory understanding of collective variables
and correlations  behaviours in multiparticle production in the examined 
classes of collisions where experimental data were available.

Accordingly, clan structure analysis, in view also of its suggestive QCD
roots,  became an interesting tool in multiparticle dynamics.
Furthermore clan existence itself was demanded by a convincing description
of forward-backward multiplicity correlations c.m.\ energy dependence
in \pp\ collisions. 

The suppression in the same reaction of the average number 
of clans  of the semi-hard component in the TeV energy region  in the QCD 
inspired scenario as well as in the scenario with strong KNO scaling violation 
(both obtained by extrapolating our knowledge on data in the GeV energy domain),
together with the huge increase of the average number of particle per clan,
was surprising. 
It suggested the onset  of  a new scenario in the TeV energy
domain characterised by the appearance of a  
third class of presumably quite hard  events (to be added to the soft and 
semi-hard ones)  with few clans, maybe just one, with the characteristics 
of high particle density fireballs. 
The possible  anticipation of this new 
class of events at 14 TeV would be easily experimentally 
detectable and its finding  would  shed some light also on the link of
\pp\ collisions with  heavy ion collisions, where fireballs are expected
to be much larger.

Along the years, clans as groups of particles exhausting all correlations
inside each clan and  generated by independent
intermediate particle sources (the clan ancestors) revealed additional
properties which helped to make their physical  nature more precise,
up to the extreme thought that clans might be real physical objects
controlling the hadronization process and whose partonic partners might be
QCD parton showers originated by intermediate independent gluon  sources.

Clan concept turns out to be in this perspective more general than jet 
concept, in that no kinematical cuts are needed for its definition,
and it is also not a purely statistical concept like a standard cluster
in cluster expansion in statistical mechanics (as thought at the time
of its introduction)  in that  we learned that a mass can be attributed
to clans and that these masses could be  quite different 
in each component and vary from one collision to another one.
Assuming that clan existence would be also verified experimentally,
a new intriguing question would possibly concern  other quantum numbers
of these intermediate massive objects occurring  in the production process.

In conclusion, it is not clear how far we went in clarifying  the
enigma of multiparticle dynamics
evoked in the introduction to the present paper. 
This was not indeed our main goal.
We wanted to point out how successful and inspiring  has been 
in our search the continuous dialog between  theory and experiment
and to focus the attention of the reader on the development of  a series
of experimental  facts and theoretical ideas which might, hopefully, transform 
an enigma in an Arianna thread in the labyrinth of multiparticle 
dynamics in its awkward journey toward QCD and open new perspectives
in \pp\ and heavy ion collisions in the TeV energy domain.

\newpage
%%%%%%%%%%%%% %%%%%%%%%%% %%%%%%%%%%%%%%

\appendix
%%%% Appendices can ONLY use \section and \subsection!

\section{Physical and mathematical properties of the NB (Pascal) MD}

Cornerstone of the present approach to multiparticle dynamics has been
the search for regularities of collective variables behaviour in the
components or substructures of the various collisions.
In view of the relevance in this framework of the NB (Pascal) MD,
after some historical notes on its origin and its
use, its main mathematical and physical properties, with particular
emphasis on clan structure analysis and particle shower development,
are summarised for the benefit of the reader.
All these properties have been used indeed in the present paper 
in various occasions and are at the basis of the interpretation 
of the NB (Pascal) regularity itself.

\subsection{Historical  notes}\label{sec:A.history}

A few remarks will help to illustrate the main
motivations of  the wide use of the negative binomial (Pascal)
multiplicity distribution (NB (Pascal) MD) and
in more general terms of the class of compound Poisson multiplicity 
distributions  in science.

The NB (Pascal) MD with integer parameter $k$ was known already   
to Blaise Pascal \cite{Pascal}
(that is the reason why it was decided with P. Carruthers 
\cite{Carruthers:Pascal}
few months before his passing, 
to add the name `Pascal' to the distribution
so extensively  used in multiparticle dynamics  by  of all of  us.)
The distribution appeared fifty years later  in the Volume 
\textit{Essay d'analyse sur le jeux de hazard} 
\cite{Montmort}. The front page 
of the Volume points out that the Volume  has been  published with the 
permission of the king Louis XIV and describes the negative binomial
as the  
probability distribution of the number of tosses of a coin necessary to get 
a fixed number of heads \cite{kotz}.
Gambling was indeed the favourite game of the nobility in that century and
the discovery of its statistical rules a real achievement. 

The NB (Pascal) MD 
is used in modern times in many fields (biology, econometrics,
medicine\dots). 
Its first applications to the spreading of a disease in terms of 
sickness proneness of various groups of individuals goes back to  1920 and is 
due to M. Greenwood and G.U. Yule \cite{Yule}.

The distribution appears almost in the same years  in quantum statistical 
mechanics in order to describe $n$ Bose particle multiplicity 
distribution in $k$ 
identical systems, each containing on average $\nbar/k$ particles
\cite{Planck}.

The distribution becomes later of fundamental importance in quantum optics 
in the study of  $n$-photon MD   from a partially  coherent source  
\cite{Mandel}; it is remarkable 
that the  photon  distribution is Poissonian when the source is coherent 
($k = \infty$)  and it reduces to a NB (Pascal) MD with $k=1$  
(geometric, i.e., Bose-Einstein) when the source is thermal.

After its      introduction   in  high energy  collisions  in 
the accelerator region in 1972, the NB (Pascal) MD became quickly ---as already
pointed out--- a stimulating    phenomenological tool for describing  
multiparticle production general trends in all classes of collisions.

Many are the reasons of this success.

Firstly, the NB (Pascal) MD  is a compound Poisson distribution and
represents  one of the the simplest two-parameter MD 
correcting the independent   $n$-particle production process suggested 
for instance by
the multi-peripheral model prediction, i.e., a Poisson multiplicity  
distribution in the average number of particles $\nbar$. The generating 
function of the NB (Pascal) MD  can indeed be written as an 
exponential of a logarithmic distribution  with an  intriguing positive  
numerical  factor in front whose interpretation in terms 
of clan concept is one of the main subject of this review.
The reduction of the average population within each clan to one unit,
i.e., the equality of the average number of particle of the full distribution
with the average number of clans, leads  to the Poisson MD  of the
full $n$-particle distribution. Deviations from unity in the average number of 
particles per clan represent the correction to the Poissonian 
behaviour (see  Section \ref{sec:I.4}.)

Secondly, the NB (Pascal) MD contains a set of   multiplicity distributions 
in correspondence to  different  limiting values of its standard parameters,
as shown in Section \ref{sec:A.3}, below.

Thirdly, the NB (Pascal) MD is a hierarchical distribution, i.e., all
multiplicity correlations can be  expressed  in terms 
of second order moments, controlling two-particle correlations
(see  Section \ref{sec:I.4}.)

Fourth, NB (Pascal) MD's as solutions of the differential QCD evolution
equation in LLA   can be interpreted as  Markov branching processes of 
two (parton) populations evolving with different strengths
(corresponding to the dominant QCD vertices) down to the final (parton)
configuration (see  Section \ref{sec:II.leading-log}.)

\newcounter{mypara}
\setcounter{mypara}{0}
\renewcommand{\themypara}{\alph{mypara}}

\newcommand{\parax}{\addtocounter{mypara}{1}
\paragraph{\thesubsection.\themypara}}

\subsection{Different parametrisations of the distribution used in multiparticle
dynamics}\label{sec:A.2}

\setcounter{mypara}{0}
\parax 
The standard parametrisation of $n$ charged particle multiplicity
distribution, $P_n$, is usually given in terms of the average charged 
multiplicity, $\nbar$,  and the positive parameter $k$, which is linked to 
the dispersion $D$ of the distribution 
($D =\sqrt{\bar{n^2}  - \nbar^2}$) by the 
relation  $k = \nbar^2  / ( D^2 - \nbar )$. 
$D^2$ is therefore larger than $\nbar$.
Accordingly
\begin{equation}
	P_n (\nbar, k) = P_0 (\nbar, k ) \frac{k (k+1)...(k+n-1)}{n!}  
  \left(\frac{\nbar}{\nbar + k}\right)^n ,
\end{equation}
with 
\begin{equation}
	P_0 (\nbar, k) =  \left(\frac{k}{\nbar + k}\right)^k  .
\end{equation}

The corresponding generating function is 
\begin{equation}
	G (\nbar, k ; z) = \sum_{n=0}^\infty  P_n (\nbar, k) z^n = 
    \left[ 1 - \frac{\nbar}{k} (z-1)\right]^{-k}
\end{equation}
with $G (\nbar,k ; 1 ) = 1$ as normalisation condition.

\parax
The distribution has been also defined in terms of the coefficients
$a$ and $b$ of a linear recurrence relation between  
$P_{n+1}$ and $P_n$ multiplicities, i.e., 
\begin{equation}
	(n+1) \frac{P_{n+1}(a,b)}{P_n(a,b)}  = a + b n  , \label{eq:193}
\end{equation}
and $P_n (\nbar, k)$ from above becomes
\begin{equation}
	P_n (a, b) = P_0 (a, b) \frac{a ( a+b) ... (a+ b(n-1))}{ n! \, b^n}
\end{equation}
with
\begin{equation}
	P_0 (a, b) =  ( 1-b) ^{a/b}.
\end{equation}

The corresponding generating function is
\begin{equation}
	G (a, b; z) = \sum_{n=0}^\infty P_n (a,b) z^n = \left( 
		\frac{1-b}{1-bz}\right)^{a/b} .
\end{equation}
Notice that 
\begin{equation}
	k = a/b \qquad\text{and}\qquad \nbar = a/(1-b).
\end{equation}
This definition is particularly suitable in an approach to
multiparticle dynamics based on stimulated emission where the
parameter $b/a$ corresponds to the fraction of particles already
present stimulating the emissions of an additional one \cite{AGLVH:1}.

\parax
Being the NB (Pascal) MD a compound Poisson distribution, it can be
written by using another set of parameters, i.e., the parameters of
clan structure analysis $\Nbar$ (the average number of clans) and $\nbar_c$
(the average number of particle per  clan, $\nbar_c = \nbar / \Nbar$)
and the corresponding generating function  turns out to be 
\begin{equation}
	G ( \Nbar,\nbar_c ; z) = \exp [ \Nbar ( G_{\text{log}} (\nbar_c ; z) -
	1 ) ]
\end{equation}
with 
\begin{equation}
	\Nbar= k \ln \left( 1+ \frac{\nbar}{k} \right)  = 
      - \frac{a}{b}  \ln (1-b)
\end{equation}
and
\begin{equation}
	\nbar_c = \frac{\nbar }{  k \ln ( 1 + \nbar/k)  }= 
  \frac{- b }{(1-b) \ln (1-b)}.
\end{equation}
$G_{\text{log}} (\nbar_c ; z )$ 
is the logarithmic multiplicity generating function 
and is equal to   $ \ln (1-bz)/\ln (1-b)$.
In this framework, $1/k$ is an aggregation parameter, i.e.,
the ratio between the probability that two particle belong to the same
clan to the probability that they belong to different clans
\cite{AGLVH:1}.

\parax
Notice that $G_{\text{log}} (\nbar_c ; 0 ) = 0$  and that
$\exp(-\Nbar) = G (\nbar, k ; 0)$.
It follows that the zero particle probability is given by
\begin{equation}
	\exp( -\Nbar) = P_0 (\nbar, k) = 
			 \left(\frac{ k}{k+\nbar } \right)^k= (1-b)^{a/b}.
\end{equation}

This result can be extended to any interval, $\Delta w$,
of a generic variable $w$, 
(e.g., $w$ could be (pseudo)-rapidity or transverse momentum or
coordinate in the real 
space). In this case the probability to find zero particles in $\Delta w$, 
$P_0 (\nbar(\Delta w), k ( \Delta w) )$,  
is defined and an intriguing connection
with clan structure parameters in $\Delta w$ is established, i.e.,
\begin{equation}
	\Nbar (\nbar(\Delta w) , k (\Delta w) ) =
    - \ln P_0 (\nbar(\Delta w), k ( \Delta w) )
\end{equation}
and
\begin{equation}
	\nbar_c (\nbar(\Delta w), k ( \Delta w)) = 
    \frac{1}{V_0 (\Delta w)} ,
\end{equation}
$V_0$ being the void function of Section \ref{sec:I.4}.

Of particular interest is indeed the following theorem:
factorial cumulant structure is hierarchical, i.e., $n$-order factorial
cumulants are controlled by second order factorial cumulants, iff the
function  $V_0(\Delta w)$ scales with energy and $\Delta w$ as  a function
of $\nbar (\Delta w)$ multiplied by the second order factorial cumulant.
Being for the NB (Pascal) MD the second order factorial cumulant equal
to $1/ k(\Delta w)$, it follows that 
\begin{equation}
	\nbar_c \left(\frac{\nbar(\Delta w)}{k ( \Delta w)} \right) 
    = \frac{1}{V_0 (\Delta w)} , 
\end{equation}
i.e., the distribution is hierarchical, as $K_n = f(K_2)$.

The use of $V_0 (\Delta w)$ and $P_0 (\nbar(\Delta w), k ( \Delta w)
)$ is of particular interest in the study of rapidity gaps in the
various collisions \cite{pzero}.

\parax
The NB (Pascal) MD can be obtained also by the weighted superposition
of Poissonian MD's with average multiplicity $\nbar_\lambda = \lambda \nbar$
and gamma weight
\begin{equation}
	f (\lambda, k) = \frac{ k (\lambda k )^{k-1} }{ (k-1)!} \exp (- k
	\lambda) .
\end{equation}
It follows
\begin{equation}
\begin{split}
	P_n (\nbar, k) &= \int_0^\infty 
	  \frac{ (\lambda\nbar )^{n} }{ n!} e^{- \nbar\lambda}
		f(k, \lambda) d \lambda\\ &=
  \left(\frac{k}{\nbar + k}\right)^k \frac{k (k+1)...(k+n-1)}{n!}
    \left(\frac{\nbar}{\nbar + k}\right)^n .
\end{split}
\end{equation}
This representation of the distribution has been used in Section
\ref{sec:III.bootstrap} in order to describe corrections to the
multi-peripheral model predictions.
It is known in quantum optics as ``Mandel's equation'' 
\cite{Mandel} and in
mathematics as ``Poisson transform'' with gamma weight 
\cite{CarrShih}.

\parax
The convolution of two NB (Pascal) MD's  with different parameter $k$
but the same parameter $b$,
$P_n(k, b)$ and $P_{n'} (k', b')$ 
is again a NB (Pascal) MD \cite{ChangChang}.
In fact
\begin{equation}
	\begin{split}
		\sum_{n+n'=N}   P_n(k, b) P_{n'} (k' , b) &=
	  (1-b)^{k+k'} b^N \frac{ (N+k+k'-1)! }{N ! (k+k'-1 )! } \\
		&= P_N(k+k',b) .
	\end{split}
\end{equation}

\subsection{The one-dimensional limits of the parameters of the
	  distribution}\label{sec:A.3} 

\setcounter{mypara}{0}
\parax
The limit $b\to 0$ of $P_n (a, b)$ leads to a Poissonian MD:
\begin{equation}
	\lim_{b\to 0} P_n (a, b) = P_0 \frac{a^n }{  n!} ,
\end{equation}
with  $P_0 = e^{-a}$ and in this limit $a = \nbar$.
i.e. 
\begin{equation}
	P_n (\nbar) = e^{- \nbar}  \frac{\nbar ^n }{  n! }.
\end{equation}

The corresponding generating function (GF) is 
\begin{equation}
	G (a ; z) = \exp [a (z-1)].
\end{equation}
Notice that the limit $b\to 0$  
corresponds to $a\to \nbar$ and $k\to \infty$.

\parax
The limit $a\to 0$, with constant $b$, leads to the logarithmic MD:
\begin{equation}
	\lim_{a\to 0} P_n (a, b) = P_1 (b) \frac{b^{n-1}}{ n } ,
\end{equation}
with $P_1(b)= - b / \ln (1-b)$ ($n$ is always $\ge 1$).
Therefore 
\begin{equation}
	P_n (b) = - b^n / n \ln (1-b) .
\end{equation}
Notice that this limit corresponds to $k\to 0$, with constant $b$.

\parax
The limit $a = b$, i.e., $k = 1$, leads to the geometric MD:
\begin{equation}
	P_n (b, b) = P_0 (b) b^n ,
\end{equation}
with $P_0 (b) = (1-b)$,
and  therefore
\begin{equation}
	P_n (b) = (1-b) b^n .
\end{equation}
The average multiplicity of the geometric distribution is $b/(1-b)$
and the corresponding GF:
\begin{equation}
	G(b;  z) = (1-b)/(1-bz)  .
\end{equation}

\parax
 For $\nbar \gg k$ one gets the gamma distribution
 \begin{equation}
	 \lim_{\nbar \gg k}
	 P_n (\nbar , k) = \frac{1}{\nbar}  
	 \frac{k^k }{ \Gamma(k)} (n/\nbar)^{k-1} \exp (- k n /\nbar) .
 \end{equation}                                      
The distribution is concave for $k>1$ and convex for $k<1$. It becomes an
exponential for $k=1$. Notice that  $D^2 /\nbar^2$ 
is approximately equal to $1/k$
and that the maximum value of the gamma MD is obtained for
$n/\nbar = 1 - 1/k$, see \cite{RU:NewPhysics}.

\parax
The (positive) binomial limit:
For $a  > 0$ and $b < 0$, and $a/b$ an integer number,
 the negative binomial MD
becomes a standard binomial MD and $-k$ (now positive)
is the maximum multiplicity.
$-1/k$ can be seen as an anti-aggregation parameter, in that particles
like to stay far apart from each other.

\subsection{Physical informations contained  in the differential of
	the generating function of the multiplicity distribution with
	respect to its parameters considered as independent variables}\label{sec:A.4}

\setcounter{mypara}{0}
Let us consider the differential of the NB (Pascal) MD GF, 
\begin{equation}
	G(a, b; z)  = \left( \frac{1-b}{1-bz} \right)^{a/b} , 
\end{equation}
with respect to its independent variables after the introduction of
$\nu = 1/(1-b)$ and remembering that  $k =a/b$, i.e., the differential of
\begin{equation}
  G(a, b; z)  = \left( \frac{1}{ \nu  + z - \nu z } \right)^k  ,
\end{equation}
with  respect to $z$, $\nu$  and $k$ variables, it follows
\begin{equation}
	d \ln G(a, b; z) = 
  \frac{a}{1-bz} dz + \ln G(a, b; z) \frac{dk}{k} + 
	k \left[G_{\text{geom}}(\nu ; z)-1\right] \frac{d\nu}{\nu - 1} ,
\end{equation}
where $G_{\text{geom}}(\nu ;z) = (\nu +z-\nu z)^{-1}$ 
is the GF of the geometric MD.

\parax
By taking $k$ and $b$ ($\nu$) constant, the above equation becomes
\begin{equation}
	(1 - bz) \frac{\partial  }{ \partial z }  G(a, b; z) = a G(a, b; z) .
\end{equation}
The new equation is the differential form in the generating function
language of the linear relation between $n+1$ and $n$ particle MD's, 
Eq.~(\ref{eq:193}).

\parax
For $z$ and $b$ ($\nu$) constant one obtains that $\ln G(\nbar, k; z)$ is 
proportional to $k$:
\begin{equation}
	\ln G(\nbar, k; z) = k \ln G(\nbar/k , 1; z)
\end{equation}
The new equation  reminds us in the GF language that the corresponding
MD is a CPMD.

\parax
By taking $z$ and $k$ constant and integrating the differential between
$1$ and $\nu$ one has
\begin{equation}
	\ln G(\nbar, k; z)= k  \int_1^\nu \left[ G(\nu', z) - 1 \right]
	\frac{d\nu' }{\nu'} .
\end{equation}
Notice that for $b=0$ ($\nu=1$) the GF is equal to one.
This result is strictly related to the clan structure expressed by the
relation
\begin{equation}
	G(\nbar, k; z) = G_{\text{Poisson}}\left(\Nbar ; 
	          G_{\text{log}} (b; z) \right),
\end{equation}
with
\begin{equation}
	\Nbar= k \ln \nu 
\end{equation}
and
\begin{equation}
	G_{\text{log}}(b; z) = 
  \frac1{\ln \nu} \int_1^\nu G_{\text{geom}}(\nu' ;z ) d \nu' / \nu' .
	\label{eq:225}
\end{equation}
The NB (Pascal) MD is therefore generated by independent emission of geometric
clans with mean multiplicity $\nu'$ in the interval ($1$, $\nu$) and the
average number of geometric  clans in ($\nu'$ , $\nu' + d \nu'$) is 
$k d\nu'/\nu'$.

It should be pointed out that the GF of the geometric MD, 
$G_{\text{geom}}(\nu ;z )$, obeys the differential equation
\begin{equation}
	\nu \frac{\partial}{\partial \nu}  G_{\text{geom}}(\nu; z) = 
   G_{\text{geom}}(\nu; z)( G_{\text{geom}}(\nu; z) - 1) .
\end{equation}
In terms of probabilities, it describes a typical self-similar branching 
process   as that found in the Markov process version of the KUV model
of gluon shower  when the $g\to q\bar q$ branching is neglected, as
discussed in Section \ref{sec:II.leading-log}.

\subsection{Informations contained in the differential of the
	logarithmic MD GF}\label{sec:A.5}

Let us  now consider the GF of the logarithmic MD:
\begin{equation}
	G_{\text{log}}(b; z) = \frac{\ln (1  - bz)}{ \ln (1-b)}.
\end{equation}

A simple calculation shows that
\begin{equation}
  \begin{split}
	-(1-b)\ln(1-b) \frac{d G_{\text{log}} (b ; z) }{d b} 
	&=  - G_{\text{log}}(b; z) + \frac{(1-b)z}{1-bz} \\
	&= -G_{\text{log}}(b;z) + z G_{\text{geom}}(b; z) ,
	\end{split}
\end{equation}
or by using the variable $\nu = 1/(1-b)$:
\begin{equation}
	\ln \nu \frac{d G_{\text{log}} (b; z)}{d \ln \nu } =
  - G_{\text{log}}(b; z) + z G_{\text{geom}}(b; z) .
\end{equation}

For an infinitesimal change, the MD of an average clan, which is
logarithmic, evolves by addition of a geometric distribution.
The first term in the equation is  ensuring the normalisation condition 
$G_{\text{log}}(b; 1)= 1$.
Accordingly, the logarithmic clan can be taken as an average over geometric
clans as shown explicitly in Eq.~(\ref{eq:225}).

\newpage
%%%%%%%%%%%%% %%%%%%%%%%% %%%%%%%%%%%%%%

\newpage

\section*{}
\addcontentsline{toc}{section}{References}
\bibliographystyle{prstyR}
\bibliography{abbrevs,bibliography}

\begin{thebibliography}{100}

\bibitem{Fermi+Fermi:1+Fermi:2}
{E. Fermi}, Prog.\ Theor.\ Phys. {\bf 5},  100  (1950);
Phys.\ Rev.\ {\bf 81},  115  (1951);
Phys.\ Rev.\ {\bf 92},  452  (1953).

\bibitem{Landau}
{L. Landau}, Izv.\ Akad.\ Nauk.\ USSR {\bf 17},  51  (1953).

\bibitem{Carlson}
{J.F. Carlson and J.R. Oppenheimer}, Phys.\ Rev.\ {\bf 51},  220  (1937).

\bibitem{Bhabha}
{H.J. Bhabha and W. Heitler}, Proc.\ Roy.\ Soc. {\bf 159 A},  432  (1937).

\bibitem{Furry}
{W.H. Furry}, Phys.\ Rev.\ {\bf 52},  569  (1937).

\bibitem{Oppenheimer}
{J.R.\ Oppenheimer, H.W.\ Lewis and S.H.\ Wouthuysen}, Phys.\ Rev.\ {\bf 73},
  127  (1948).

\bibitem{Arley:1+Arley:2}
{N. Arley}, Proc.\ Roy.\ Soc. {\bf 168 A},  569  (1937);
{\em On the theory of stochastic processes and their application to
  cosmic rays} (John Wiley \& Sons, New York, 1931).

\bibitem{Heisenberg:1+Heisenberg:2+Heisenberg:3+Heisenberg:4}
{W.\ Heisenberg}, Z. Phys.\ {\bf 113},  61  (1939);
{W.\ Heisenberg et al.}, {\em {Vortr\"age \"uber Kosmische Strahlung}}
  (Springer Verlag, Berlin, 1943);
{W.\ Heisenberg}, Z. Phys.\ {\bf 126},  569  (1949);
Z. Phys.\ {\bf 133},  65  (1952).

\bibitem{Fubini}
{S. Fubini, D. Amati and A. Stanghellini}, Il Nuovo Cimento {\bf 26},  896
  (1962).

\bibitem{Wataghin}
{G. Wataghin}, Anais da Academia Brasileira de Ci\^encias  129  (1942).

\bibitem{Cosmic}
{P.K.\ MacKeown and A.W.\ Wolfendale}, Proc.\ Phys.\ Soc.\ {\bf 89},  553
  (1966).

\bibitem{Derrick:1972}
{M. Derrick et al.}, Phys.\ Rev.\ Lett.\ {\bf 29},  515  (1972).

\bibitem{AGCim:2+AGCim:5}
{A.\ Giovannini}, Lett. al Nuovo Cimento {\bf 6},  514  (1973);
Lett.\ al Nuovo Cimento {\bf 7},  35  (1973).

\bibitem{AGCim:3+AGCim:4}
{A.\ Giovannini et al.}, Il Nuovo Cimento {\bf 24A},  421  (1974);
{M.\ Garetto et al.}, Il Nuovo Cimento {\bf 38A},  38  (1977).

\bibitem{AGCim:1+AGCim:6}
{A.\ Giovannini}, Il Nuovo Cimento {\bf 10A},  713  (1972);
Il Nuovo Cimento {\bf 15A},  543  (1973).

\bibitem{UA5:4}
{G.J.\ Alner et al.\ (UA5 Collaboration)}, Phys.\ Lett.\ {\bf B160},  193
  (1985).

\bibitem{NA22}
{M.\ Adamus et al., NA22 Collaboration}, Phys.\ Lett.\ {\bf B177},  239
  (1986).

\bibitem{HRS:1}
{M.\ Derrick et al., HRS Collaboration}, Phys.\ Lett.\ {\bf B168},  299
  (1986).

\bibitem{ISR:1}
{A.~Breakstone et al.}, Il Nuovo Cimento {\bf 102A},  1199  (1989).

\bibitem{TASSO}
{W.\ Braunschweig et al., TASSO Collaboration}, Z. Phys.\ {\bf C45},  193
  (1989).

\bibitem{EMC}
{M.\ Arneodo et al., EMC Collaboration}, Z. Phys.\ {\bf C35},  335  (1987).

\bibitem{AGLVH:0}
{L.\ Van Hove and A.\ Giovannini},  in {\em XVII International Symposium on
  Multiparticle Dynamics}, edited by {M.\ Markitan et al} (World Scientific,
  Singapore, 1987), p.\ 561.

\bibitem{AGQCD}
{A.\ Giovannini}, Nucl.\ Phys.\ {\bf B161},  429  (1979).

\bibitem{KUV}
{K.\ Konishi, A.\ Ukawa and G.\ Veneziano}, Nucl.\ Phys.\ {\bf B157},  45
  (1979).

\bibitem{KittelNB}
{W.\ Kittel},  in {\em Workshop on Physics with Future Accelerators}, edited by
  {J.\ Mulvay} (CERN, Yellow Rep. 87-7, 1987), Vol.~II, p.\ 424.

\bibitem{AGLVH:2}
{L. Van Hove and A. Giovannini}, Acta Phys.\ Pol.\ B {\bf 19},  917  (1988).

\bibitem{UA5:3}
{R.E.\ Ansorge et al.\ (UA5 Collaboration)}, Z. Phys.\ C {\bf 43},  357
  (1989).

\bibitem{DEL:1}
{P.\ Abreu et al.\ (DELPHI Collaboration)}, Z. Phys.\ C {\bf 50},  185  (1991).

\bibitem{DEL:2}
{P.\ Abreu et al.\ (DELPHI Collaboration)}, Z. Phys.\ C {\bf 52},  271  (1991).

\bibitem{DEL:4}
{P.\ Abreu et al.\ (DELPHI Collaboration)}, Z. Phys.\ {\bf C56},  63  (1992).

\bibitem{Elba}
{A.\ Giovannini, S.\ Lupia and R.\ Ugoccioni}, Nucl.\ Phys.\ (Proc.\ Suppl.)
  {\bf B25},  115  (1992).

\bibitem{hqlett:2}
{A.~Giovannini, S.~Lupia and R.~Ugoccioni}, Phys.\ Lett.\ B {\bf 374},  231
  (1996).

\bibitem{RU:FB}
{A. Giovannini and R. Ugoccioni}, Phys.\ Rev.\ D {\bf 66},  034001  (2002).

\bibitem{rint:th}
S.~J. Lee and A.~Z. Mekjian, Nucl.\ Phys.\ A {\bf 730},  514  (2004);
H. Nakamura and R. Seki, Phys.\ Rev.\ C {\bf 66},  024902  (2002);
C.~E. Aguiar and T. Kodama, Physica A {\bf 320},  371  (2003);
A.~Z. Mekjian, Phys.\ Rev.\ C {\bf 65},  014907  (2002);
M. Martinis and V. Mikuta-Martinis, Fizika B {\bf 5},  269  (1996);
G. Calucci and D. Treleani, Phys.\ Rev.\ D {\bf 57},  602  (1998).

\bibitem{rint:nucl}
D. Ghosh, A. Deb, S. Bhattacharyya, and J. Ghosh, Phys.\ Rev.\ C {\bf 69},
  027901  (2004);
{M.M. Aggarwal et al.\ (WA98 Collaboration)}, Phys.\ Rev.\ C {\bf 65},  054912
  (2002);
T. Abbott {\it et~al.}, Phys.\ Rev.\ C {\bf 63},  064602  (2001);
A.~Z. Mekjian, B.~R. Schlei, and D. Strottman, Phys.\ Rev.\ C {\bf 58},  3627
  (1998).

\bibitem{rint:ee}
T. Osada, M. Maruyama, and F. Takagi, Phys.\ Rev.\ D {\bf 59},  014024  (1999);
D.-w. Huang, Phys.\ Rev.\ D {\bf 55},  5845  (1997);
P. Achard {\it et~al.}, Phys.\ Lett.\ B {\bf 577},  109  (2003);
P. Abreu {\it et~al.}, Phys.\ Lett.\ B {\bf 457},  368  (1999).

\bibitem{rint:other}
{K. Boruah}, Phys.\ Rev.\ D {\bf 56},  7376  (1997);
S.~V. Chekanov, Eur.\ Phys.\ J. C {\bf 6},  331  (1999);
S. Aid {\it et~al.}, Z. Phys.\ C {\bf 72},  573  (1996);
E. Kokoulina, Acta Phys.\ Pol.\ B {\bf 35},  295  (2004).

\bibitem{Koba}
Z. Koba,  in {\em 1973 CERN - JINR School of Physics} (CERN, Yellow Rep.\
  73-12, 1973), p.\ 171.

\bibitem{Sergio:thesis}
{S. Lupia}, Ph.D. thesis, University of Torino, 1995.

\bibitem{DeWolf:rep}
{E.A.~De Wolf, I.M.~Dremin and W.~Kittel}, Physics Reports {\bf 270},  1
  (1996).

\bibitem{correl}
{P. Carruthers}, Phys.\ Rev.\ {\bf A43},  2632  (1991).

\bibitem{Fug}
{C.\ Fuglesang},  in {\em Multiparticle Dynamics: Festschrift for L\'eon Van
  Hove}, edited by {A.\ Giovannini and W.\ Kittel} (World Scientific,
  Singapore, 1990), p.\ 193.

\bibitem{Dremin:1993}
{I.M. Dremin}, Phys.\ Lett.\ B {\bf 313},  209  (1993).

\bibitem{kendall}
{M.\ Kendall and A.\ Stuart}, {\em The advanced theory of statistics} (4th
  edition, Griffin \& Co.\ Lim., London).

\bibitem{kotz}
{N.L.~Johnson, S.~Kotz and A.W.~Kemp}, {\em Univariate discrete distributions}
  (2nd edition, J.~Wiley \& Sons, New York).

\bibitem{Feller}
{W. Feller}, {\em An introduction on probability theory and its applications}
  (Vol.\ I, 3rd ed., J. Wiley \& Sons).

\bibitem{Hegyi+Hegyi:1993}
{S. Hegyi}, Phys.\ Lett.\ {\bf B271},  214  (1992);
Phys.\ Lett.\ {\bf B318},  642  (1993).

\bibitem{DreminHwa2}
{I.M.~Dremin and R.C.~Hwa}, Phys.\ Rev.\ {\bf D49},  5805  (1994).

\bibitem{Peebles}
{P.J.E. Peebles}, {\em The large scale structure of the universe} (Princeton
  University Press, New York, 1980).

\bibitem{Void}
{S. Lupia, A. Giovannini and R. Ugoccioni}, Z. Phys.\ {\bf C59},  427  (1993).

\bibitem{Carrut}
{P. Carruthers and I. Sarcevic}, Phys.\ Rev.\ Lett.\ {\bf 63},  1562  (1989).

\bibitem{LVH:3}
{L. Van Hove}, Phys.\ Lett.\ B {\bf 242},  485  (1990).

\bibitem{Bjorken2}
{J.D.~Bjorken}, Phys.\ Rev.\ {\bf D47},  101  (1993).

\bibitem{pzero}
{S. Lupia, A. Giovannini and R. Ugoccioni}, Z. Phys.\ {\bf C66},  195  (1995).

\bibitem{Huang:1998gd}
D.-w. Huang, Phys.\ Rev.\ D {\bf 58},  017501  (1998).

\bibitem{pQCD}
{Yu.L. Dokshitzer, V.A. Khoze, A.H. Mueller and S.I. Troyan}, {\em Basics of
  perturbative QCD} (Editions Fronti\`eres, Gif-sur-Yvette, 1991).

\bibitem{Dremin:2000ep}
I.~M. Dremin and J.~W. Gary, Physics Reports {\bf 349},  301  (2001).

\bibitem{Khoze:1997dn}
V.~A. Khoze and W. Ochs, Int. J. Mod. Phys. {\bf A12},  2949  (1997).

\bibitem{KNO}
{Z.\ Koba, H.B.\ Nielsen and P.\ Olesen}, Nucl.\ Phys.\ {\bf B40},  317
  (1972).

\bibitem{Dokshitzer}
{Yu.L.~Dokshitzer}, Phys.\ Lett.\ {\bf B305},  295  (1993).

\bibitem{hqlett}
{R.~Ugoccioni, A.~Giovannini and S.~Lupia}, Phys.\ Lett.\ B {\bf 342},  387
  (1995).

\bibitem{SLD}
{K. Abe et al., SLD Collaboration}, Phys.\ Lett.\ {\bf B371},  149  (1996).

\bibitem{SPS}
{R.\ Ugoccioni and A.\ Giovannini}, Z. Phys.\ {\bf C53},  239  (1992).

\bibitem{GSPS}
{R. Ugoccioni, A. Giovannini and S. Lupia}, Z. Phys.\ {\bf C64},  453  (1994).

\bibitem{GSPS:2}
{A.~Giovannini, S.~Lupia and R.~Ugoccioni}, Z. Phys.\ {\bf C70},  291  (1996).

\bibitem{AGLVH:1}
{A. Giovannini and L. Van Hove}, Z. Phys.\ C {\bf 30},  391  (1986).

\bibitem{Single}
{F. Bianchi, A. Giovannini, S. Lupia and R. Ugoccioni}, Z. Phys.\ C {\bf 58},
  71  (1993).

\bibitem{AGLVH:3}
{A. Giovannini and L. Van Hove}, Acta Phys.\ Pol.\ B {\bf 19},  931  (1988).

\bibitem{LundModel}
{B. Andersson}, {\em The Lund Model} (Cambridge University Press, Cambridge,
  U.K., 1996).

\bibitem{Marchesini:1988cf}
{G. Marchesini and B.R. Webber}, Nucl.\ Phys.\ B {\bf 310},  461  (1988).

\bibitem{Becattini:1996if+Becattini:1997rv}
{F. Becattini}, Z. Phys.\ C {\bf 69},  485  (1996);
{F. Becattini and U.W. Heinz}, Z. Phys.\ C {\bf 76},  269  (1997).

\bibitem{LVH:1}
{L. Van Hove}, Physica {\bf 147A},  19  (1987).

\bibitem{Finkel}
{J. Finkelstein}, Phys.\ Rev.\ {\bf D37},  2446  (1989).

\bibitem{AGBecattini}
{F. Becattini, A. Giovannini and S. Lupia}, Z. Phys.\ C {\bf 72},  491  (1996).

\bibitem{Kittel:Fest}
{W.\ Kittel},  in {\em Multiparticle Dynamics: Festschrift for L\'eon Van
  Hove}, edited by {A.\ Giovannini and W.\ Kittel} (World Scientific,
  Singapore, 1990), p.\ 323.

\bibitem{CarrShih}
{P.~Carruthers and C.C.~Shih}, Int.\ J.\ Mod.\ Phys.\ {\bf A2},  1447  (1987).

\bibitem{UA5:5}
{G.J.\ Alner et al.\ (UA5 Collaboration)}, Phys.\ Lett.\ {\bf B160},  199
  (1985).

\bibitem{UA5:rep}
{G.J.\ Alner et al.\ (UA5 Collaboration)}, Physics Reports {\bf 154},  247
  (1987).

\bibitem{Singapore}
{L.\ Van Hove and A.\ Giovannini, invited talk},  in {\em 25th International
  Conference on High Energy Physics}, edited by {K.K.~Phua and Y.~Yamaguchi}
  (World Scientific, Singapore, 1991), p.\ 998.

\bibitem{Giacomelli}
{G.~Giacomelli and M.~Jacob}, Physics Reports {\bf 55},  1  (1979).

\bibitem{Schmitz}
{N.\ Schmitz},  in {\em Multiparticle Dynamics: Festschrift for L\'eon Van
  Hove}, edited by {A.\ Giovannini and W.\ Kittel} (World Scientific,
  Singapore, 1990), p.\ 25.

\bibitem{NA22:b}
{M.\ Adamus et al., NA22 Collaboration}, Phys.\ Lett.\ B {\bf B177},  239
  (1986);
Phys.\ Lett.\ {\bf B205},  401 (1988);
Z. Phys.\ {\bf C32},  475  (1986);
Z. Phys.\ {\bf C37},  215  (1988).

\bibitem{Jetset}
{T.~Sj\"ostrand and M.\ Bengstsson}, Computer Physics Commun.\ {\bf 82},  74
  (1994).

\bibitem{Dengler+Bailly:1988bx}
{F. Dengler et al.}, Z. Phys.\ C {\bf 33},  187  (1986);
{J.L. Bailly et al.\ (EHS-RCBC Collaboration)}, Z. Phys.\ C {\bf C40},  215
  (1988).

\bibitem{NA35}
{J. B\"achler at el., NA35 Collaboration}, Z. Phys.\ {\bf C57},  541  (1993).

\bibitem{Tannenbaum}
{M.~Tannenbaum},  in {\em XXIII International Symposium on Multiparticle
  Dynamics}, edited by {M.~Block and A.~White} (World Scientific, Singapore,
  1994), p.\ 261;
{M.~Tannenbaum}, Mod.\ Phys.\ Lett.\ {\bf A9},  89  (1994).

\bibitem{Gianini:1}
{I.M.~Dremin et al.}, Phys.\ Lett.\ {\bf B336},  119  (1994).

\bibitem{DEL:single}
{P. Abreu et al., DELPHI Collaboration}, Z. Phys.\ C {\bf 70},  179  (1996).

\bibitem{OPAL:FB}
{R.~Akers et al.\ (OPAL Collaboration)}, Phys.\ Lett.\ {\bf B320},  417
  (1994).

\bibitem{UA1}
{G. Ciapetti (UA1 Collaboration)},  in {\em 5th Topical Workshop on
  Proton-Antiproton Collider Physics}, edited by {M.~Greco} (World Scientific,
  Singapore, 1986), p.\ 488.

\bibitem{UA1:minijets}
{C.~Albajar et al., UA1 Collaboration}, Nucl.\ Phys.\ {\bf B309},  405  (1988).

\bibitem{NijmegenAG}
{A.~Giovannini, S.~Lupia e R.~Ugoccioni},  in {\em 7th International Workshop
  ``Correlations and Fluctuations''}, edited by {R.C. Hwa, W. Kittel, W.J.
  Metzger and D.J. Schotanus} (World Scientific, Singapore, 1997), p.\ 328.

\bibitem{combo:prd}
{A.~Giovannini and R.~Ugoccioni}, Phys.\ Rev.\ D {\bf 59},  094020  (1999).

\bibitem{combo:eta}
{A.~Giovannini and R.~Ugoccioni}, Phys.\ Rev.\ D {\bf 60},  074027  (1999).

\bibitem{walker}
{T. Alexopoulos et al. (E735 Coll.)}, Phys.\ Lett.\ {\bf B435},  453  (1998).

\bibitem{two-comp}
{J. Dias de Deus and R. Ugoccioni}, Phys.\ Lett.\ {\bf B469},  243  (1999).

\bibitem{Walker:p}
{S.G.~Matinyan and W.D.~Walker}, Phys.\ Rev.\ {\bf D59},  034022  (1998).

\bibitem{DelFabbro:2001rs}
A. Del~Fabbro and D. Treleani, Nucl. Phys. Proc. Suppl. {\bf 92},  130  (2001).

\bibitem{CDF:soft-hard}
{D. Acosta et al.\ (CDF Collaboration)}, Phys.\ Rev.\ D {\bf 65},  072005
  (2002).

\bibitem{RU:FBasymm}
{A. Giovannini and R. Ugoccioni}, J. Phys.\ G {\bf 28},  2811  (2002).

\bibitem{jdd:FB}
{J. Dias de Deus, J. Kwiecinski and M. Pimenta}, Phys.\ Lett.\ {\bf B202},  397
   (1988).

\bibitem{Carruthers:FB}
{P. Carruthers and C.C. Shih}, Phys.\ Lett.\ {\bf B165},  209  (1985).

\bibitem{RU:FBproblems}
{A. Giovannini and R. Ugoccioni}, Phys.\ Lett.\ B {\bf 558},  59  (2003).

\bibitem{Chou:FB+Chou:geometrical}
{T.T. Chou and C.N. Yang}, Phys.\ Lett.\ {\bf B135},  175  (1984);
Phys.\ Rev.\ D {\bf 32},  1692  (1985).

\bibitem{Gutay:plasma}
{T. Alexopoulos et al.}, Phys.\ Lett.\ B {\bf 528},  43  (2002).

\bibitem{Bialas:clanmass}
{A. Bia{\l}as and A. Szczerba}, Acta Phys.\ Pol.\ B {\bf 17},  1085  (1986).

\bibitem{RU:clanmass}
{A. Giovannini and R. Ugoccioni}, J. Phys.\ G {\bf 29},  777  (2003).

\bibitem{RU:NewPhysics}
{A. Giovannini and R. Ugoccioni}, Phys.\ Rev.\ D {\bf 68},  034009  (2003).

\bibitem{Bj:energy}
{J.D.~Bjorken}, Phys.\ Rev.\ {\bf D27},  140  (1983).

\bibitem{Pascal}
{B. Pascal}, {\em Varia Opera Mathmatica D. Petri de Fermat} (Tolossae,
  1679).

\bibitem{Carruthers:Pascal}
{P. Carruthers}, University of Arizona preprint No.~AZPH-TH-94-10.

\bibitem{Montmort}
{P.R. Montmort}, {\em Essay d'analyse sur le jeux de hazard} (Quillou, Paris,
  1713).

\bibitem{Yule}
{M. Greenwood and G.U. Yule}, J. Royal Statistical Society {\bf A
  83},  255  (1920).

\bibitem{Planck}
{M. Planck}, Sitzungber.\ Deutsch.\ Akad.\ Wiss.\ Berlin {\bf 33},  355
  (1923).

\bibitem{Mandel}
{L. Mandel}, Proc.\ Phys.\ Soc.\ London  233  (1959).

\bibitem{ChangChang}
{L.-N. Chang and N.-P. Chang}, Phys.\ Rev.\ D {\bf 9},  660  (1974).

\end{thebibliography}

\newpage

\section*{}
\addcontentsline{toc}{section}{Contents}
\tableofcontents

\end{document}